\documentclass{iopart}
\usepackage{graphicx,epsfig}
\usepackage{bm}
\usepackage{iopams}
\newcommand{\ba}{\begin{eqnarray}}
\newcommand{\ea}{\end{eqnarray}}
\newcommand{\bse}{\numparts}
\newcommand{\ese}{\endnumparts}
\newcommand{\ACal}{{\cal{A}}}

\newcommand{\DD}{{\cal {D}}}

\newcommand{\W}{{\cal {W}}}
\newcommand{\bbq}{\begin{quote}}
\newcommand{\eeq}{\end{quote}}
\newcommand{\RR}{{}^3{\cal{R}}}

\newcommand{\T}{{}^3{\cal{T}}}

\newcommand{\EE}{{\cal{E}}}
\newcommand{\FF}{{\cal{F}}}

\newcommand{\VV}{{\cal{V}}}
\newcommand{\Vp}{{\cal{V}}_{(p)}}
\newcommand{\Vq}{{\cal{V}}_{(q)}}
\newcommand{\HH}{{\cal{H}}}
\newcommand{\KK}{{\cal{K}}}

\newcommand{\QQ}{{\cal{Q}}}
\newcommand{\Q}{{\tt{Q}}}
\newcommand{\A}{{\tt{A}}} 
\newcommand{\mav}{\langle \rho\rangle}
\newcommand{\kav}{\langle \KK\rangle}
\newcommand{\HHav}{\langle\HH\rangle}

\newcommand{\Sav}{\langle S\rangle}

\newcommand{\hOm}{\Omega_q}

\newcommand{\qef}{{\textrm{q}}_{\tiny{\textrm{eff}}}}
\newcommand{\Ommef}{\Omega_{\tiny{\textrm{eff}}}^{(m)}}
\newcommand{\OmQef}{\Omega_{\tiny{\textrm{eff}}}^{({\cal{Q}})}}
\newcommand{\hq}{\hat{\tt{q}}}
\newcommand{\OmC}{\hat\Omega^{(\tt{\tiny{Q}})}}

\newcommand{\Oml}{\hat\Omega^{(m)}}

\newcommand{\Dim}{\delta_0^{(\rho)}}
\newcommand{\Dik}{\delta_0^{(\KK)}}
\newcommand{\Da}{\delta^{(S)}}

\newcommand{\Dh}{\delta^{(\HH)}}

\newcommand{\Dm}{\delta^{(\rho)}}

\newcommand{\Dk}{\delta^{(\KK)}}

\newcommand{\dd}{{\rm{d}}}
\newcommand{\tbb}{t_{\textrm{\tiny{bb}}}}

\newcommand{\tcoll}{t_{\textrm{\tiny{coll}}}}

\newcommand{\tmax}{t_{\textrm{\tiny{max}}}}

\newcommand{\rtv}{r_{\rm{tv}}}

\newcommand{\rmax}{r_{\textrm{\tiny{max}}}}
\newcommand{\Hty}{\HH_{_\infty}}
\newcommand{\mty}{\rho_{_\infty}}
\newcommand{\kty}{\KK_{_\infty}}
\newcommand{\hOmty}{\Omega_{q_{\infty}}}
\newcommand{\hqty}{\hat{\textrm{q}}_{_\infty}}

\newcommand{\Fty}{\FF_{_\infty}}
\newcommand{\rr}{\bar r}

\usepackage{pdfsync}

\begin{document}

\title[Back--reaction and effective acceleration in generic LTB dust models.]{Back--reaction and effective acceleration in generic LTB dust models.}


\author{Roberto A Sussman}
  \address{Instituto de Ciencias Nucleares, UNAM, AP 70--543, M\'exico DF, 04510, M\'exico}



%
\ead{sussman@nucleares.unam.mx}
\date{\today}

\begin{abstract}
We provide a thorough examination of the conditions for the existence of back--reaction and an ``effective'' acceleration (in the context of Buchert's averaging formalism) in regular generic spherically symmetric Lema\^\i tre--Tolman--Bondi (LTB) dust models. By considering arbitrary spherical comoving domains, we verify rigorously the fulfillment of these conditions expressed in terms of suitable scalar variables that are evaluated at the domains' boundaries. Effective deceleration necessarily occurs in all domains in: (a) the asymptotic radial range of models converging to a FLRW background, (b) the asymptotic time range of non--vacuum hyperbolic models, (c) LTB self--similar solutions and (d) near a simultaneous big bang. Accelerating domains are proven to exist in the following scenarios: (i) central vacuum regions, (ii) central (non--vacuum) density voids, (iii) the intermediate radial range of models converging to a FLRW background, (iv)  the asymptotic radial range of models converging to a Minkowski vacuum and (v) domains near and/or intersecting a non--simultaneous big bang. All these scenarios  occur in hyperbolic models with negative averaged and local spatial curvature, though scenarios (iv) and (v) are also possible in low density regions of a class of elliptic models in which local spatial curvature is negative but its average is positive. Rough numerical estimates between $-0.003$ and $-0.5$ were found for the effective deceleration parameter. While the existence of accelerating domains cannot be ruled out in models converging to an Einstein de Sitter background and in domains undergoing gravitational collapse, the conditions for this are very restrictive. The results obtained may provide important theoretical clues on the effects of back--reaction and averaging in more general non--spherical models.                
\end{abstract}



\section{Introduction.}

The dominant cosmological paradigm ( ``concordance'' or ``$\Lambda$--CDM'' model) provides an excellent fit to observational data provided an elusive ``dark energy'' source is assumed to dominate an accelerated late time cosmic evolution \cite{CMreview}. Alternative models have been proposed that challenge this paradigm under the assumption that cosmic dynamics follows from fully inhomogeneous solutions of Einstein's equations instead of perturbations on a FLRW background~\cite{celerier,wiltshire1,wiltshire2,buchrev,buchcar,clarmaar,wiltshireF,ellisR1,ellisR2}. Since observed cosmic structure is dominated by low density regions (voids), numerous empirical models of inhomogeneous voids have been proposed \cite{voids1,voids2,voids3,voids4} that can fit observational data without assuming the existence of dark energy (see specially more recent literature \cite{voids5,voids6,voids7}). Almost all examined void configurations are based on the Lema\^\i tre--Tolman--Bondi (LTB) models \cite{LTB_orig,kras1,kras2,KH1,KH2,KH3,KH4,ltbstuff,suss02}, which form a well known class of spherically symmetric exact solutions of Einstein's equations for a dust source. Although LTB models yield a very idealized description of cosmic voids, they have been quite useful to explore, as a rough first approximation, the effects of inhomogeneity in cosmic observations without the need to employ complicated numerical methods. As shown in \cite{bolsus}, the description of cosmic voids can be greatly enhanced by using non--spherical models.

An important line of theoretical work among the alternative proposals to the concordance model is to  consider  the possibility that different interpretations of observational data could emerge from a suitable averaging procedure applicable to inhomogeneous sources~\cite{celerier,wiltshire1,buchrev,buchert,buchlet,zala,colpel}. While a fully covariant averaging procedure acting on proper tensors has already been proposed by Zalaletdinov~\cite{zala,colpel}, the  foliation dependent averaging formalism developed by Buchert, being widely used in the literature~\cite{buchrev,buchert,buchlet}, is restricted to scalars. This may be, however, sufficient to address the effective evolution of cosmological parameters (see ~\cite{buchrev,buchcar2,marozzi}). However, devising a well posed averaging procedure in General Relativity is still  an open issue subjected to development and debate (see \cite{buchcar2,ParSin1,ParSin2,Par1,marozzi}). In Buchert's formalism the averaging of scalar evolution equations for inhomogeneous dust sources yields a variance term (the so--called ``kinematic back--reaction'') involving the squared fluctuations of the expansion and shear scalars. While this back--reaction term could effectively mimic the dynamical effect of a sort of dark energy source in the averaged Raychaudhuri equation, leading to an ``effective'' cosmic acceleration in this context, the interpretation of this effect in terms of actual observations is still a subject of debate  \cite{buchrev,buchlet,kolbetal,buchobs} (see \cite{Par2,ParSin4,Wald,ras1,zibin} for a critical view and \cite{antiWald} for a counter reply). 

Given their widespread use to describe cosmic voids, there is an extensive literature \cite{LTBave2,ras2,ras3,ras4,ParSin3,BolAnd,LTBave3,mattsson,sussBR,sussIU} that considers LTB models as natural candidates to explore the effects and predictions of Buchert's formalism. The following is a quick summary of this literature. Conditions for the existence of effective acceleration were given by Paranjape and Singh \cite{LTBave2} in the asymptotic late time evolution of hyperbolic models (as we prove in section 12, these conditions are incorrect). A simplified toy model was proposed by R\"as\"anen \cite{ras2,ras3} for the interpretation of back--reaction and effective acceleration in a cosmological perturbative scenario and in the context of gravitational collapse. This work was further improved and corrected by R\"as\"anen \cite{ras4} and Paranjape and Singh \cite{ParSin3}. Numeric estimations of the deceleration parameter associated with Buchert's formalism for particular models were presented by Bolejko and Andersson \cite{BolAnd} and Chuang {\it et al} \cite{LTBave3}.  More recently, the role of shear in estimating the magnitude of back--reaction was examined by Mattsson and Mattsson for central voids in hyperbolic models in \cite{mattsson}, while conditions on the sign of the back--reaction were given by Sussman for generic LTB models in \cite{sussBR} and on the effective acceleration in restricted situations in \cite{sussIU}. 

In the present article we provide a continuation and broad generalization of \cite{sussBR} and \cite{sussIU}. Hence, we  extend, enhance and generalize all previous literature,  since we examine by rigorous, qualitative and numeric arguments the fulfillment of sufficient conditions for an effective acceleration for domains in generic models, as opposed to looking at excessively simplified toy models (as in \cite{ras2,ras3,ParSin3}) or particular cases defined by observational constraints (as in \cite{BolAnd}) or suitable ansatzes (as in \cite{LTBave3,mattsson}). We consider a wide range of scenarios in regular hyperbolic and open elliptic LTB models: different comoving averaging domains (central region, intermediate radial ranges, radial and time asymptotic ranges, near an expanding or collapsing singularities). In particular, we examine the existence of effective acceleration in the asymptotic radial range of models converging to a vacuum Minkowskian state, which was not considered previously (notice that all models examined in previous articles \cite{ras2,ras3,ras4,ParSin3,BolAnd,LTBave3,mattsson} are radially asymptotic to a FLRW background). However, we do exclude from consideration parabolic models or regions containing a symmetry center (because back--reaction vanishes identically \cite{LTBave2,LTBave3}) and models whose space slices have, either spherical topology (``closed'' elliptic models), or lack symmetry centers (wormholes \cite{kras2} and LTB self--similar solutions \cite{kras2,carr}).

The section by section content of the article is summarized as follows. Generic features of LTB models are briefly listed in section 2. In particular, we highlight the fact that sections of Minkowski or Schwarzschild--Kruskal spacetimes can be formally considered as  vacuum particular cases of the models. We introduce in section 3 a scalar proper volume average functional acting on spherical comoving domains of space slices orthogonal to the 4--velocity. We also introduce an auxiliary ``weighed'' functional (the ``quasi--local'' average), together with the real functions associated to the average functionals, fluctuations and their mathematical properties. Buchert's dynamical equations and the sufficient conditions for the existence of a positive back--reaction and a negative effective deceleration parameter are given in section 4. Since Buchert's scalar averaging is defined by proper volume integrals along the rest frames of the 4--velocity, we use extensively various results of previously published work \cite{RadAs,RadProfs} that deals with the behavior of scalars (expansion, density and spatial curvature) along radial rays of space slices. We examine the sign of the back--reaction term in section 5, proving that it is positive in the radial asymptotic range for all hyperbolic models and for a wide class of elliptic models. In section 6 we provide qualitative guidelines describing several scenarios that should favor the existence of effective acceleration, identifying in section 7 specific situations in which direct rigorous proof of this existence (or non--existence) can be given readily without further assumptions (other than standard regularity).  Following these guidelines we examine the existence of accelerating domains for central non--vacuum density voids (section 8), in the asymptotic radial range of models converging to a section of Minkowski spacetime (``non--standard'' and Milne sections, section 9), in low density regions in the intermediate transitional radial range of density clump profiles (section 10), in domains near (or intersecting) the curvature singularities (section 11) and in the context of the ``spherical collapse'' model (section 12).  Table 4 provides a summary of these scenarios. In section 13 we prove that non--vacuum hyperbolic models effectively decelerate in their asymptotic time range and in section 14 we examine the relation between accelerating domains and the sign of the averaged and local spatial curvature. Section 15 provides a summary and a final discussion of our results, together with a brief discussion of their theoretical implications and connections with previous literature. 

In order to keep this article as self--contained and complete as possible, we have included three appendices containing necessary background and support material: analytic solutions and regularity conditions are given in Appendix A, while rigorous proof of various formal results introduced in the main text are given in Appendices B and C.

\section{LTB dust models.}

  Spherically symmetric inhomogeneous dust sources are described by the well known Lema\^\i tre--Tolman--Bondi metric in a comoving frame~\cite{LTB_orig} (see \cite{kras1,kras2} for comprehensive reviews)
\begin{equation}\dd s^2 = -\dd t^2+ \frac{R'{}^2}{\FF^2}\dd r^2+R^2\left(\dd\theta^2+\sin^2\theta\dd\phi^2\right).\label{ltb}\end{equation}
where $R=R(t,r)$,\, $R'=\partial R/\partial r,\,\FF=\FF(r)$ (we use geometric units $G=c=1$ and $r$ has length units). The field equations for (\ref{ltb}) and the dust energy--momentum tensor $T^{ab} =\rho\,u^au^b$ with rest matter--energy density $\rho=\rho(t,r)$ and $u^a=\delta^a_0$ reduce to
\ba \dot R^2 &=& \frac{2M}{R}+\FF^2-1,\label{eqR2t}\\
M' &=& 4\pi\,\rho\,R^2R',\label{eqrho1}\ea
where $M=M(r)$ and $\dot R=u^a\nabla_a R=\partial R/\partial t$. The sign of $\FF^2-1$ determines the zeroes of $\dot R$ and thus classifies LTB models in terms of the following kinematic classes: $ \FF^2 = 1$ (parabolic), $\FF^2 \geq 1$ (hyperbolic) and $\FF^2\leq 1$ (elliptic). Following another important classification we denote the models admitting symmetry centers by ``open'' and ``closed'', according to the topological equivalence class ($\mathbb{R}^3$ and  $\mathbb{S}^3$ respectively) of the hypersurfaces $\T[t]$, orthogonal to $u^a$ and marked by arbitrary constant $t$. 

Besides $\rho$ and $R$ given above, other covariant objects of LTB spacetimes are the expansion scalar $\Theta$, the Ricci scalar $\RR$ of the hypersurfaces $\T[t]$, the shear tensor $\sigma_{ab}$ and the electric Weyl tensor $E^{ab}$: 
\ba  \fl \Theta &=& \tilde\nabla_au^a=\frac{2\dot R}{R}+\frac{\dot R'}{R'},\qquad
\RR = \frac{2[(1-\FF^2)\,R]'}{R^2R'},\label{ThetaRR}\\
\fl \sigma_{ab} &=& \tilde\nabla_{(a}u_{b)}-(\Theta/3)h_{ab}=\Sigma\,\Xi_{ab},\qquad
E^{ab}=  u_cu_d C^{acbd}=\EE\,\Xi^{ab},\label{SigEE}\ea
where $h_{ab}=u_au_b+g_{ab}$,\, $\tilde\nabla_a = h_a^b\nabla_b$,\, and $C^{abcd}$ is the Weyl tensor, $\Xi^{ab}=h^{ab}-3\eta^a\eta^b$ with $\eta^a=\sqrt{h^{rr}}\delta^a_r$ being the unit radial vector orthogonal to $u^a$ and to the orbits of SO(3).  The scalars $\EE$ and $\Sigma$ in (\ref{SigEE}) are
\begin{equation}\Sigma = \frac{1}{3}\left[\frac{\dot R}{R}-\frac{\dot R'}{R'}\right],\qquad
\EE = -\frac{4\pi}{3}\rho+ \frac{M}{R^3}.\label{SigEE1}\end{equation}
Since the dynamics of LTB models can be fully characterized by the local covariant scalars $\{\rho,\,\Theta,\,\Sigma,\,\EE,\,\RR\}$, the evolution of these models can be completely determined by a ``fluid flow''  description of scalar evolution equations for these scalars as in \cite{1plus3} (see \cite{suss10a,zibin,dunsbyetal} for the LTB case).

We can formally include as LTB models the following two vacuum sub-cases that follow if the local rest mass matter--energy density vanishes:  
\begin{description}
\item[Vacuum-LTB models.] These are hyperbolic models that follow by taking $M=0$ and $\FF\geq 1$ in (\ref{eqR2t}), so that we have $\rho=0$ from (\ref{eqrho1}). This class of models are the solutions ``[$s_2$]'' in \cite{ltbstuff} (see also \cite{LTBave2}) and describe sections of Minkowski spacetime in non--standard coordinates. The choice $\FF^2 \propto 1+ r^2$ defines the particular case of the Milne Universe. We will denote by ``{\it vacuum LTB} '' all Minkowski sections that are different from (and generalize) the Milne Universe.
\item[Schwarzschild--Kruskal models.] If we take $M=M_{\textrm{\tiny{schw}}}=$ constant and $\FF$ arbitrary in (\ref{eqR2t}), then $\rho=0$ follows from (\ref{eqrho1}), though a solution of this equation can also be given in which density is a formal Dirac delta distribution of a point mass associated with the ``Schwarzschild mass'' $M_{\textrm{\tiny{schw}}}$. These models are representations of the Schwarzschild--Kruskal manifold in comoving coordinates constructed with its radial timelike geodesics (since (\ref{eqR2t}) with $M=M_{\textrm{\tiny{schw}}}$ is the equation of these geodesics). The parabolic and elliptic cases are respectively known as Lema\^\i tre and Novikov coordinate representation (see page 203 of \cite{kras2}).         
\end{description}
Notice that all covariant kinematic quantities (\ref{ThetaRR})--(\ref{SigEE1}) are well defined for the vacuum models. These quantities correspond to a fluid flow description of sections of Minkowski and Schwarzschild--Kruskal spacetimes defined by a 4--velocity field associated by specific classes of test observers (not dust layers).

In general, LTB models can contain regions with different kinematics (parabolic, hyperbolic or elliptic), and can also be matched (along a comoving boundary) to regions of the vacuum cases above (see \cite{ltbstuff} for a summary of these ``mixed'' configurations). The analytic solutions of (\ref{eqR2t}) and regularity conditions for the models in general are summarized in Appendix A.

\section{Proper volume average and quasi--local average.}

\begin{figure}[htbp]
\includegraphics[width=3.5in]{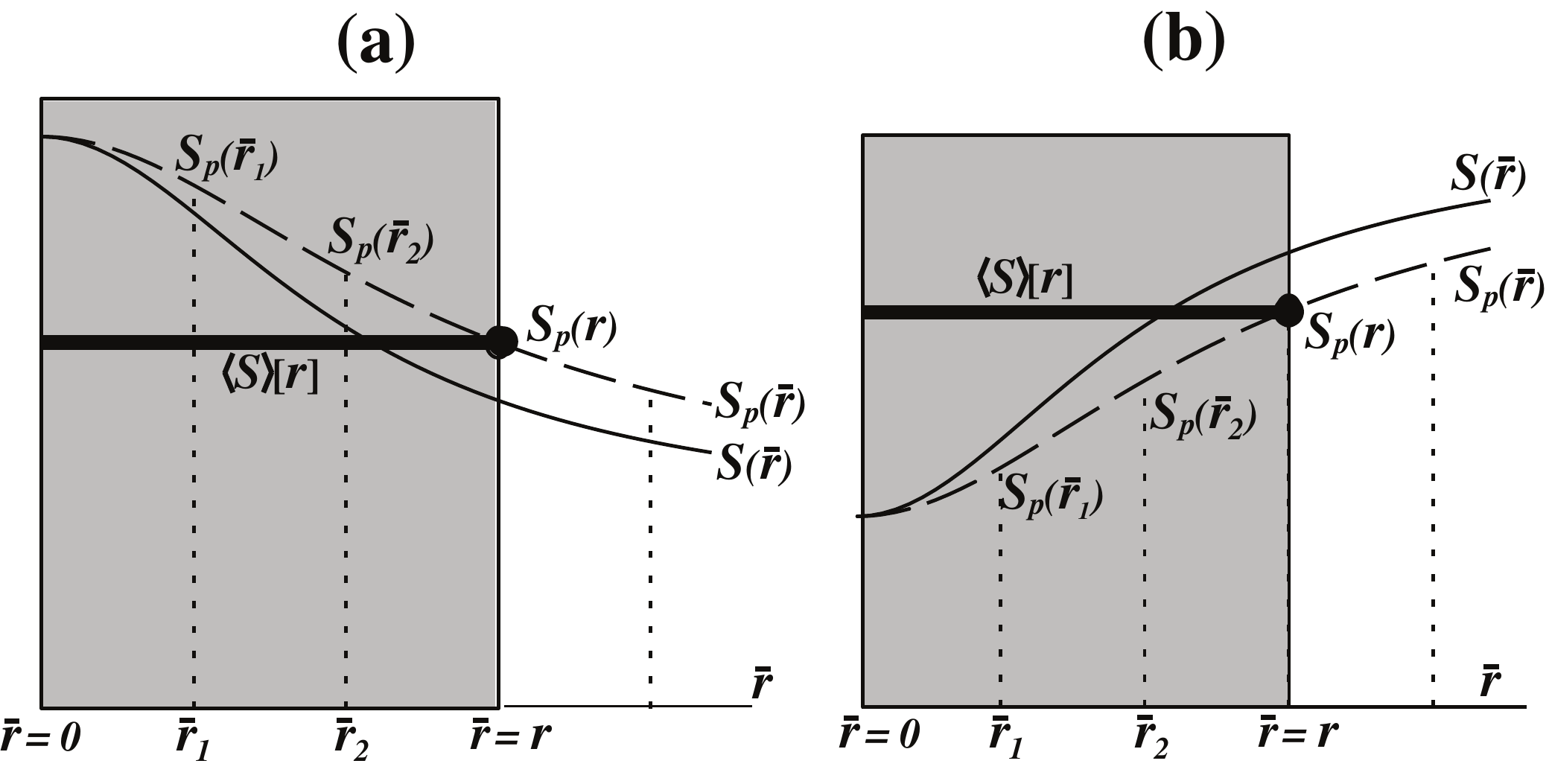}
\caption{{\bf The difference between $S_p$ and $\Sav_p$.} The figure displays the radial profile of a scalar function $S(\rr)$ (solid curve) along a regular hypersurface $\T(t)$,  together with its dual p--function $S_p(\rr)$ (dashed curve) defined by (\ref{pqfun}). Panels (a) and (b) respectively display the cases when $S'\leq 0$ (``clump'') and $S'\geq 0$ (``void''). The average functional (\ref{avedef}) assigns the real number $\Sav[r]$ to the full domain (shaded area) marked by $\vartheta[r]=\{\rr\,|\, 0\leq \rr\leq r\}$, whereas the function $S_p$ varies along this domain. Hence, $S_p$ and $\Sav_p$ are only equal at the domain boundary $\rr=r$ (for every domain), and so they satisfy the same differentiation rules (\ref{Apdot})--(\ref{Apr}) locally, {\it i.e.} $\dot S_p(r)=\Sav\,\dot{}_p[r]$ and $S_p'(r)=\Sav'_p[r]$, but behave differently when integrated along any domain. Notice, from (\ref{propp}) and (\ref{Apr}), that if $S'\leq 0$ in all $\vartheta[r]$ then $S-\Sav_p\leq 0$ and the opposite situation occurs if $S'\geq 0$. This figure also applies for the quasi--local functions and averages.}
\label{f1}
\end{figure}

The hypersurfaces $\T[t]$ orthogonal to a geodesic 4--velocity provide a natural and covariant time slicing for the LTB models. Let $X(\DD[r])$ be the set of all scalar functions in any compact spherical comoving domain $\DD[r]=\mathbb{S}^2\times\vartheta[r]\subset \T[t]$, where $\mathbb{S}^2$ is the unit 2--sphere, $t$ is an arbitrary constant parameter and $\vartheta[r]\equiv \{\rr\,|\,0\leq \rr \leq r\}$ is a semi--open set of a radial ray (spacelike geodesic) with $\rr=0$ marking a symmetry center~\footnote{Our notation reflects the fact that spherical domains are characterized by a radial range and are univocally labeled by the comoving radius of the boundary $\rr =r$. This notation is easily compared with the standard notation: $\langle S\rangle[r]=\langle S\rangle_{\DD[r]}$. We consider domains not enclosing a symmetry center in sections 10 and 11.}. We define:
\begin{description}
\item[The proper volume average (``p--average'')] is the linear functional $\langle\hskip 0.2 cm\rangle_p[r] :\,X(\DD[r])\to \mathbb{R}$ that assigns to any $S\in X(\DD[r])$ the real number 
\begin{equation}\Sav_p[r]=\frac{\int_{\DD[r]}{S\dd \Vp}}{\int_{\DD[r]}{\dd \Vp}}=\frac{\int_0^r{S\,\FF^{-1}R^2R'\dd\rr}}{\int_0^r{\FF^{-1}R^2R'\dd\rr}},\label{avedef}\end{equation}
where $\dd \Vp=\sqrt{{\rm{det}}(h_{ab})}\,\dd^3x=\FF^{-1}R^2R'\sin\theta\,\dd r\dd\theta\dd\phi$ is the proper volume element and $\int_0^r{..\dd\rr}=\int_{\rr=0}^{\rr=r}{..\dd\rr}$.  
\item[The quasi--local volume average (``q--average'')] is defined in a similar way as (\ref{avedef}):  the linear functional $\langle\hskip 0.2 cm\rangle_q[r] :\,X(\DD[r])\to \mathbb{R}$ that assigns to any $S\in X(\DD[r])$ the real number 
\begin{equation}\Sav_q[r]=\frac{\int_{\DD[r]}{S\FF\dd \Vp}}{\int_{\DD[r]}{ \FF\dd\Vp}}=\frac{\int_0^r{S\,R^2R'\dd\rr}}{\int_0^r{R^2R'\dd\rr}}.\label{aveqdef}\end{equation}
\item[p and q functions] The functionals (\ref{avedef}) and (\ref{aveqdef}) can be used as correspondence rules of functions whose argument is the radial coordinate marking (for an arbitrary domain) the domain boundary $r$. We define the ``p--functions'' and ``q--functions'' as the real valued local functions $S_p:\mathbb{R}^+\to\mathbb{R}$ and $S_q:\mathbb{R}^+\to\mathbb{R}$ such that for all $z\geq 0$   
\begin{equation}S_p(z)=\Sav_p[z],\qquad S_q(z)=\Sav_q[z].\label{pqfun}\end{equation}
To simplify notation, and whenever there is no risk of confusion, we will use the symbol $r$ as argument of these functions. The difference between the average functionals (\ref{avedef}) and (\ref{aveqdef}) and their associated functions (\ref{pqfun}) is illustrated in figure \ref{f1} (see also \cite{sussBR,sussIU}).
\end{description}
It is straightforward to show directly from (\ref{avedef}) and (\ref{aveqdef}) that $S_p$ and $S_q$ satisfy the following properties (valid also for averages by replacing $S_p$ and $S_q$ below with $\Sav_p$ and $\Sav_q$):
\bse\ba \fl S(r)-S_p(r)=\frac{1}{\Vp(r)}\,\int_0^r{S'\,\Vp \dd\rr},\qquad S(r)-S_q(r)=\frac{1}{\Vq(r)}\,\int_0^r{S'\,\Vq \dd\rr},\label{propp}\\
\fl \dot S_p=[S_p]\,\dot{}=(\dot S)_p+(\Theta S)_p-\Theta_p\,S_p,\qquad \dot S_q=[S_q]\,\dot{}=(\dot S)_q+(\Theta S)_q-\Theta_q\,S_q,\label{Apdot}\\
\fl S_p' =[S_p]'= \frac{\Vp'}{\Vp}\,\left[S-S_p\right],\qquad S_q' = [S_q]'=\frac{\Vq'}{\Vq}\,\left[S-S_q\right],\label{Apr}\\
\fl \frac{\Vp'}{\Vp} =\frac{3R'}{R}\,\frac{\FF_p}{\FF},\qquad \frac{\Vq'}{\Vq}=\frac{3R'}{R},\qquad  \frac{\Vq(r)}{\Vp(r)}=\FF_p(r)\label{VVr}\ea\ese
where $\FF_p$ is the p--function associated with the scalar $\FF$ and the proper volume $\Vp$ and ``quasi--local'' volume $\Vq$ are given by
\ba\Vp(r) = \int_{\DD[r]}{\dd \Vp} = 4\pi\int_0^r{\FF^{-1}R^2\,R'\,\dd\rr},\label{Vpdef}\\
\Vq(r) = \int_{\DD[r]}{\FF\,\dd \Vp} = 4\pi\int_0^r{R^2\,R'\,\dd\rr} =\frac{4\pi}{3}R^3(r).\label{Vqdef}\ea
While functions and functionals satisfy the same derivation laws, they behave differently under integration (see figure 1): only the p and q functionals satisfy $\langle\Sav[r]\rangle[r] = \Sav[r]$, and as a consequence, they can be considered as average distributions complying with the variance and covariance moment definitions for continuous random variables~\cite{sussBR,sussIU}.

Applying (\ref{aveqdef}) to the scalars $\rho,\,\Theta$ and $\RR$ in (\ref{eqR2t}),\, (\ref{eqrho1}) and (\ref{ThetaRR}) we obtain the following scaling laws in closed analytic form (see Appendix A):
\begin{equation} \frac{4\pi}{3} \rho_q = \frac{M}{R^3},\qquad  \KK_q =\frac{1-\FF^2}{R^2}.\label{qvars}\end{equation}
\begin{equation} \HH_q =\frac{\dot R}{R},\qquad \HH_q^2 = \frac{8\pi}{3} \rho_q -\KK_q,\label{Hqsq}\end{equation}
where we are using (and will use henceforth) the notation 
\begin{equation} \HH\equiv \frac{\Theta}{3},\qquad \KK\equiv \frac{\RR}{6}.\label{mkH}\end{equation}
so that $\HH_q=\Theta_q/3$ and $\KK_q=\RR_q/6$ hold.
By applying (\ref{aveqdef}) to (\ref{SigEE1}) the remaining covariant scalars $\Sigma$ and $\EE$, associated with the shear and electric Weyl tensor, can be expressed as deviations or fluctuations of $\rho$ and $\Theta$ with respect to their q--duals:
\begin{equation} \Sigma = -\left[\HH-\HH_q\right],\qquad
\EE = -\frac{4\pi}{3}\left[\rho-\rho_q\right].\label{SE2}\end{equation}
Given the pair of scalars $\{S,\,S_q\}$, it is very useful to define their relative fluctuation as
\begin{equation} \Da \equiv \frac{S-S_q}{S_q}=\frac{S'_q/S_q}{3R'/R} =\frac{1}{S_q(r) R^3(r)}\int_0^r{S'\,R^3\,\dd\rr},\label{Dadef}\end{equation}
where we used the properties (\ref{propp}) and (\ref{Apr}). Radial differentiation of $\HH_q^2$ in (\ref{Hqsq}) and using (\ref{Apr}) and (\ref{Dadef}) to eliminate radial gradients of $\HH_q,\,\rho_q,\,\KK_q$ in terms of $\Dh,\,\Dm,\,\Dk$, yields the following useful constraint among the fluctuations: 
\begin{equation}  2\Dh = \hOm\,\Dm+(1-\hOm)\,\Dk.\label{slaw_Dh}\end{equation}
with $\hOm$ and $1-\hOm$ given by 
\begin{equation} \hOm \equiv \frac{8\pi\rho_q}{3\HH_q^2},\qquad 1-\hOm=-\frac{\KK_q}{\HH_q^2}.\label{Omdef}\end{equation}
where we remark that $\Omega_q$ is also a q--scalar for being a function of the q--scalars $\rho_q$ and $\HH_q$ (see Appendix C of \cite{szeknew}). The scalar $\hOm$  is analogous to the FLRW Omega parameter, since its value depends on the kinematic class:  $0<\hOm<1$ for hyperbolic models, $\hOm>1$ for elliptic models and $\hOm=1$ for parabolic models.

\section{Buchert's scalar averaging formalism.}

Buchert's equations \cite{buchrev,buchert} applied to LTB models (see previous literature \cite{ras2,sussBR,ras3,LTBave2,sussIU}) are the proper volume average (p--average) of the 1+3 energy balance equation, Raychaudhuri equation and Hamiltonian constraint 
\footnote{To simplify notation the subscript ${}_p$ and the domain indicator $[r]$ will be henceforth omitted unless they are needed for clarity. However, we will keep the notation $\QQ[r]$ for the back--reaction term.}
\bse\ba \langle\dot \rho+3\,\rho\,\HH\rangle &=& \mav\,\dot{}+3\mav\HHav=0,\label{ave_ev_rho}\\ 
\HHav\,\dot{}+ \HHav^2&=&-\frac{4\pi}{3}\mav+2\QQ[r],\label{ave_Raych}\\
\HHav^2 &=& \frac{8\pi}{3}\mav -\kav-\QQ[r],\label{ave_Fried}\ea\ese
plus the integrability condition between (\ref{ave_Raych}) and (\ref{ave_Fried})
\begin{equation} \dot\QQ[r]+6\HHav\QQ[r]+2\HHav\kav+\kav\dot{}=0,\end{equation}
where $\QQ[r]$ is the kinematic ``back--reaction'' term given by 
\begin{equation}\QQ[r] \equiv \langle(\HH-\HHav)^2-(\HH-\HH_q)^2\rangle,\label{QQ}\end{equation}
and we have used (\ref{SE2}) to eliminate $\sigma_{ab}\sigma^{ab}=6\Sigma^2$ in terms of $(\HH-\HH_q)^2$. Equation (\ref{ave_ev_rho}) simply expresses the compatibility between the averaging (\ref{avedef}) and the conservation of rest mass, but (\ref{ave_Raych}) and (\ref{ave_Fried}) lead to an interesting re--interpretation of cosmic dynamics because of the presence of $\QQ[r]$. 
From the Raychaudhuri equation (\ref{ave_Raych}), we have the following definition for an ``effective'' cosmic acceleration  
\begin{equation}\ACal \equiv \HHav\,\dot{}+ \HHav^2= -\frac{4\pi}{3}\mav +2\QQ[r],\label{efeacc}\end{equation}
together with (as is customary in the literature \cite{buchrev,buchert,buchlet}) a definition of an ``effective'' dimensionless deceleration parameter:
\footnote{Since $3\HHav= \dot\Vp/\Vp$ with $\Vp$ defined by (\ref{Vpdef}), a scale factor associated with an averaging domain $\DD$ is often introduced \cite{buchrev,buchert} as  $a_\DD=\Vp^{1/3}$, hence we can formally write: $\HHav=\dot a_\DD/a_\DD$ and $\ACal = \ddot a_\DD/a_\DD$. While this scale factor is a theoretically appealing definition, it will not be used here, as it is not helpful in practice to provide analytic conditions for the sign of the effective acceleration.}
\begin{equation} \fl \qef \equiv -\frac{\ACal}{\HHav^2}=\frac{1}{2}\Ommef-\OmQef,\qquad \Ommef\equiv \frac{8\pi\mav}{3\HHav^2},\quad \OmQef=\frac{2\QQ[r]}{\HHav^2},\label{qef}\end{equation}
which (bearing in mind (\ref{mkH}) and (\ref{efeacc})) mimics the local deceleration parameter in FLRW dust cosmologies, with $\OmQef$ playing the kinematic role analogous to that of the Omega factor of a dark energy source. Therefore, considering (\ref{efeacc}) and (\ref{qef}), we shall adopt the following conventions to hold at any domain $\vartheta[r]$:
\ba \hbox{Effective acceleration:}\qquad \ACal>0\quad\hbox{or}\quad\qef<0,\label{effe_acc}\\\hbox{Effective deceleration:}\qquad\ACal<0\quad\hbox{or}\quad\qef>0.\label{effe_dec}\ea  
As a consequence, if (\ref{effe_acc}) holds then Buchert's dynamical equations (\ref{ave_ev_rho})--(\ref{ave_Fried}), together with (\ref{efeacc}) and (\ref{qef}), are qualitatively analogous to the dynamical equations of a $\Lambda$--CDM model in which $\QQ[r]$ is effectively mimicking the effect of a dark energy source.

\subsection{Sufficient conditions for back--reaction and effective acceleration.} 
 
Our task in this article is to examine for generic LTB models the fulfillment of condition (\ref{effe_acc}), by looking at the sign of either $\ACal$ or $\qef$. Unfortunately, without resorting to numerical methods it is practically impossible to evaluate $\ACal$ and $\qef$, as both quantities  involve integrals of the type (\ref{avedef}) whose integrands are (in general) not given by closed analytic functions of $(t,r)$. In order to proceed without having to evaluate the average integrals (\ref{avedef}), we remark that the signs of $\ACal$ or $\qef$ in a given domain $\vartheta[r]$ can be inferred by looking at the sign of the scalars inside the $\langle ...\rangle$ brackets. This can be achieved by using the following property that is valid for every scalar function $S$ and provides sufficient conditions on the sign of $\Sav$:
\begin{equation}S(\rr)\geq 0\,\,\,\forall\,\, \rr\in\vartheta[r]\,\,\Rightarrow\,\, \Sav[r]\geq 0,\label{Avpos} \end{equation}
where we notice that the converse is not necessarily true. However, condition (\ref{Avpos}) is still too strong, as it requires a given sign to hold in every point of the domain $\vartheta[r]$. In particular, the scalar inside the $\langle ...\rangle$ brackets in (\ref{QQ}) depends on points inside the domain ($\rr$) and on the boundary ($r$), which makes it very hard to apply (\ref{Avpos}). However, considering that $\HHav[r]=\HH_p(r)$ holds for every $r$, we can use the following identity (proved in Appendix B, see also \cite{sussBR,sussIU})
\begin{equation} \langle(\HH(\rr)-\HHav[r])^2\rangle[r]=\langle(\HH(\rr)-\HH_p(\rr))^2\rangle[r],\label{HHavp}\end{equation}
which holds for every domain $\vartheta[r]$, to express the sign condition (via (\ref{Avpos})) on $\QQ[r]$ {\it only} in terms of the parameter $\rr$ (which can take the value $\rr=r$). A sufficient condition for $\QQ[r]\geq 0$ (which is necessary for (\ref{efeacc})) follows by applying (\ref{Avpos}) and (\ref{HHavp}) to (\ref{QQ})
\begin{equation}\fl \Q(r)\geq 0\quad\Rightarrow\quad \QQ[r]\geq 0,\quad\hbox{with:}\quad \Q(r) \equiv [\HH(r)-\HH_p(r)]^2-[\HH(r)-\HH_q(r)]^2,\label{Cpos}\end{equation}
where $\Q(r)$ is evaluated at the domain boundary $\rr=r$. 

Since we assume that $\rho\geq 0$ and $\HHav^2\geq 0$ must hold for every domain, the sufficient condition for the fulfillment of (\ref{effe_acc}) in any domain is simply
\begin{equation}\fl \A\geq 0\quad\Rightarrow\quad \ACal\geq 0,\qquad\hbox{with:}\quad  \A(r) \equiv 2\Q(r)-\frac{4\pi}{3}\rho(r),\label{Apos}\end{equation} 
where (again) the converse implication does not hold, the scalars $\A,\,\Q,\,\rho$ are evaluated at the domains' boundary $\rr=r$ and we have used (\ref{HHavp}), (\ref{Cpos}), (\ref{Apos}) and $\HHav[r]=\HH_p(r)$. Sufficient conditions for an effective acceleration can also be given by a dimensionless deceleration parameter obtained by normalizing (\ref{Apos}) with $\HH_q^2=(\dot R/R)^2$ in (\ref{Hqsq}), which (as $\HH$) is analogous to a Hubble factor. This leads  (from (\ref{Avpos})) to the following sufficient (not necessary) condition to fulfill (\ref{effe_acc}):
\begin{equation} \fl \hq\leq 0\quad \Rightarrow\quad \qef\leq 0,\qquad \hbox{with:}\quad\hq\equiv -\frac{\A}{\HH_q^2}=  \frac{1}{2}\,\Oml-\OmC, \label{newApos}\end{equation}
where  $\OmC$ and $\Oml$ are analogous to $\OmQef$ and $\Ommef$ in (\ref{qef}): 
\ba \OmC &\equiv& \frac{2\Q}{\HH_q^2}=2\,\left[1-\frac{\HH_p}{\HH_q}\right]\left[1-\frac{\HH_p}{\HH_q}+\hOm\,\Dm+\left(1-\hOm\right)\,\Dk\right],\label{OmC}\\
 \Oml &\equiv& \frac{8\pi\rho}{3\HH_q^2}= \hOm\,\left(1+\Dm\right),\label{Om}\ea
with $\hOm$ given by (\ref{Omdef}), and we have used (\ref{slaw_Dh}) and $\rho=\rho_q(1+\Dm)$. Considering that, because of (\ref{Avpos}) and (\ref{HHavp}), sufficient conditions for an effective acceleration can be given by evaluating the scalars in (\ref{Apos}), (\ref{newApos}), (\ref{OmC}) and (\ref{Om}) at the boundary of a domain $\vartheta[r]$, it is far less complicated to verify the fulfillment of $\A\geq 0$ or $\hq\leq 0$ than the fulfillment of $\ACal\geq 0$ or $\qef\leq 0$ from (\ref{effe_acc}).

\section{Probing the sign of the back--reaction term.}

It is useful to look first at condition (\ref{Cpos}) because it is necessary (but not sufficient) for (\ref{Apos}) (which is, in turn, sufficient for (\ref{effe_acc})). Once we identify domains and models violating (\ref{Cpos}), so that $\QQ[r]\leq 0$ holds, we can state rigorously that these domains exhibit effective deceleration (\ref{effe_dec}).  The fulfillment of condition (\ref{Cpos}) is examined for domains in parabolic, hyperbolic and elliptic models separately below (see section 12 for the mixed elliptic/hyperbolic case). We provide a summary of results in table \ref{tabla1}.

\subsection{Domains in parabolic models and self--similar solutions.}

The first rigorous result concerning condition (\ref{Cpos}) is that $\Q(r)=0$ (and thus $\QQ[r]=0$) holds identically for all domains $\vartheta[r]$ in parabolic LTB models or regions containing a symmetry center (see table \ref{tabla1}). The proof is trivial \cite{LTBave2,LTBave3}, since $\FF=1$ and $\HH_p(r)=\HH_q(r)$ hold in these domains for all $r$.

Notice that $\Q(r)$ (and thus $\QQ[r]$) does not vanish in domains containing the ``external'' parabolic region of a mixed elliptic--parabolic model (see sections 12, 14 and \cite{sussBR}). We also remark that (in general) $\FF\ne 1$ holds in domains of the vacuum LTB models and of Schwarzschild--Kruskal models that are not parabolic, hence $\Q(r)$ is nonzero in these cases. 

LTB self--similar solutions comply with $\FF=\FF_0$, where $\FF_0>1$ or $\FF_0<1$ for the hyperbolic and elliptic cases \cite{carr} (see also Appendix C4 of \cite{RadAs}). From (\ref{ThetaRR}) and (\ref{qvars}) we have $\KK$ and $\KK_q$ proportional to $R^{-2}$. Although these solutions are not compatible with the existence of a center worldline, and thus  comoving domains like $\DD=\mathbb{S}^2\times \vartheta[r]$ are not compact, functions of the type $S_p$ and $S_q$ can still be  constructed as in (\ref{avedef}) and (\ref{aveqdef}) by means of improper integrals. As with the parabolic case, it is evident that $\FF=\FF_0\ne 1$ also implies that $\HH_p(r)=\HH_q(r)$ must hold for all $r$, so that $\Q(r)=0$ (and thus $\QQ[r]=0$)) hold for all domains in all slices $\T[t]$. However, while self--similarity implies $\FF=\FF_0$, the converse implication is not true: as long as $M$ (and thus $\rho_q$) remains a free function, the condition $\FF=\FF_0$  defines a wider class of regular models (which includes self--similar solutions) for which $\QQ[r]=0$ holds in every domain \cite{buchfocus}.  

\subsection{Domains in hyperbolic and elliptic models.} 

For the remaining of this article we will only consider domains in hyperbolic and elliptic models and regions and, unless specifically stated otherwise, we will assume henceforth that $\FF$ is not constant. Since the fulfillment of (\ref{Cpos}) in these models is strongly dependent on the monotonicity of the radial profiles of $\HH$ and $\FF$, we introduce the notion of  ``{\bf Turning Value}'' of a scalar $S$ (TV of $S$) as a point $\rr=\rtv\in \vartheta[r]$ such that $S'(\rtv)=0$. As a consequence, $S$ is monotonous in $\vartheta[r]$ if there is no TV of $S$ in this domain. Notice that the radial coordinate of a TV of $\FF=\FF(r)$ is fixed for all slices $\T[t]$, while the coordinate of a TV of the time depending scalar $\HH$ changes from slice to slice (see \cite{RadProfs}).

The regularity conditions (\ref{noshxGh}) for hyperbolic models require $\FF$ to be monotonous and $\FF'\geq 0$ to hold for all $r$ (see (\ref{FFr}) further ahead), hence these models admit radial profiles with either a monotonous $\HH$ or with a TV of $\HH$. In elliptic models the regularity conditions do not require a monotonous $\FF$ (see Appendix A), hence the slices $\T[t]$ can have TV's of $\FF$ and/or $\HH$.  Consider the following lemmas whose proof is given in \cite{sussBR} (see also \cite{sussIU}): \\

\noindent {\bf{Lemma 1}}. Let $\vartheta[r]$ be an arbitrary radial domain in a slice $\T[t]$ of a regular hyperbolic LTB model (including the vacuum LTB models, see section 2): 
\begin{description}
\item (a) If $\HH$ is monotonous in $\vartheta[r]$, then $\QQ[r]\geq 0$ holds for all domains $\vartheta[r_k]$ with $r_k\leq r$, including an asymptotic domain covering the whole slice.
\item (b) If there is a TV of $\HH$ at $r=\rtv$ in a slice $T[t]$, then values $r_1<\rtv$ and $r_2>\rtv$ always exist such that $\QQ[r]\geq 0$ holds for all $\vartheta[r]$ with $r<r_1$ {\bf {and}} $r>r_2$
\end{description}  
\noindent {\bf{Lemma 2}}. Consider domains $\vartheta[r]$ in arbitrary space slices $\T[t]$ of regular open elliptic LTB models: 
\begin{description}
\item (a) If $\HH$ and $\FF$ are monotonous in $\vartheta[r]$, then $\QQ[r]\leq 0$ holds for all domains $\vartheta[r_k]$ with $r_k\leq r$, including an asymptotic domain covering the whole slice.
\item(b) If there is, either a TV of $\FF$ at $\rr=y$ with monotonous $\HH$, or TVs of $\HH$ and $\FF$ at $\rr=\rtv$ and $\rr=y$, then a value $z>\hbox{max}(\rtv,y)$ always exists such that $\QQ[r]\geq 0$ holds for all $\vartheta[r]$ with $r>z$.
\item(c) If there is a TV of $\HH$ at $r=\rtv$ but $\FF$ is monotonous, then values $r_1<\textrm{min}(y,\rtv)$ and $r_2>\textrm{max}(y,\rtv)$ always exist such that $\QQ[r]\geq 0$ holds for all $\vartheta[r]$ with $r_1<r<r_2$. 
\end{description}    
The full proof of these lemmas is provided in \cite{sussBR}
\footnote{The proof of the case with TV's of $\FF$ and $\HH$ in \cite{sussBR} (proposition 7) has an error: since we need to consider domains for which $r>\hbox{max}(y,\rtv)$ holds, then the possibility that $\Q(r)>0$ in the range $r<\hbox{min}(y,\rtv)$ can be ruled out.}
and is based on using (\ref{propp}) and (\ref{VVr}) to express $\Q(r)$ in (\ref{Cpos}) as $\Q(r)=\Phi(r)\Psi(r)$ with:
\ba \fl \Phi(r)=\HH_q(r)-\HH_p(r)=\int_0^r{\HH'(\rr)\frac{\Vp(\rr)}{\Vp(r)}\left[1-\frac{\FF_p(\rr)}{\FF_p(r)}\right]\,\dd\rr},\label{Phi}\\
\fl \Psi(r)=2\HH(r)-\HH_q(r)-\HH_p(r)=\int_0^r{\HH'(\rr)\frac{\Vp(\rr)}{\Vp(r)}\left[1+\frac{\FF_p(\rr)}{\FF_p(r)}\right]\,\dd\rr},\label{Psi}\ea  
which relates the sign of $\Q(r)$ to the monotonicity of $\FF_p$ (which depends on the monotonicity of $\FF$) in the radial range $\vartheta[r]$. 
 
Regularity conditions (see Appendix A) imply that all vacuum LTB models correspond to case (a) of lemma 1 and thus $\QQ[r]\geq 0$ holds in all domains of all their slices $\T[t]$. Notice that even if there are TV's of $\FF$ and $\HH$ at $\rr=y,\rtv$, we have $\QQ[r]\leq 0$ in every elliptic domain with $r<\hbox{min}(y,\rtv)$. As shown in \cite{RadProfs}, the conditions for a monotonous $\HH$ in the radial direction are not restrictive in non--vacuum hyperbolic models: it can occur at all $\T[t]$ or for slices $\T[t]$ in a wide  evolution range.  On the other hand, the conditions for a monotonous $\HH$ are quite restrictive in elliptic models: it can only occur for some $\T[t]$ in a restricted evolution range (see \cite{RadProfs}). 

Notice that taking into consideration the full time evolution range, then, in general, the monotonicity of $\HH$ (and thus $\HH_q$) can change from one $\T[t]$ to the other in both, hyperbolic and elliptic, models (see \cite{RadProfs}). Therefore, the same model (whether elliptic or hyperbolic or mixed) can contain particular domains $\vartheta[r]$ in which $\Q(r)$ (and thus $\QQ[r]$) may have a different behavior for specific ranges of $t$.

\subsection{A necessary condition for a positive back--reaction.}

It is evident from lemmas 1 and 2 and the discussion above (see also Appendix A and \cite{sussBR,sussIU}) that a necessary (not sufficient) condition for $\QQ[r]\geq 0$ in an arbitrary domain $\vartheta[r]$ is given by evaluating at the boundary of every domain $\vartheta[r]$ the condition
\ba 
\fl \FF'(r)\geq 0\quad \hbox{with}\quad \FF' = -\frac{3r\KK_{q0}}{2\FF}\left(\frac{2}{3}+\Dik\right)=-\frac{3r\KK_q}{2\FF} a^2\Gamma\left(\frac{2}{3}+\Dk\right),\label{FFr}\ea
where the subindex ${}_0$ in $\KK_{q0},\,\Dik$ denotes evaluation at arbitrary fixed $t=t_0$, \, the scale factors $a,\,\Gamma$ are defined by (\ref{LGdef}) and we have used (\ref{Dadef}), (\ref{MF}), (\ref{slaw_mk}) and (\ref{RrFpos}). Evidently, (\ref{FFr}) is also a necessary condition for an effective acceleration. For domains in hyperbolic models ($\KK_{q0}<0$) the condition (\ref{FFr}) plus the regularity conditions (\ref{noshxGh}) imply that $\FF$ must be monotonically increasing in every domain $\vartheta[r]$ (and in every slice $\T[t]$ because of (\ref{slaw_mk})). For domains in elliptic models ($\KK_{q0}>0$), we have $\FF(0)=1$ and $\FF'\leq 0$ near $r=0$, hence a TV of $\FF$ (case (b) of lemma 2) implies that $\FF'>0$ holds for $r>\rtv$ and $\FF\to 1$ as $r\to\infty$. Hence, $\QQ[r]> 0$ necessarily holds in domains in which (\ref{FFr}) holds and spatial curvature is rapidly decreasing (see section 13).
\begin{table}
\begin{center}
\begin{tabular}{|c| c| c|}
\hline
\hline
\hline
\hline
\hline
\hline
\multicolumn{3}{|c|}{Domains in parabolic models: $\FF=1$ ($\KK_q=0$)}
\\  
\hline
\hline
{Turning values} &{Domain restrictions on $\QQ[r]$} &{Comments} 
\\
\hline
\hline
{None} &{$\QQ[r]=0$ holds for all $\vartheta[r]$} &{See \cite{LTBave2,LTBave3}}
\\
\hline
\hline  
\hline
\hline
\hline
\hline
\multicolumn{3}{|c|}{Domains in hyperbolic models: $\FF\geq 1$  ($\KK_q\leq 0$)}
\\  
\hline
\hline
{Turning values} &{Domain restrictions on $\QQ[r]$} &{Comments} 
\\
\hline
\hline
{None} &{$\QQ[r]\geq 0$ holds for all $\vartheta[r]$} &{Lemma 1(a)} 
\\
\hline
{$\HH$} &{$\QQ[r]\geq 0$ holds for $\vartheta[r]$ with} &{Lemma 1(b)} 
\\
{} &{$0\leq r\leq r_1$ {\bf{and}} $r\geq r_2$ with $r_1<r_2$} &{} 
\\
\hline 
\hline
\hline
\hline
\hline
\hline
\multicolumn{3}{|c|}{Domains in open elliptic models: $\FF^2\leq 1$ ($\KK_q\geq 0$)}
\\  
\hline
\hline
{Turning values} &{Domain restrictions on $\QQ[r]$} &{Comments} 
\\ 
\hline
\hline
{None} &{$\QQ[r]\leq 0$ holds for all $\vartheta[r]$} &{Lemma 2(a)} 
\\
\hline
{$\HH$} &{$\QQ[r]\geq 0$ holds for $\vartheta[r]$} &{Lemma 2(c)}
\\
{} &{with $r_1\leq r\leq r_2$} &{} 
\\
\hline
{$\FF$} &{$\QQ[r]\geq 0$ holds for $\vartheta[r]$} &{Lemma 2(b)} 
\\ 
{} &{with $r\geq y$ with $y>0$} &{} 
\\
\hline
{$\FF$ and $\HH$} &{as above} &{Lemma 2(b)} 
\\
\hline
\hline
\end{tabular}
\end{center}
\caption{{\bf{Summary of sufficient conditions for the sign of back--reaction in LTB models}}. The conditions are expressed (second column) in terms of the sign of $\QQ[r]$ in domains characterized by the radial range $\vartheta[r]=\{\rr\,|\,0\leq \rr\leq r\}$ for each class of models, according to the existence of turning values discussed in section 5 (first column). The third column lists the lemmas dealing with each case. The domains correspond to arbitrary regular slices $\T[t]$ not intersecting a singularity (see figures 8 and 9a).  The parameters $r_1,\,r_2$ and $y$ depend on the profiles of $\FF$ and $\HH$ for each case and are given in the proofs of lemmas 1 and 2 in  \cite{sussBR}. See section 12 for a discussion of back--reaction in mixed elliptic/hyperbolic and elliptic parabolic configurations. Back--reaction for closed elliptic models is discussed in \cite{sussBR}.}
\label{tabla1}
\end{table} 

\section{Probing the existence of effective acceleration.}

In domains in which we have proven that $\Q(r)\geq 0$ or $\OmC\geq 0$ hold, the conditions for an effective acceleration (\ref{Apos}) or (\ref{newApos}) become a comparison of two positive quantities evaluated at the boundary of any domain $\vartheta[r]$. However, it is not (in general) a simple task to verify in these cases if the positive or negative contribution dominates in this comparison. 

\subsection{Qualitative guidelines.}

Assuming domains such that $\OmC>0$ holds (positive back--reaction) in (\ref{OmC}), then the following two conditions (which are not mutually exclusive) yield the scenarios with the best possibilities for (\ref{newApos}) to occur:
\begin{description}
\item[$\Oml$ is as small as possible.] This condition occurs in the following scenarios:
\begin{enumerate}
\item {\underline {Strict density vacuum:}} $\rho=\rho_q=0\,\,\Rightarrow\,\, \hOm=\Oml=0$. See section 7.2. 
\item {\underline {Schwarzschild density vacuum:}} $\rho=0\,\,\hbox{but}\,\,\rho_q>0\,\,\Rightarrow\,\,1+\Dm=0$, hence we have $\Oml=0$ but $\hOm>0$. See section 7.2. 
\item {\underline {Very low (but nonzero) density:}} $\KK_q<0\,\,\hbox{and}\,\, \rho_q\ll |\KK_q|\,\,\Rightarrow\,\,\hOm\ll 1$ (see (\ref{Omdef}), (\ref{slaw_mk}) and (\ref{slaw_H})). Domains in hyperbolic or asymptotically hyperbolic models in the following conditions:
\bse\ba
\fl t\;\;\hbox{finite: \quad central void region (see section 8).}\label{lowdens0}\\ 
\fl t\;\hbox{finite: \quad asymptotic radial range (see sections 9 and 10).}\label{lowdens1}\\   
\fl t\to \infty\qquad \hbox{asymptotic time range (see section 13).} \label{lowdens2}
\ea\ese 
\item $1+\Dm \to 0$. This scenario is compatible with hyperbolic and elliptic models or regions.  From (\ref{Dadef}) applied to $S=\rho$, it occurs if $\rho\ll \rho_q$ under the following conditions of very small and very large density:
\bse\ba
\fl \{\rho, \rho_q\}\to 0\quad \hbox{radial asymptotic range (see section 9).}\label{lowdens3}\\
\fl \{\rho, \rho_q\}\to \infty \quad \hbox{near curvature singularities  (see sections 11 and 12).}\label{lowdens4} 
\ea\ese  
\end{enumerate}
\item[$\OmC$ is as large as possible.] This condition implies the largest possible growth of back--reaction, which  couples to density and spatial curvature \cite{buchrev,buchlet} through their radial gradients. The necessary condition for $\QQ[r]\geq 0$ in (\ref{FFr}) shows this coupling with the gradients of spatial curvature through the relative fluctuation $\Dk$, which relates to the gradient $\KK'_q$ through (\ref{Dadef}). The relation between the growth of back--reaction and the existence of large radial gradients of density and/or spatial curvature can also be appreciated from the fact that $\Q(r)$ in (\ref{Cpos}) involves the fluctuations $\HH-\HH_p$ and $\HH-\HH_q$, which relate to the gradients $\HH'_p$ and $\HH'_q$ by means of (\ref{Apr}). In terms of these gradients we obtain (see \cite{sussIU}):
\begin{equation}\OmC =2\left(\frac{\HH'_p/\HH_p}{\Vp'/\Vp}\right)^2\frac{\HH_p^2}{\HH_q^2}-2\left(\frac{\HH'_q/\HH_q}{\Vq'/\Vq}\right)^2,\label{Cgrads}\end{equation}
with $\HH'_p$ and $\HH'_q$ related to gradients of density and curvature by means of the constraints (\ref{Hqsq}) and (\ref{ave_Fried}).  However, the coupling between gradients of density, spatial curvature and expansion can be complicated, and thus there may be several scenarios in which  these gradients could ``cooperate'' so that $\OmC$ increases and $\Oml$ decreases. A favorable scenario may be a low density region (small $\Oml$) in which fluctuations of spatial curvature are larger than density fluctuations (see sections 8 and 10), so that the growth of $\Q(r)$ is mostly driven by the gradients of spatial curvature in that region.  As commented by \cite{buchrev,buchlet,ras2,ras3}), large fluctuations of spatial curvature occur when the functional form of $\langle \KK\rangle = \KK_p$ is different from a ``FLRW--like'' dependence ($\propto a_\DD^{-2}$ where $a_\DD^3=\Vp$).  
\end{description}
It is important to remark that the scenarios we have described above simply provide the guidelines that {\it may favor} the existence of an effective acceleration in terms of sufficient (not necessary) conditions that are evaluated in the domain boundary, and thus we cannot rule out such acceleration arising under different conditions. Notice that even when these guidelines are satisfied the actual existence of this acceleration must be verified by looking carefully at the behavior of the scalars in (\ref{Apos}) or (\ref{newApos}) in  the appropriate domains.

\section{Scenarios allowing for a direct proof of effective deceleration and acceleration.}

Consider the following limits along radial rays of a fully regular and complete slice $\T[t]$ with arbitrary finite $t$: 
\bse\ba
\fl \mathop {\lim }\limits_{r \to 0 } S(r) = S_c = S(0) \quad  \Leftrightarrow \quad \mathop {\lim }\limits_{r \to 0 } S_p (r) = \mathop {\lim }\limits_{r \to 0 } S_q (r) = S_c,
\label{Ac}\\ 
\fl \mathop {\lim }\limits_{r \to \infty } S(r) = S_{_\infty} \quad  \Leftrightarrow \quad \mathop {\lim }\limits_{r \to \infty } S_p (r) = \mathop {\lim }\limits_{r \to \infty } S_q (r) = S_{_\infty},\label{A0}\ea\ese
where $S_c$ and $S_{_\infty}$ are real constants. If $S=S(t,r)$ and $S_c$ and $S_{_\infty}$ are nonzero, then the latter are (in general) different constants at different slices $\T[t]$ ({\it{i.e.}} we have: $S_c=S_c(t)$ and $S_{_\infty}=S_{_\infty}(t)$). The subindex ${}_c$ will henceforth denote evaluation at $r=0$, while ${}_{_\infty}$ will denote the asymptotic limit as $r\to\infty$. By applying (\ref{Ac}) and (\ref{A0}) to the definitions (\ref{Apos}) and (\ref{newApos}) we obtain direct proof of existence of effective deceleration or acceleration in specific situations.

\subsection{Effective deceleration.}  
\begin{description} 
\item[Non--vacuum center and central region of elliptic models.] If $\rho_c>0$ then (\ref{Hqsq}) implies $\HH_c\ne 0$ (with $\KK_c<8\pi\rho_c/3$). From (\ref{Dadef}) we have $\Dm(0)=\Dk(0)=\Dh(0)=0$ and (\ref{Ac}) implies $\HH_p(0)=\HH_q(0)=\HH_c$, hence we have an effective deceleration in the center worldline of every regular model with a non--vacuum center:
\begin{equation}\fl \Q(0)=0,\qquad  \A(0) = -\frac{4\pi}{3}\rho_c<0, \qquad \hq(0)=\frac{\hOm(0)}{2}>0.\label{nonvac}\end{equation}
Since $\QQ[r]<0$ holds for domains sufficiently close to the center in all elliptic models (all cases of lemma 2), then (\ref{efeacc}) and (\ref{effe_dec}) imply effective deceleration in these domains. In domains of hyperbolic models $\QQ[r]>0$ holds for $r>0$ near the center (in the two cases of lemma 1), hence it is not immediately evident if (\ref{efeacc}) and (\ref{effe_acc}) hold for $r>0$. We examine this possibility in sections 7 and 9.  
\item[Radial asymptotic range in models converging to FLRW.] LTB models can be radially asymptotic to spatially flat (Einstein de Sitter) or negatively curved FLRW models, but not to positively curved FLRW models (see details in \cite{RadAs}). LTB Models with this asymptotic behavior
are characterized by  $\rho_{_\infty}>0$, and thus (\ref{Hqsq}), (\ref{Dadef}) and (\ref{A0}) imply for expanding models $\HH_{_\infty}>0$ and $\Dm_{_\infty}=\Dh_{_\infty}=0$, while $\HH_p\to \HH_{_\infty}$ and $\HH_q\to \HH_{_\infty}$ as $r\to\infty$ (see \cite{RadAs}). Hence
\begin{equation} \fl \mathop {\lim }\limits_{r \to \infty } \Q (r) = 0,\quad \mathop {\lim }\limits_{r \to \infty } \A (r) = -\frac{4\pi}{3}\rho_{_\infty} < 0,\quad \mathop {\lim }\limits_{r \to \infty }\hq =\hqty=\frac{\hOmty}{2}>0,\label{radasFLRW}\end{equation}
where $\mty$ and $\hOmty$ correspond to the density and Omega factor of the ``background'' FLRW state reached asymptotically along the radial rays \cite{RadAs}. Evidently, (\ref{radasFLRW}) implies an effective deceleration (\ref{effe_dec}) in the radial asymptotic range of these models. The same situation occurs in the radial range of ``Swiss Cheese'' configurations in which the FLRW background is not reached asymptotically but is forced by a matching with an LTB region in a comoving boundary marked by finite $r$. We remark that all models examined in previous literature  \cite{ras2,ras3,ras4,ParSin3,BolAnd,LTBave3,mattsson} are radially convergent to FLRW cosmologies, hence there is effective deceleration in domains extending into their  asymptotic radial range.         
\end{description}
\begin{figure}[htbp]
\begin{center}
\includegraphics[width=3.5in]{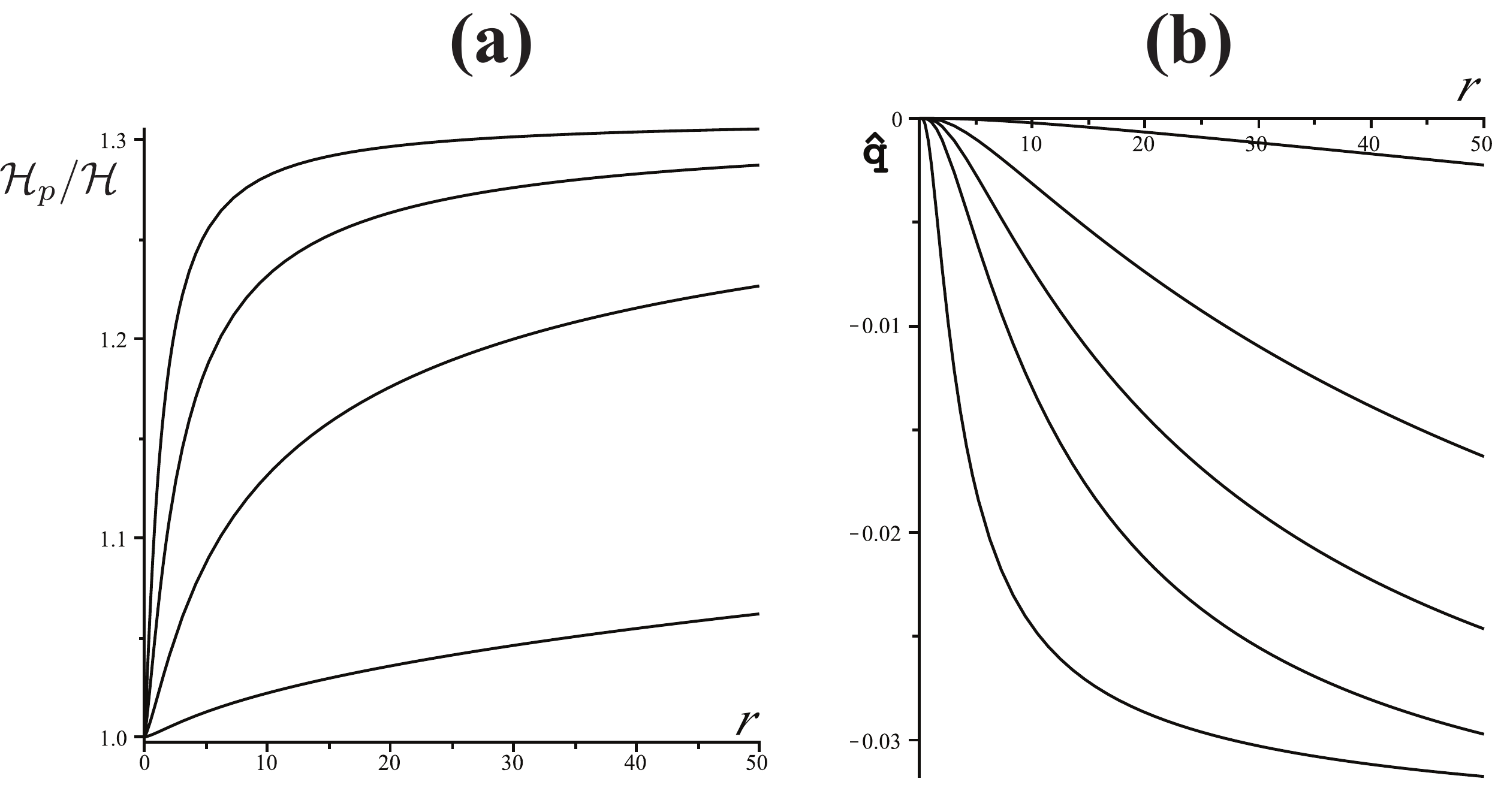}
\caption{{\bf LTB vacuum region.} The figure displays the ratio $\HH_p/\HH$ (panel (a)) and $\hq$ defined by (\ref{newApos}) (panel (b)) for an extended central vacuum LTB region. The region can be matched to an expanding non--vacuum hyperbolic model at any fixed $r$. The curves (from top to bottom in (a) and bottom to top in (b)) correspond to $t=0,5,10,20,100$, with $t=t_0=0$ marking the fiducial slice where we define the initial value functions $\rho_{q0} =0$ and $\KK_{q0} = 1/(1+r^{5/4})$. Notice how both plotted quantities tend to zero for large $t$ and tend to an asymptotic value for large $r$. Panel (a) shows that the averaged Hubble factor (whose magnitude is $\HH_p$) can be up to $30\,\%$ larger than the local Hubble factor.}
\label{f2}
\end{center}
\end{figure}
\begin{figure}[htbp]
\begin{center}
\includegraphics[width=2in]{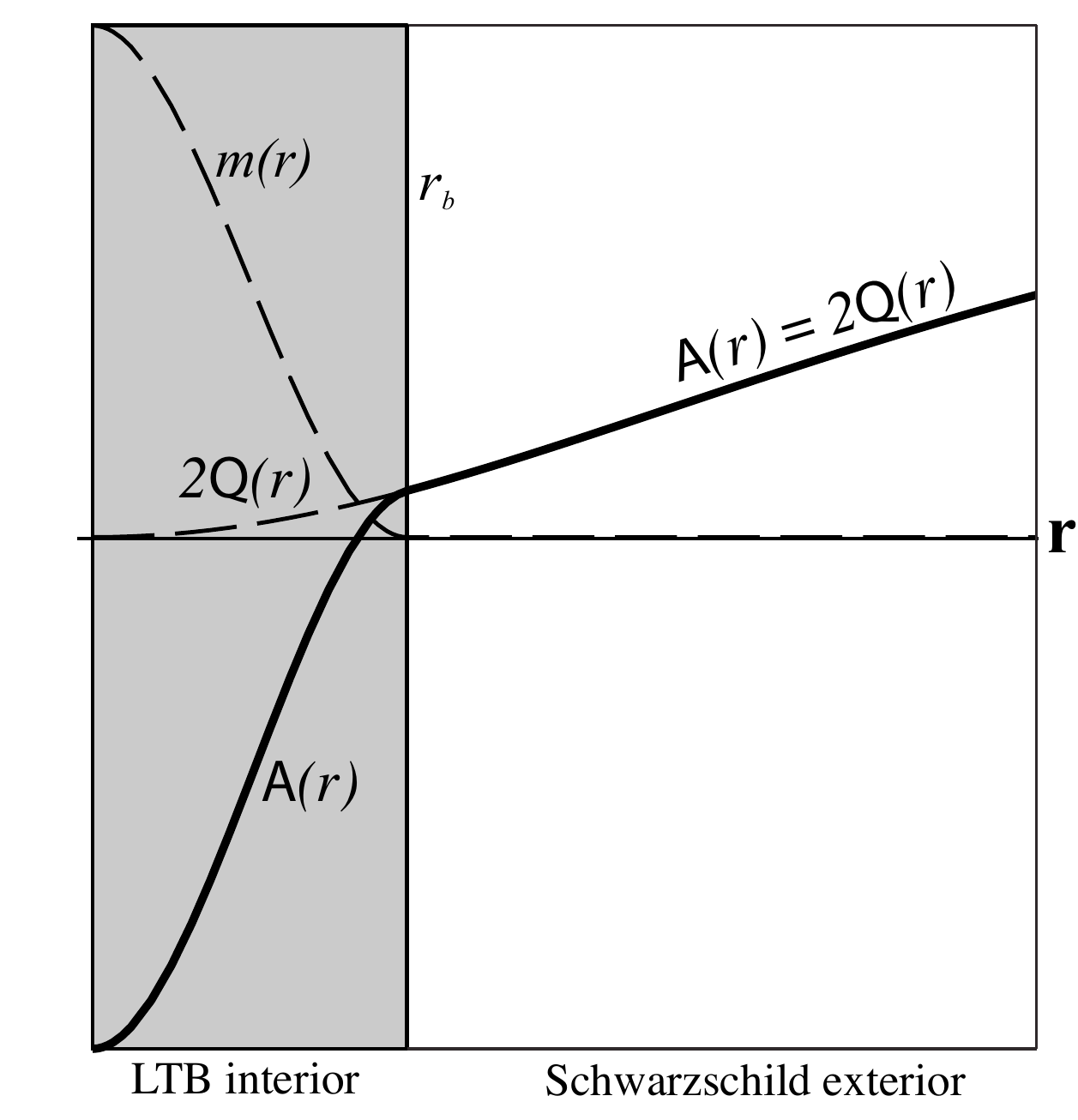}
\caption{{\bf LTB inner region matched to a Schwarzschild exterior.} The figure displays qualitative plots of the functions $2\,\Q(r)$,\,$m(r)=4\pi\rho/3$ and $\A(r)$, appearing in (\ref{Apos}), at an arbitrary slice $\T[t]$ of a configuration made of a region of an LTB non--vacuum model (shaded area) complying with $\Q(r)\geq 0$ matched with a Schwarzschild exterior at $r=r_b$. The local density complies with $\rho=0$ for $r\geq r_b$, hence for this radial range we have $\A=2\,\Q>0$, though $\A$ may be negative in most of the inner region where $\rho>0$.}
\label{f3}
\end{center}
\end{figure}
\subsection{Effective acceleration in vacuum regions imposed by matchings.}
\begin{description} 
\item[Vacuum central region.] This is the strict vacuum scenario of case (i) in the guidelines above: a vacuum LTB model extending (see section 2) between $r=0$ and a given finite $r=r_b$, can be smoothly matched with a section of a non--vacuum hyperbolic model extending for $r>r_b$ (see \cite{ltbstuff}). Matching and regularity conditions imply that $\rho(r_b)=0$ must hold and $\rho$ must be continuous \cite{ltbstuff}, which implies that density must have a  radial void profile ({\it i.e.} $\rho'>0$) for $r>r_b$ for all the time evolution. However, as shown in \cite{RadProfs}, a void profile for all the time evolution is only compatible with absence of shell crossings in non--vacuum hyperbolic models if they have a simultaneous big bang ($\tbb'=0$), and this feature necessarily implies (from \cite{RadAs}) radial asymptotic convergence to a FLRW model.  Since $\rho(r)=0$ and (from lemma 1) we have $\Q(r)\geq 0$ in the vacuum region, then $\A(r)=2\Q(r)\geq 0$ and $\hq(r)=-\OmC(r)\leq 0$ hold in this region, and thus, (\ref{effe_acc}) holds by vitue of (\ref{Apos}). While $\A(r)$ may remain positive near $r_b$, the radial convergence to FLRW implies that $\A<0$ holds in the asymptotic radial range (as in (\ref{radasFLRW})). We plot in figure \ref{f2} the deceleration parameter $\hq$ and the ratio $\HH_p/\HH$ for a central vacuum region (the matching with a non--vacuum region is not displayed). As shown by the figure, both quotients tend asymptotically to constants whose maximal values are $\hq \approx -0.03$ and $\HH_p/\HH\approx 1.35$.   
\item[Schwarzschild exterior.] This is the Schwarzschild vacuum scenario of case (ii) in the guidelines above: a spherical comoving region of any LTB model enclosing a symmetry center (the ``interior'') can be smoothly matched at $r=r_b$ with a section of a Schwarzschild--Kruskal manifold (the ``exterior'') (see page 332 of \cite{kras2}). Notice that $\rho_q=0$ holds only for vacuum LTB models ($M=0$), but not for Schwarzschild--Kruskal models: $(4\pi/3)\rho_q=M_{\textrm{\tiny{schw}}}/R^3$. In particular, we can obtain this configuration by choosing the local density so that $\rho(r_b)=0$, with $\rho(r)=0$ for $r>r_b$. Hence we have $\Dm(r_b)=-1$, and $\rho=0$ together with $\Dm=-1$ for $r>r_b$, while (\ref{qvars}) implies that $(4\pi/3)\rho_q(r_b)=2M(r_b)/R^3(r_b)>0$ holds. However, the smoothness of the match \cite{ltbstuff} does not require the interior density vanishing at $r_b$,  so that $\rho$ and $\Dm$ can have a step discontinuity at the interface, though $\rho_q,\,\KK_q$ and $\HH_q$ must remain continuous at $r_b$. If $\Q(r)\geq 0$ holds in the whole inner region (see lemmas 1 and 2 and condition (\ref{FFr})), then depending on the density, $\A$ may not be positive for all $r<r_b$. However, since we have $\rho=0$ in the Schwarzschild exterior, then $\A(r)=2\Q(r)>0$ will necessarily hold at least for parts of the range $r>r_b$ (and thus (\ref{effe_acc}) holds by vitue of (\ref{Apos})). Evidently, there will be an effective acceleration in radial ranges of the exterior where $\Q(r)>0$ holds, and thus (following lemma 1 and lemma 2), these ranges depend on whether the exterior is hyperbolic or elliptic (we omit the case of a parabolic interior because $\Q(r)=0$).  Figure \ref{f3} provides a qualitative depiction of $\A(r)$ for an LTB region with $\Q(r)>0$ matched to a Schwarzschild exterior.   
\end{description}
It is evident that an inner vacuum LTB region or an outer Schwarzschild exterior are rather artificial configurations. Probing the existence of accelerating domains in less artificial cases is obviously more difficult. 
\begin{figure}[htbp]
\begin{center}
\includegraphics[width=3.5in]{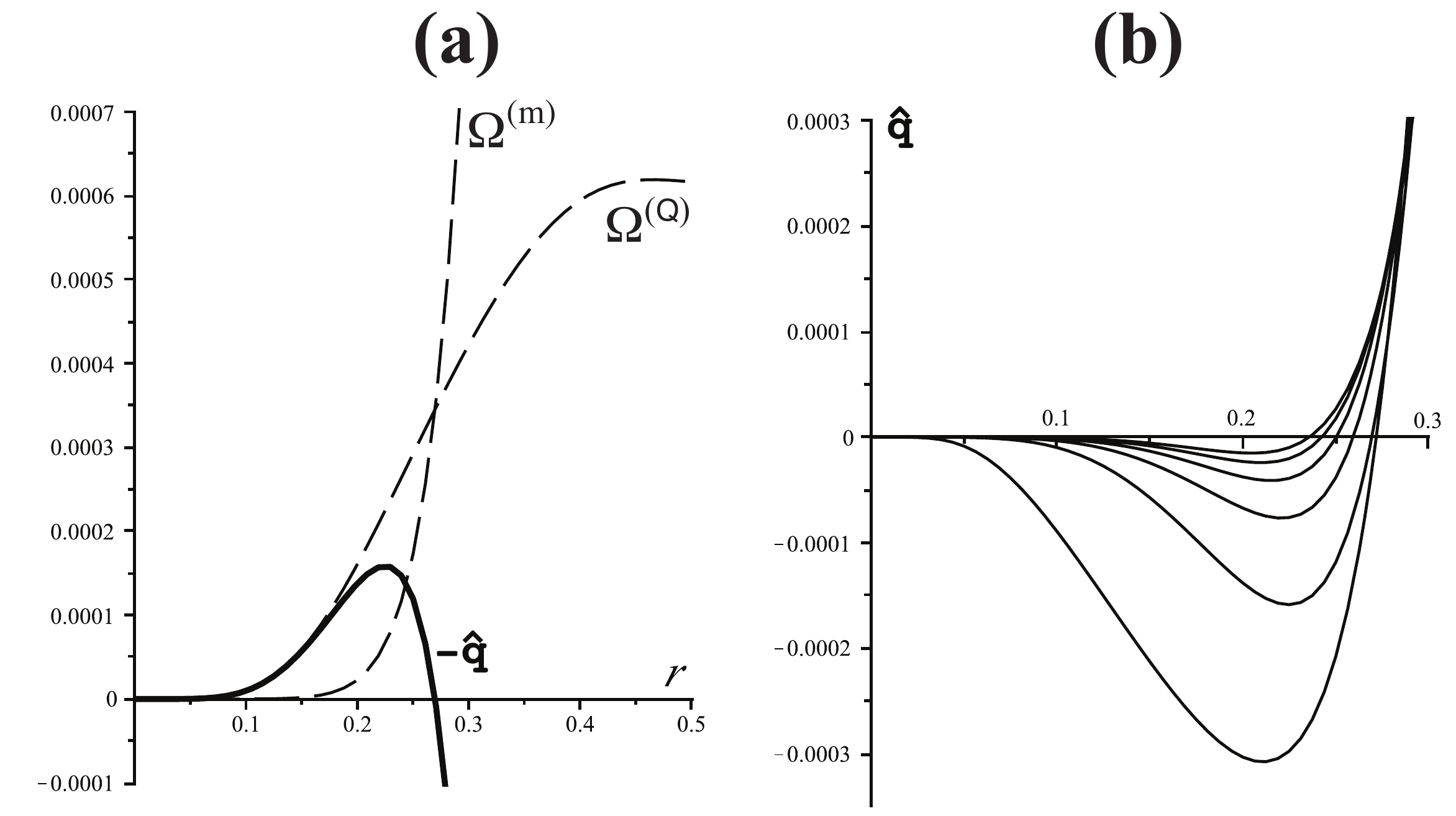}
\caption{{\bf Expanding non--vacuum void.} Pannel (a) displays the functions $\OmC,\,\Oml$ and $-\hq$, appearing in (\ref{newApos}), for varying $r$ in the initial slice $t=t_0=0$ for a non--vacuum central void in a hyperbolic model. The functions were obtained from a numeric integration of the implicit solution (\ref{cthyp}) for initial value functions that yield an evolution free from shell crossings (see Appendix A). Notice that as $r$ grows back--reaction tends asymptotically to a small value ($\OmC\approx 0.0006$), while local density ($\Oml$) grows much faster. Hence, $\hq$ is negative only in a small radial range. Panel (b) displays $\hq$ in (\ref{newApos}) for the values (bottom to top) $t =-1,0,1,2,3,4$. Notice that $\hq\to 0$ as $t$ grows.}
\label{f4}
\end{center}
\end{figure}

\section{Under dense void regions of hyperbolic models.}   

While (\ref{nonvac}) implies a positive $\hq=\qef$ at the center worldline if $\rho_c>0$
\footnote{The possibility of having $\rho_c=0$ only in the central world line is equivalent to reducing the vacuum LTB region of the previous section to this single worldline. For non--vacuum models this situation  can be considered as the limit $\rho_c\to 0$ of very low (but nonzero) central density. While $H_c=0$ can arise mathematically if either $0<\KK_c=(8\pi/3)\rho_c$ or $\KK_c=\rho_c=0$ occur, both possibilities are incompatible with regularity conditions and will not be considered.},
it is possible to find conditions that facilitate a negative $\hq$ for finite $r>0$ in non--vacuum regions with low density. Since regularity and radial void density profiles are only compatible for hyperbolic models \cite{RadProfs}, we consider an expanding hyperbolic central void region having a low (but nonzero) central density, so that $\hOm(0)\ll 1$ (case (\ref{lowdens0})). 

Following the qualitative guidelines outlined in section 6, a plausible set of conditions that may favor the joint fulfillment of a small $\Oml$ and a large $\OmC$ in a range $0<r_1<r<r_2$ are:  
\bse\ba \hOm\approx \epsilon \ll 1,\qquad 1-\hOm\approx 1,\qquad |\Dk|\gg |\Dm|\approx 0,\label{Void1}\\
 1-\frac{\HH_p}{\HH_q} < 0\quad \hbox{and}\quad \Dk< 0.\label{Void2}\ea\ese
and clearly correspond to the low density scenario (\ref{lowdens0}), together with a growth of $\OmC$ that is mostly driven by the variation (gradients) of spatial curvature ($|\Dk|$). On the other hand, (\ref{Void2}) follows from the regularity conditions, since a regular void profile requires \cite{RadProfs} that $\Dm>0$ and $\Dk<0$ hold, and is also compatible with the fact that (\ref{Void1}) implies $2\Dh\approx \Dk<0$ (which implies $\HH'<0$, leading from (\ref{Phi}) to a negative $1-\HH_p/\HH_q$).

Under the assumptions (\ref{Void1}) and (\ref{Void2}) condition (\ref{newApos}) takes in this radial range the approximated form
\begin{equation}\hq\approx -2\left| 1-\frac{\HH_p}{\HH_q}\right|\left|1-\frac{\HH_p}{\HH_q}+\Dk\right|+\frac{\epsilon}{2}\leq 0,\label{WposVoid}\end{equation}
which may hold if $\epsilon/2$ is effectively smaller than the other term. Evidently, the chances for (\ref{WposVoid}) to hold improve if the non--vacuum void is ``deeper'' (smallest possible $\epsilon\ll 1$) and the larger we have the ratio between the gradients of density and curvature ($|\Dim/\Dik|\approx 0$). The ``wider''  we can set up the void ((\ref{Void1}) and (\ref{Void2}) holding for the largest possible radial range), the larger radial range may exhibit $\hq <0$. In the limit $\rho_c\to 0\;\;\Rightarrow\;\;\hOm(0)\to 0$, the void region that fulfills (\ref{WposVoid}) extends all the way to $r=0$ (where $\hq\to 0$). Since it is very difficult to obtain analytic constraints on $1-\HH_p/\HH_q$, we provide in figure \ref{f4} a numerical example of a central void region in which $\hq<0$ holds in an intermediate region close to the center.

Models of non--vacuum voids were examined by Mattsson and Mattsson in \cite{mattsson}, who explored the relation between the growth of back--reaction, the average of the shear scalar and the smoothness of the gradients of density. As  argued in this reference, the use of discontinuous step function density radial profiles made by joining regions with constant density (for example glueing FLRW regions as in \cite{ras2,ras3}) yields a gross error in the estimation of the magnitude of the back--reaction in comparison with smooth profiles, since shear is zero in each constant density region. The connection  between back--reaction and shear follows from the fact that the shear scalar $\sigma_{ab}\sigma^{ab}=6\Sigma^2=6(\HH_q\Dh)^2$ (from (\ref{SigEE}) and (\ref{SigEE1})) is related to $\HH'_q$ through (\ref{SE2}) and (\ref{Dadef}). Hence, shear may favor an effective acceleration in a given radial range in low density non--vacuum voids if it is associated with large gradients of spatial curvature.

\section{Effective acceleration in the radial asymptotic range.}

Since there is no effective acceleration in domains extending into the radial asymptotic range of LTB models that converge to non--vacuum FLRW spacetimes (section 7.1), we consider in this section only models converging in their radial asymptotic range to asymptotic states that are not FLRW~\cite{RadAs}: 
\begin{description} 
\item[Models radially asymptotic to non--Milne sections of Minkowski (vacuum LTB).] These hyperbolic and elliptic models are characterized by the asymptotic limits $\{\mty,\,\kty,\,\Hty\}= 0$ \cite{RadAs}. Since we have $\HH_q\to 0$ and $\HH_p\to 0$, we cannot use (\ref{Dadef}) and (\ref{A0}) to obtain the asymptotic values of $\Dm$ and $1-\HH_p/\HH_q$ in this limit. The asymptotic limit of $\hOm$ depends then on the asymptotic behavior of the ratio $x$:
\begin{equation}\fl x\equiv\frac{|\KK_q|}{\frac{4\pi}{3}\rho_q}  \to \left\{ \begin{array}{l}
 0,\qquad\hOm\to 1,\qquad \hbox{Matter dominated (MD)} \\ 
 x_{_\infty},\qquad\hOm\to \hOmty,\qquad \hbox{Generic (G)}  \\ 
 \infty,\qquad\hOm\to 0,\qquad \hbox{Vacuum dominated (VD)}  \\ 
 \end{array} \right.\label{x} \end{equation}
where in the G case $\hOmty>1$ if $\KK_q>0$ and $0<\hOmty<1$ if $\KK_q<0$. Notice that for all elliptic models we must have $\KK_q\leq (8\pi/3)\rho_q$, hence $x\leq 2$ holds and all these models are either MD ($x_{_\infty}<2$) or G ($x_{_\infty}=2$). On the other hand, hyperbolic models can be MD, G or VD.  
\item[Models radially asymptotic to Milne.]
These models must be hyperbolic and VD, corresponding to~\cite{RadAs} $\rho_q\to\mty=0$ and $\KK_q\to \kty<0$, and thus $\Hty=|\kty|^{1/2}$ and $\hOm\to 0$.     
\end{description}
While some results can be obtained without making specific assumptions on the radial asymptotic convergence of the involved scalars, it is useful (and necessary) in most cases to consider these assumptions to prove the existence and an estimated numerical value of an effective acceleration in this range. Following reference \cite{RadAs}, the radial asymptotic forms of all scalars can be obtained from the asymptotic behavior of the initial value functions $\rho_{q0}$ and $\KK_{q0}$ (evaluated at an arbitrary $t=t_0$). In particular, we will consider convergence to the following power law trial functions (the same results were obtained with other trial functions in \cite{RadAs}):
\ba \fl \rho_q\sim \rho_{q0} \sim m_0\,r^{-\alpha}\qquad\Dm\sim -\frac{\alpha}{3},\qquad (0<\alpha\leq 3),\label{asconvm}\\
\fl \KK_q\sim \KK_{q0} \sim k_0\,r^{-\beta}\qquad \Dk\sim -\frac{\beta}{3},\qquad \left\{ \begin{array}{l}
 0<\beta\leq 2,\quad \hbox{(hyperbolic models)} \\  
 \beta\geq 2,\qquad \hbox{(elliptic models)}  \\
 \beta=0,\quad \hbox{(asymptotically Milne)}\\ 
 \end{array} \right.,\label{asconvk}
\ea
where the symbol $\sim$ indicates uniform convergence (see \cite{RadAs} for the precise definition),  $m_0,\,k_0$ are nonzero real constants with inverse square length units, and the ranges of the parameters $\alpha$ and $\beta$ follow from compliance with the regularity conditions (\ref{noshxGh}) and (\ref{noshxGe}). In particular, the range of $\beta$ in elliptic models follows from $\FF$ with $\KK_{q0}>0$ in (\ref{MF}). Models asymptotic to self--similar LTB solutions are the G models complying with $\alpha=\beta=2$, so that the limit (\ref{limHpHqF0}) holds, while models converging to Schwarzschild  are characterized by $\alpha=3$ (so that $M\to M_{\textrm{\tiny{schw}}}$, see \cite{RadAs}).

Since the free parameters commonly used in the literature are the functions $M$ and $\FF$ (commonly given as ``$\FF=\sqrt{1+2E}$'') in (\ref{eqR2t}) and (\ref{eqrho1}), it is useful to provide the asymptotic form of these parameters under the assumptions (\ref{asconvm})--(\ref{asconvk}):
\begin{equation} M\sim m_0\,r^{3-\alpha},\qquad \FF \sim \left[1-k_0\,r^{2-\beta}\right]^{1/2},\label{asMF}\end{equation}
while the bang time $\tbb$ follows by applying (\ref{asconvm}) and (\ref{asconvk}) to (\ref{tbb}), leading to $\tbb \sim 1/|k_0|^{1/2}$ for models converging to Milne, while for models converging to a vacuum LTB state we have
\begin{equation} \tbb \sim t_0 - m_0|k_0|^{-3/2}r^{3\beta/2-\alpha}Z\left(|k_0|\,r^{\alpha-\beta}/m_0\right),\end{equation}
where $Z$ stands for $Z_e$ or $Z_h$ defined by (\ref{hypZ1a}) and (\ref{ellZ1a}) for hyperbolic or elliptic models, so that $\tbb\to -\infty$ as $r\to\infty$ \cite{RadAs}.

As opposed to models converging to FLRW, finding the asymptotic behavior of $\hq$ now requires further examination, as we cannot use the limits (\ref{A0}) and (\ref{nonvac})--(\ref{radasFLRW}) to infer the  asymptotic limit of the  ratio $1-\HH_p/\HH_q$. Whether we consider the assumptions (\ref{asconvm})--(\ref{asconvk}) or not, we remark that the difference between $\HH_q$ and $\HH_p$ is the presence of the function $\FF$ in their definitions (\ref{avedef}) and (\ref{aveqdef}) (see also (\ref{Phi})), hence the limit of this ratio depends on the asymptotic behavior of this function. As we prove in Appendix C, the following two cases arise  
\ba\mathop {\lim }\limits_{r \to \infty } \FF=\Fty\quad \Rightarrow \quad \mathop {\lim }\limits_{r \to \infty } 1 - \frac{{\HH_p }}{{\HH_q }} = 0.\label{limHpHqF0}\\ 
\mathop {\lim }\limits_{r \to \infty } \FF=\infty\quad  \Rightarrow  \quad \mathop {\lim }\limits_{r \to \infty } 1 - \frac{{\HH_p }}{{\HH_q }} =-\xi < 0\label{limHpHq}\ea
where the form of $\xi>0$ depends on the asymptotic convergence of $\rho_q,\,\KK_q,\,\HH_q$ (see (\ref{xias})). Notice that both limits (\ref{limHpHqF0}) and (\ref{limHpHq}) are compatible with the necessary condition for $\QQ[r]\geq 0$ in (\ref{FFr}). We can now infer the existence of an effective acceleration associated with $\hq\to\hqty<0$ according to the following cases:
\begin{figure}[htbp]
\begin{center}
\includegraphics[width=2in]{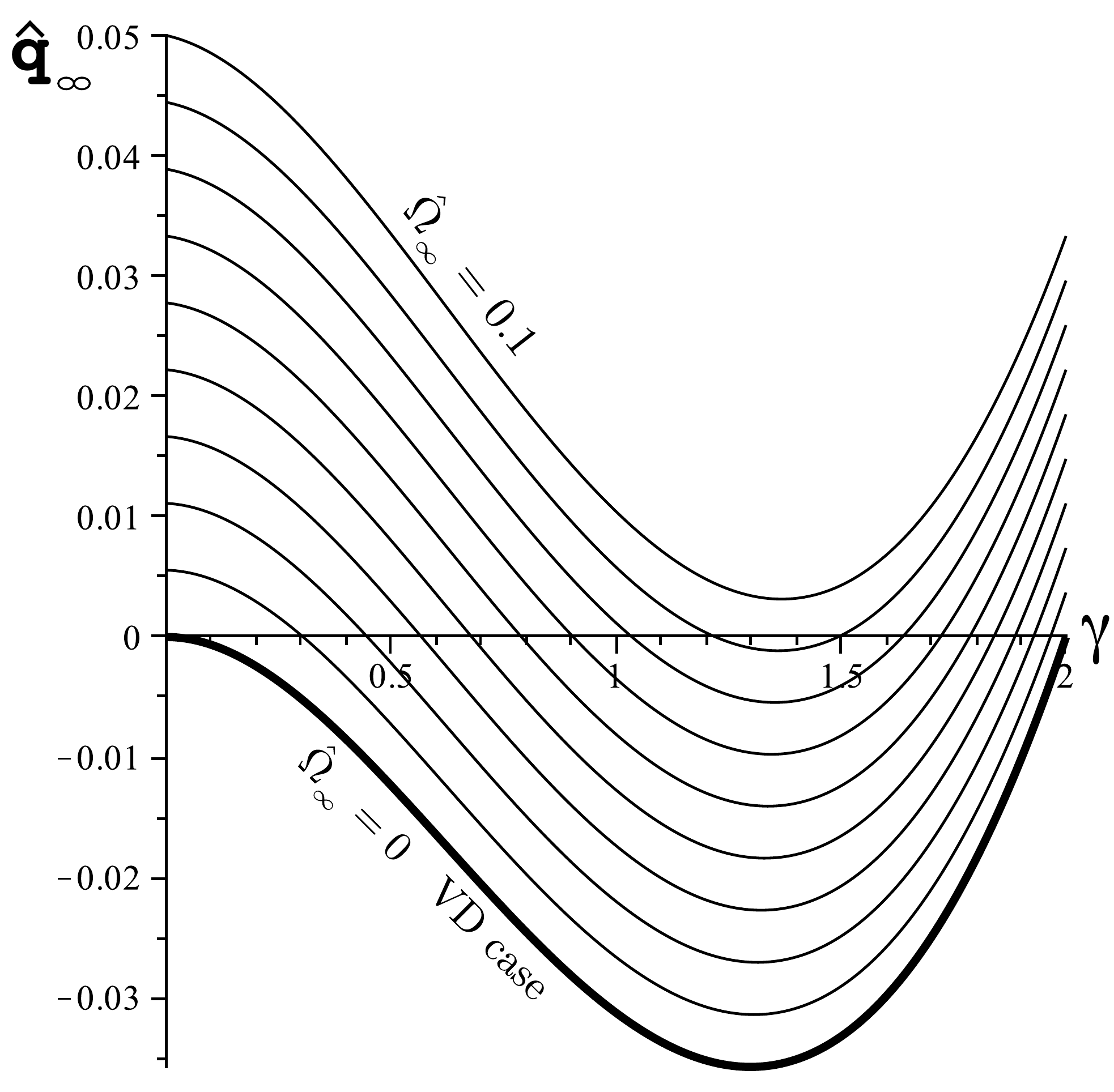}
\caption{{\bf Value of the effective acceleration in the radial asymptotic limit of G and VD hyperbolic models.} The asymptotic limit $\hqty$ of the function $\hq$ in (\ref{W01G}) for hyperbolic G models (case (C)) is displayed as a function of the parameter $\gamma$ given by (\ref{W01G}) for values (bottom to top) $\hat\Omega_{\infty}\equiv \hOmty=0,0.011,0.022,0.033,0.044,0.055,0.066,0.077,0.088,0.1$. The thick curve is the case $\hOmty=0$. Since (\ref{W01G}) reduces to (\ref{W00}), this case corresponds to VD models (case (A)). Notice that a very small value $\hOmty<0.09$ is needed for having $\hq<0$ in the G models, while for the VD models $\hq$ reaches a maximal value $\hq\approx -0.035$ for $\gamma=\beta\approx 1.25$. }
\label{f5}
\end{center}
\end{figure}
\begin{description}
\item [Case (A): there is effective acceleration.] The only case in which a conclusive proof of existence of an effective acceleration can be given without making assumptions on asymptotic convergence are the class of LTB models complying with:
\begin{equation}\fl \KK_q<0\;\;\hbox{(hyperbolic)},\qquad \FF\to \infty\;\;\hbox{[eq. (\ref{limHpHq})]},\qquad \hOm\to 0\;\; \hbox{(VD)}.\label{modsApos}\end{equation}
Independently of the assumptions (\ref{asconvm})--(\ref{asconvk}), condition (\ref{newApos}) takes the limit
\begin{equation} \mathop {\lim }\limits_{r \to \infty }\hq = \hqty=-2\xi\left(\xi+|\Dk_{_\infty}|\,\right)<0, \label{W0}\end{equation}
where $\Dk_{_\infty}$ is the asymptotic limit of $\Dk$, which must be negative since $\Dk<0$ must hold asymptotically because  $|\KK_q|\to 0$ and thus $|\KK_q|' < 0$. This case corresponds to the asymptotic forms (\ref{asconvm}) and (\ref{asconvk}) with $3\geq \alpha>2>\beta$ (models converging to Schwarzschild are the particular case $\alpha=3$). Considering (\ref{modsApos}), we have $n=\beta/2$ from (\ref{xias}), so that (\ref{newApos}) and (\ref{asconvm})--(\ref{asconvk}) imply the following form for the limit (\ref{W0}):
\begin{equation} \hqty = -\frac{\beta^2(2-\beta)(10-\beta)}{288},\label{W00}\end{equation}
whose plot in figure \ref{f5} (the thick curve) reveals that it has a maximum of $\hqty\approx -0.035$ for $\beta\approx 1.25$.
\item [Case (B): there is no effective acceleration.] Conclusive proof that $\hq\to\hqty>0$ holds can be readily given for LTB models that are:  
\begin{enumerate}
\item asymptotic to FLRW dust cosmologies (see section 7.1).
\item asymptotic to vacuum LTB that are elliptic and violate the necessary condition (\ref{FFr}). Hence, (from (\ref{Dadef})) $\KK_q$ decays as $R^{-2}$ ($\beta=2$), and thus $\FF\to\Fty<1$, so that $\FF'<0$ and $\Q<0$ hold asymptotically (case (a) of lemma 2). Models asymptotic to elliptic self--similar solutions are contained in this class.
\item asymptotic to vacuum LTB that are elliptic and comply with (\ref{FFr}), so that $\KK_q$ decays faster than $R^{-2}$ ($\beta>2$), but $\rho_q$ decays slower than $R^{-3}$ ($\alpha<3$) from (\ref{qvars})). Hence $1-\HH_p/\HH_q\to 0$ and $\hOm\to\hOmty\geq 1$ hold, but we have $\Dm\to\Dm_{_\infty}>-1$ (from (\ref{Dadef})).  
\item asymptotic to vacuum LTB that are hyperbolic with $\FF\to\Fty>1$ (case (\ref{limHpHqF0})) and that are MD and G, since $1-\HH_p/\HH_q\to 0$ with $\hOmty= 1$ (MD) and $0<\hOmty<1$ (G), and $\Dm\to\Dm_{_\infty}>-1$ because $\rho_q$ decays slower than $\KK_q$ (since $x\to 0$ or $x\to x_{_\infty}$ in (\ref{x})). Models asymptotic to hyperbolic self--similar solutions are contained in this class. 
\end{enumerate}
\item [Case (C): there may be effective acceleration] (Limit of $\hq$ is the difference of two positive quantites) in models complying with
\begin{equation}\fl \KK_q<0\;\;\hbox{(hyperbolic)},\qquad \FF\to \infty,\qquad \hOm\to \left\{ \begin{array}{l}
 0<\hOmty<1,\quad\;\; \hbox{(G)}, \\  
 \hOmty=1, \qquad \hbox{(MD)},  \\ 
 \end{array} \right.\label{modsApos2}\end{equation}
which are the MD and G models of case (b) above, and lead to the exact asymptotic limit:
\begin{equation} \hq\to \hqty=-2\xi\,\left(\xi+|\Dk_{_\infty}|\right)+\frac{\hOmty}{2}\,\left(1-|\Dm_{_\infty}|\right),\label{W01}\end{equation}
where  $\Dk_{_\infty}$ and $\Dm_{_\infty}$ must be negative because $\rho'_q<0$ and $\KK'_q<0$ must hold asymptotically to reach $\rho_q\to 0$ and $\KK_q\to 0$. The sign of the limit of $\hq$ in (\ref{W01})  depends on the models being MD or G:
\begin{description}
\item[G models:] we have $\alpha=\beta=\gamma<2$, with (\ref{W01}) taking the form
\ba\fl  \hqty = -\frac{\gamma^2(2-\gamma)(10-\gamma)}{288}+\frac{\hOmty}{2}\left(1-\frac{\gamma}{6}\right),\qquad \hOmty =\frac{2m_0}{2m_0+|k_0|}<1,\nonumber\\\fl\label{W01G}\ea
which contains the VD models as the limiting case $\hOmty\to 0$. The graph of $\hqty$ as a function of $\gamma$ for various values of $\hOmty$ is displayed by figure \ref{f5}, revealing that $\hqty<0$ holds only for $\hOmty<0.09$ and taking the largest values (and ranges of $\gamma$) in the limit $\hOmty\to 0$ (case (A) above). 
\item[MD models:] equation (\ref{W01}) takes the form:
\begin{equation}\fl \hqty = -\frac{\alpha^2(\beta-2)(\beta-2\alpha+10)}{18(\beta-\alpha+4)^2}+\frac{1}{2}\left(1-\frac{\alpha}{6}\right),\qquad \alpha<\beta<2.\label{W01MD}\end{equation}
whose sign can be found by plotting the right hand side of this equation as a function of $\alpha,\beta$. Such plot (not displayed) reveals that there is no effective acceleration for these MD models, since $\hqty$ is negative only in a range of values for which $\alpha>\beta$. 
\end{description}
\item [Case (D): there may be effective acceleration] if $\hq<0$ as $\hq\to 0$. This occurs in LTB models that are:
\begin{enumerate}
\item asymptotic to the Milne Universe, since $1-\HH_p/\HH_q\to 0$ but $\hOm\to 0$.
\item elliptic asymptotic to Schwarzschild. These models comply with (\ref{FFr}), with $\KK_q$ decaying faster than $R^{-2}$ ($\beta>2$), so that there is a TV of $\FF$ and $\FF\to 1,\,\,\Q>0$ hold asymptotically (case (b) of lemma 2), but $\rho_q$ decays as $R^{-3}$ ($\alpha=3$), so that  $1+\Dm\to 0$ holds asymptotically. Notice that, from (\ref{qvars}), (\ref{asconvm}) and (\ref{MF}), we have in this limit $M\to m_0R_0^3 =$ const, which can be identified with an asymptotic Schwarzschild mass $M_{\textrm{\tiny{schw}}}$~\cite{RadAs}.  
\item hyperbolic with $\FF\to\Fty>1$ (case (\ref{limHpHqF0})) and that are VD, since we have $1-\HH_p/\HH_q\to 0$ but also $\hOm\to 0$ holds. 
\end{enumerate}
\end{description}

\subsection{A closer look at models of Case (D).}
We examine below the possibility that $\hq<0$ holds as $\hq\to 0$ for models asymptotic to Milne (case (D)(i)) and elliptic models converging to Schwarzschild (the case (D)(ii)). 
\begin{description}
\item[Models converging to Milne.] These models are characterized by asymptotic forms in (\ref{asconvm}) and (\ref{asconvk}):  $\rho_q\sim m_0r^{-\alpha}$ and $|\KK_q|\sim k_0>0$ ($\beta=2$) up to the first leading term. We consider their asymptotic forms up to a second order term:
\begin{equation}\fl \rho_q \sim m_0 r^{-\alpha}+m_1 r^{-\alpha_1},\quad \alpha<\alpha_1,\qquad
|\KK_q| \sim k_0 + k_1 r^{-\beta_1},\end{equation}
where $k_1$ and $\beta_1$ are positive. Following the guidelines outlined in section 6.1, the best conditions for $\hq<0$ (lowest possible near constant density and significant gradients of $\KK_q$), we assume that $\alpha>\beta_1$. Hence, we get from (\ref{Hqsq}), (\ref{slaw_Dh}), (\ref{Omdef}) and (\ref{asMF})
\bse\ba \fl \HH_q \approx \sqrt{k_0}\left[1+\frac{k_1}{2k_0}r^{-\beta_1}\right],\quad \Dh \approx -\frac{\beta k_1}{3k_0}r^{-\beta_1},\quad \hOm\approx \frac{m_0}{m_0}r^{-\alpha},\\
\fl \HH = \HH_q(1+\Dh) \approx \sqrt{k_0}\left[1+\frac{k_1}{2k_0}\left(1-\frac{\beta}{3}\right)r^{-\beta_1}\right],\quad  \FF\approx \sqrt{k_0}r,\ea\ese
where we are using the symbol $\approx$ instead of $\sim$ because the convergence may not be uniform. In order to compute the asymptotic form of $\HH_p$, we follow the proof of the limit (\ref{limHpHq}) in Appendix E to evaluate the following integrals from (\ref{avedef}) and (\ref{Vpdef}):
\ba \fl \Vp(r) \sim \tilde\Vp(y)+\frac{2\pi}{\sqrt{k_0}}r^2,\quad \tilde\Vp(y)=\Vp(y)-\frac{2\pi y^2}{\sqrt{k_0}},\label{asMilne1}\\
\fl 4\pi\int_0^r{\frac{\HH \Vq'}{\FF}\dd\rr}\sim I_p(y) +2\pi r^2\left[1+\frac{k_1(1-\beta_1/3)}{2k_0(1-\beta_1/2)}r^{-\beta_1/2}\right],\label{asMilne2}\\
\fl\hbox{where:}\quad I_p(y)=4\pi\int_0^y{\frac{\HH \Vq'}{\FF}\dd\rr}-2\pi r^2\left[1+\frac{k_1(1-\beta_1/3)}{2k_0(1-\beta_1/2)}y^{-\beta_1/2}\right].\label{asMilne3}\ea
From (\ref{avedef}) and considering $r\gg y$ we have 
\begin{equation} \fl \HH_p \approx \sqrt{k_0}\left[1+\frac{k_1(1-\beta_1/3)}{2k_0(1-\beta_1/2)}r^{-\beta_1}\right] \quad\Rightarrow\quad 1-\frac{\HH_p}{\HH_q}\approx -\frac{k_1\beta_1\,r^{-\beta_1}}{2k_0(1-\beta_1/2)},\end{equation}
which leads to 
\begin{equation}\fl \hq \approx -\frac{k_1\beta_1(3-\beta_1)}{6k_0(2-\beta_1)^2}\,r^{-2\beta_1}+\frac{m_0}{2k_0}r^{-\alpha}\left[1-\frac{\alpha}{3}-\frac{m_1(\alpha_1-\alpha)}{m_0}r^{\alpha-\alpha_1}\right].\end{equation}
Hence, condition (\ref{newApos}) is fulfilled in this case if $\beta_1<3$ and $\alpha=3$ hold (since $\alpha_1>\alpha$). It is also fulfilled for $\alpha<3$ as long as $\beta_1<3$ and $\alpha>2\beta_1$ hold, which implies that $\beta_1<3/2$ must also hold (since $\alpha< 3$). 

\item[Elliptic models converging to Schwarzschild.] In this case $\Dm\to -1$ implies $\rho_q \propto r^{-3}$, we consider the following asymptotic convergence forms up to second order
\begin{equation}\fl \rho_q \sim m_0 r^{-3} - m_1 r^{-\alpha_1}, \quad \alpha_1>3,\qquad \KK_q \sim k_0r^{-\beta},\quad \beta>\alpha_1,\end{equation}
leading to
\bse\ba \fl \HH_q \approx \sqrt{2m_0}\, r^{-3/2}\left[1-\frac{m_1}{4m_0}\,r^{3-\alpha_1}\right],\qquad \FF \approx 1-\frac{k_0R_0^2}{2}\,r^{2-\beta},\label{asell1}\\
\fl \Dm \approx -1 + \frac{m_1(\alpha_1-3)}{3m_0}\,r^{3-\alpha_1},\qquad \Dk\sim -\frac{\beta}{3},\label{asell2}\\ \fl 2\Dh \approx \Dm,\qquad  \hOm \approx 1-\frac{k_0}{2m_0}r^{3-\beta}.\label{asell3}\ea\ese
Considering that $\HH=\HH_q(1+\Dh)$ and using (\ref{slaw_Dh}), we construct the same type of integrals as in (\ref{asMilne1})--(\ref{asMilne3}) to compute $\HH_p$, leading to
\begin{equation} \HH_p \approx  \sqrt{2m_0}\, r^{-3/2}\left[1-\frac{m_1(\alpha_1-15/4)}{2m_0(9/2-\alpha_1)}\,r^{3-\alpha_1}\right],\end{equation}
so that using (\ref{asell1})--(\ref{asell3}) and (\ref{newApos}), and bearing in mind that $\beta >\alpha_1$, we get
\begin{equation} \hq \approx -\frac{m_1(\alpha_1-3)^2}{6m_0(9/2-\alpha_1)}\,r^{3-\alpha_1}, \end{equation}
which is negative if $\alpha_1<9/2$. If we assume that $3<\beta <\alpha_1$ we obtain $\hq\propto -r^{3-\beta}>0$. In either case, the value of $\hq$ is much smaller than in the hyperbolic models in which $\hqty<0$, as can be appreciated by comparing the curves of $\hq$ in figures 7a and 7b. It is important to remark that the fact that $\beta>3$ implies that $\Dik<-1$ must hold asymptotically (from (\ref{FFr})), hence as we show in section 13 (see figure 11b), local spatial curvature $\KK=\RR/6$ is asymptotically negative for these elliptic models. 
\end{description}
\begin{figure}[htbp]
\begin{center}
\includegraphics[width=3.5in]{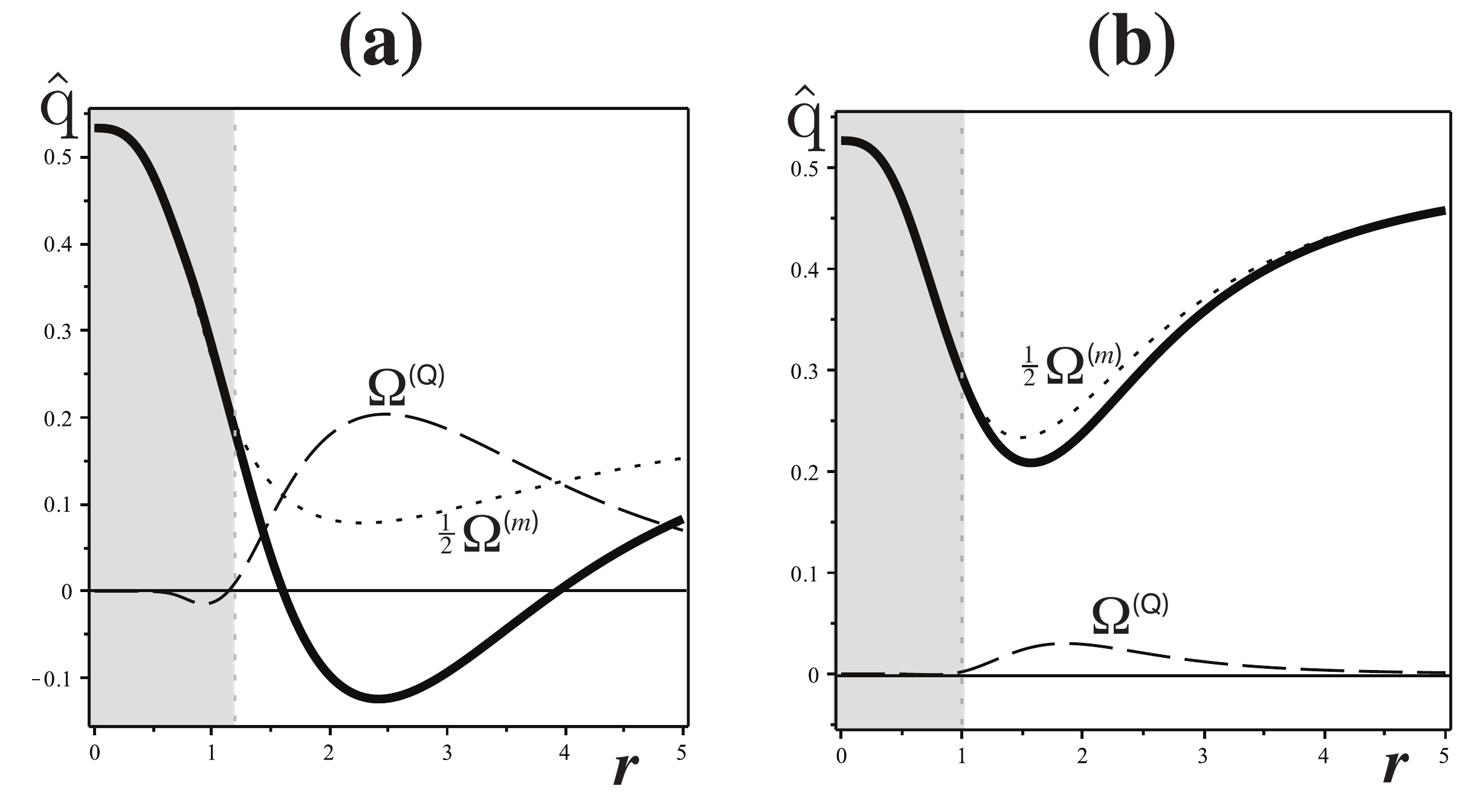}
\caption{{\bf Effective acceleration in the intermediate radial range: models converging to FLRW.} The figures depict the deceleration parameter  $\hq$ defined in (\ref{newApos}) (thick solid curve), together with the Omega factors for back--reaction, $\OmC$, and local density, $\Oml$ (dashed and dotted curves), for a fiducial initial slice $t=t_0=0$ of a mixed elliptic/hyperbolic models that converge in the asymptotic radial range to negatively curved FLRW (panel (a)) and Einstein de Sitter (panel (b)) backgrounds. Both models have the same central density and comparable density gradients in the elliptic core (the shaded area). The curves correspond to the initial value functions: $(4\pi/3)\rho_{q0} = 0.1+9.9/(1+r^3),\,\, \KK_{q0} = -0.75 -1.55/(1+r^3)$ (panel (a)) and $(8\pi/3)\rho_{q0} = 1.0+9.0/(1+r^3),\,\, \KK_{q0} = (1-r^2)/(1+r^3)$ (panel (b)). Notice that $\hq$ tends as $r\to\infty$ to positive values in both cases (deceleration) but becomes negative ($\hq \approx -0.1$) in the intermediate radial range of the model converging to negatively curved FLRW. To obtain the same negative value of $\hq$ for the model asymptotic to Einstein de Sitter its central density must be about 100 times larger. }
\label{f6}
\end{center}
\end{figure}
\begin{figure}[htbp]
\begin{center}
\includegraphics[width=3.5in]{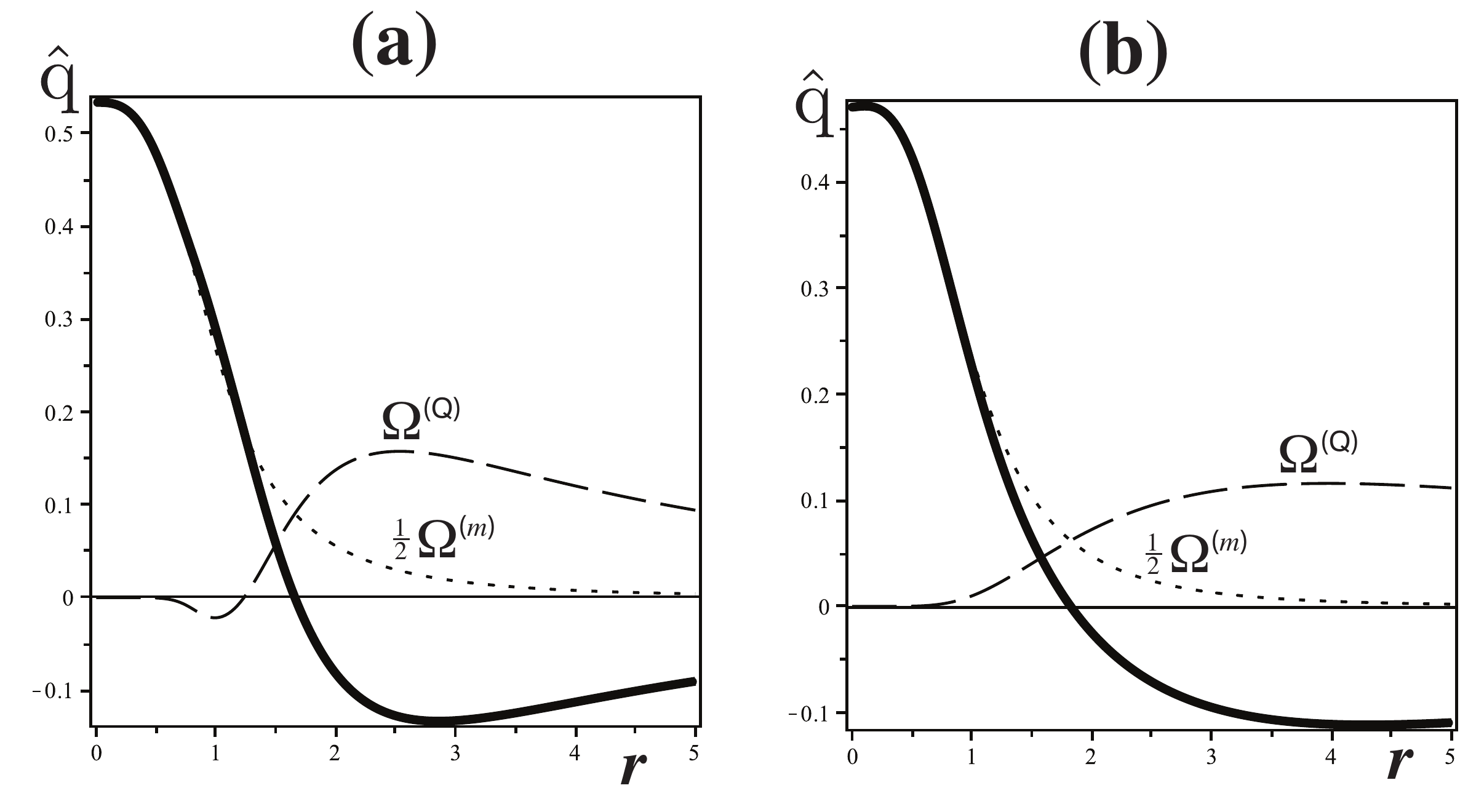}
\caption{{\bf Effective acceleration in the intermediate radial range: models converging to vacuum LTB.} The figure depicts the deceleration parameter $\hq$ (thick curve), together with $\OmC$ and $\Oml$, for an initial slice $t=t_i=0$ for two models radially converging to a vacuum LTB state: the elliptic model asymptotic to Schwarzschild (panel (a)) and the VD hyperbolic model of class (A) of section 8 (panel(b)). The parameters of the models are: $(4\pi/3) \rho_{q0} = 10/(1+r^3),\,\, \KK_{q0} = 1.8/(1+r^8)$ (panel (a)) and $(4\pi/3) \rho_{q0} = 10/(1+r^3),\,\, \KK_{q0} = -1.25/(1+r^{5/4})$ (panel (b)). Notice that $\hq$ is negative in the intermediate and asymptotic radial range, and in the limit $r\to\infty$ we have:  $\hq\to 0$ for the elliptic model and $\hq\to \hqty<0$, with $\hqty$ given by (\ref{W00}), for the hyperbolic model. }
\label{f7}
\end{center}
\end{figure}
\section{Effective acceleration in the radial intermediate range of density clumps.} 

If we have a clump density profile, either in domains of hyperbolic or elliptic models or mixed  elliptic/hyperbolic configurations, we can think of an ``intermediate'' radial range associated with the transition between ``near center'' and asymptotic conditions. Evidently, the radial gradients of the basic scalars $\rho,\,\KK,\,\HH$ take their largest value in this region, and thus we can apply the guidelines specified in section 6.1 that favor accelerating domains like (\ref{Void1}) and (\ref{Void2}).  Following the same assumptions described in section 8 for void profiles, the combination of small $\Oml$ and large $\OmC$ occurs in this range if there are large density and/or curvature gradients, but now the low density follows as a sharp density decay  from a large central value, remaining sufficiently small in the full range $r>r_2$. Hence, the possibilities for mid--range accelerating domains strongly depend on asymptotic convergence of the models. We examine these possibilities for models converging to FLRW and vacuum LTB states. 

\begin{description}

\item [Models converging radially to FLRW.] 
While $\hq$ eventually becomes positive for sufficiently large $r$ (from (\ref{radasFLRW})), it can be negative in a low density region in the intermediate radial range. The conditions for a small $\Oml$ are sensitive to both $1+\Dm$ and $\hOm$. We have the following cases (see table 3 and \cite{RadAs}): 
\begin{enumerate}    
\item {\underline{Elliptic and hyperbolic models converging to Einstein de Sitter}}. These models must be MD in the classification (\ref{x}) and are characterized by the asymptotic limit $\hOmty=1$. Conditions for a mid-range $\hq<0$ are far more restrictive for elliptic models because $\hOm\geq 1$ must hold in the whole radial range, and thus $\Oml\ll 1$ can only occur if $1+\Dm\to 0$ (very large density gradient: scenario (\ref{lowdens3})). The conditions are similar but less stringent for hyperbolic models because $0<\hOm<1$ holds in the whole radial range. 
\item {\underline{Hyperbolic models converging to negatively curved FLRW}}. These must be G models in the classification (\ref{x}). The conditions for a mid--range effective acceleration are far more favorable than for models asymptotic to Einstein de Sitter. Since $0<\hOmty<1$ holds, then it is easier to obtain $\hOm\ll 1$ in the intermediate radial range if we choose $\hOmty\ll 1$). 
\end{enumerate}
The comparison between the deceleration parameter $\hq$ in hyperbolic models with similar density profiles that are asymptotic to these FLRW backgrounds is displayed in figures 6a and 6b. See sections 13 and 15 for further discussion on the mid--range effective acceleration in models asymptotic to FLRW.
\item [Models converging radially to LTB vacuum.]  The conditions for a mid--range effective acceleration are easier to meet because of the asymptotic low density: 
\begin{enumerate}
\item {\underline{Elliptic models.}} Since we have $\hOm>1$ for all $r$, then $\Oml\ll 1$ can only be achieved if $1+\Dm\ll 1$ holds (scenario (\ref{lowdens3})). This is far more likely to  occur (from (\ref{qvars}), (\ref{mkH}) and (\ref{Dadef})) in the models of case (D)(ii) in section 8, in which $\rho_q\to 0$ sharply decays as $R^{-3}$ ($\alpha=3$), so that: $\Dm\to -1$,\,\, $M\to M_{\textrm{\tiny{schw}}}$ and $\rho\ll \rho_q$ hold (convergence to Schwarzschild \cite{RadAs}).  Figure 7a shows an effective acceleration in the mid radial range of a complete slice of this elliptic model. As shown by the figure, we have $\hq \approx -0.1$ with $\hq$ becoming less negative as $r$ grows so that $\hq\to 0$ asymptotically (in agreement with the results of section 9).
\item {\underline {Hyperbolic models.}} The fact that we have $\hqty<0$ given by (\ref{W00}) and (\ref{W01G}) in the VD models and in some of the G models with $\FF\to\infty$ clearly indicates that $\hq<0$ should hold in this range and tend asymptotically to (\ref{W00}) and (\ref{W01G}).  As depicted in figure 7b, there is effective acceleration in a complete slice in the intermediate radial range of a VD model corresponding to the case (A) of section 9, with $\hq\approx -0.1$ and $\hq$ becoming less negative as $r\to \infty$, reaching the asymptotic value $\hqty\approx -0.03$ obtained in (\ref{W0}) and (\ref{W00}). For other models the possibilities for $\hq<0$ depend on their being MD, G or VD in the classification given by (\ref{x}). In general, since $\hOm\to 1$ and $1+\Dm>0$ hold asymptotically for MD models, the conditions for them are analogous to those of models converging to an Einstein de Sitter background.      
\end{enumerate}        
\end{description}
\begin{figure}[htbp]
\begin{center}
\includegraphics[width=3.5in]{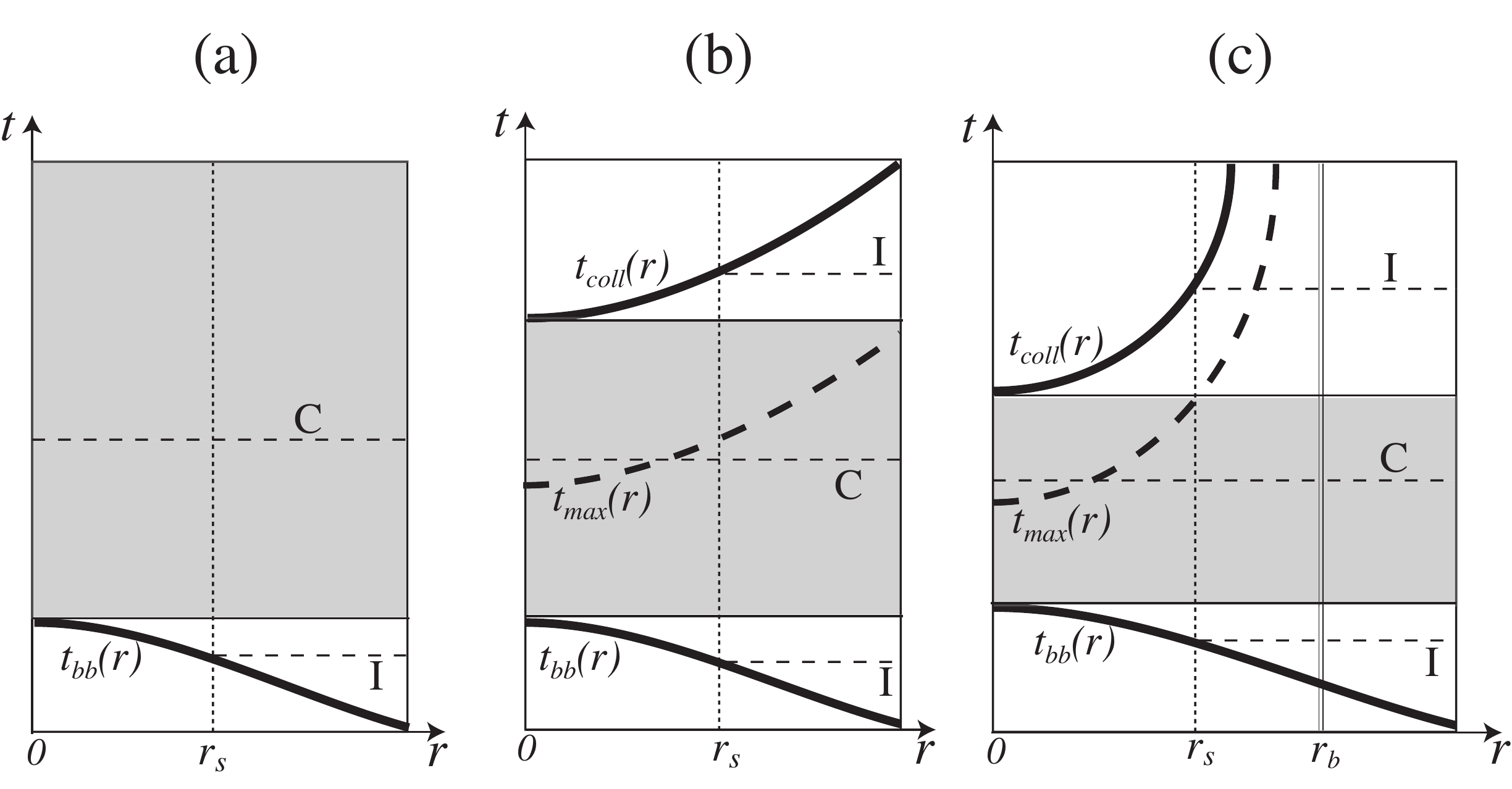}
\caption{{\bf  Expanding and collapsing singularities.} The figure depicts the curves $[\tbb(r),r]$ and $[\tcoll(r),r]$ in the $(t,r)$ plane, which provide the coordinate representation of non--simultaneous expanding (big bang) and collapsing (big crunch) singularities  for regular hyperbolic (panel(a)), elliptic (panel (b)) and mixed elliptic/hyperbolic (panel (c)) models. The shaded regions correspond to values of $t$ allowing for complete slices $\T[t]$, given by $t>\tbb(0)$ for hyperbolic models and $\tbb(0)<t<\tcoll(0)$ for  elliptic and mixed models. The letters ``C'' and `I'' denote individual complete and incomplete slices, so that the latter are only defined for $r>r_s$. For the elliptic and mixed elliptic/hyperbolic models the locus of the bounce surface where $\HH_q$ changes sign is depicted by the curve $\tmax(r)$, which in general (see \cite{RadAs}) diverges as $r\to\infty$ (elliptic models) and as $r\to r_b$ (mixed models). The comoving radius $r=r_b$ in panel (c) marks the boundary of the elliptic core, so that $\KK_q>0$ for $0\leq r<r_b$ and $\KK_q<0$ for $r>r_b$ (hyperbolic exterior). Notice that in this case, $\tcoll$ is only defined in the elliptic core, so that all layers collapse in this region but expand perpetually in the hyperbolic exterior. }
\label{f8}
\end{center}
\end{figure}
\section{Effective acceleration near the expanding and collapsing singularities.}

We examine separately the cases where these singularities are simultaneous and non--simultaneous.

\subsection{Effective deceleration near a simultaneous big--bang.}  

Regular hyperbolic and elliptic models exist in which $\tbb=t_s$ holds with $t_s$ constant and $\tbb$ given by (\ref{tbb}). In elliptic models we can have either $\tbb'=0$ or a simultaneous collapse $\tcoll'=0$ (but not both) \cite{kras2,KH4,RadAs,RadProfs,suss10a}. The behavior of scalars approaching these singularities can be examined in the limit $t\to t_s$ for a fixed arbitrary $r$. The following asymptotic limits hold in this limit when $\tbb'=0$ (the same limits hold for a simultaneous collapse, see \cite{RadProfs,suss10a} for detail):
\begin{equation} \hOm\to 1,\qquad \Dm\to 0,\qquad \Dh\to 0,\label{simbb1}\end{equation}
while imposing $\tbb'=0$ in (\ref{tbb}) and expanding (\ref{slaw_H}), (\ref{cthyp}), (\ref{ctell}) (expanding phase) and (\ref{Gamma}) for  $a\approx 0$ (which is equivalent to $t\approx t_s$) we obtain
\begin{equation}\fl a^{3/2} \approx \frac{3}{2}\sqrt{2m_{q0}} (t-t_s),\qquad \Gamma \approx 1+\Dim,\qquad \HH_q \approx \frac{2}{3 (t-t_s)}.\label{simbb2}\end{equation}
where $2m_{q0}=(8\pi/3)\rho_{q0}$. Since (\ref{simbb1}) and (\ref{simbb2}) imply $\HH\approx \HH_q$, then considering (\ref{Ridef}) and (\ref{LGdef})  the integrals in the definition (\ref{avedef}) of $\HH_p$ are
\bse\ba \fl \int_0^r{\frac{\HH R^2R'\dd\rr}{\FF}}=\int_0^r{\frac{\HH a^3\Gamma\rr^2\dd\rr}{\FF}}\approx \frac{3}{2}(t-t_s)\int_0^r{\frac{2m_{q0}(1+\Dim)\rr^2\dd\rr}{\FF}},\\ 
\fl \int_0^r{\frac{R^2R'\dd\rr}{\FF}}=\int_0^r{\frac{ a^3\Gamma\rr^2\dd\rr}{\FF}}\approx \frac{9}{4}(t-t_s)^2\int_0^r{\frac{2m_{q0}(1+\Dim)\rr^2\dd\rr}{\FF}},\ea\ese
and clearly imply that  $\HH_p\approx \HH_q$ for $t\approx t_s$. Hence, we have $\HH_p/\HH_q\to 1$ as $t\to t_s$. As a consequence, and considering (\ref{simbb1}), we have in (\ref{newApos}): $\OmC\to 0$ and $\Oml\to 1/2$, and thus $\hq\to 1/2>0$ holds in this limit. Therefore, there is effective deceleration near a simultaneous bang singularity in hyperbolic and elliptic models.~\footnote{This result was inferred but not proved in reference \cite{LTBave3}.} The same result occurs near a simultaneous crunch ($\tcoll'=0$) in elliptic models with this feature.

\subsection{Non--simultaneous singularities.}

As shown in figure 8, the coordinate locii of non--simultaneous curvature singularities are curves $[\tbb(r),r]$ (bang) and $[\tcoll(r),r]$ (crunch) in the $(t,r)$ plane, with the bang and crunch functions, $\tbb$ and $\tcoll$, given by (\ref{tbb}) and (\ref{tcoll}). The slices $\T[t]$ of hyperbolic models marked by $t<\tbb(0)$, and in elliptic or mixed models those marked by $t<\tbb(0)$ and $t>\tcoll(0)$ (slices ``I'' in figure 8), do not cover the radial range $0\leq r\leq r_s$, where $r_s$ is given generically by $t=\tbb(r_s)$ or $t=\tcoll(r_s)$ for these values of $t$. The integrals in the definitions (\ref{avedef}) and (\ref{aveqdef}) for these slices must be evaluated in the reduced domains $\vartheta_s[r]\equiv \{\rr\,|\,r_s<\rr\leq r\}\subset\vartheta[r]$. Notice that even if the scalars $\{\rho,\,\HH,\,\KK\}$ diverge as $r\to r_s$, their corresponding integrands in (\ref{avedef}) and (\ref{aveqdef}) are bounded (see (\ref{expbb2}) further ahead).

Since  we have $t-\tbb\approx \tbb'(r_s)(r-r_s)$ and $t-\tcoll\approx \tcoll'(r_s)(r-r_s)$ in domains with $r\approx r_s$, the limits $t\to\tbb$ and $t\to\tcoll$ are (in general) equivalent to limits $r\to r_s$. However, this is not the case for domains in which $\tbb'(r_s)\approx 0$ and $\tcoll'(r_s)\approx 0$, such as very near the center ($r_s\approx 0$) and in  the radial asymptotic range ($r_s\to\infty$) of LTB models asymptotically converging to FLRW or Milne (see \cite{RadAs}). In these cases the conditions of a simultaneous singularity that were examined previously apply, and thus we have in them effective deceleration even close to singularities. 

Excluding domains in which $\tbb'(r_s)\approx 0$ and $\tcoll'(r_s)\approx 0$ hold, the following strict limits hold as $r\to r_s$ in both $t\to\tbb$ and $t\to\tcoll$ (see \cite{RadProfs}): 
\begin{equation} \mathop {\lim }\limits_{r \to r_s } \Dm = -1,\qquad \mathop {\lim }\limits_{r \to r_s } \hOm = 1. \label{DmOmsing}\end{equation}
Since (\ref{DmOmsing}) implies $\Oml\to 0$ and $\hq\to \OmC$ in (\ref{newApos}), we obtain a nonzero limit of $\hq$ as $r\to r_s$ only if the limit of the ratio $1-\HH_p/\HH_q$ is nonzero, as both $\HH_p$ and $\HH_q$ diverge positively (negatively) as $t\to\tbb$ ($t\to\tcoll$). Using the definitions (\ref{avedef}) and (\ref{aveqdef}) for $A=\HH$ in the integration range $\vartheta_s[r]$ together with (\ref{LGdef}) we obtain the following rigorous general result:
\begin{equation} \fl \mathop {\lim }\limits_{r \to r_s }\frac{\HH_p(r)}{\HH_q(r)}=\mathop {\lim }\limits_{r \to r_s } \frac{\int_{r_s}^r{R^2R'\dd\rr}}{\int_{r_s}^r{\FF^{-1} R^2R'\dd\rr}}\, \frac{\int_{r_s}^r{\HH\FF^{-1}R^2R'\dd\rr}}{\int_{r_s}^r{\HH R^2R'\dd\rr}}=1 \quad \Rightarrow\quad \mathop {\lim }\limits_{r \to r_s }\hq =0,\label{qsing}\end{equation}
were we applied L'H\^opital's rule to each quotient above (since each one of the integrals vanishes separately as $r\to r_s$).

The strict limit (\ref{qsing}) provides no information on the sign of $\hq$ for domains in slices $\T[t]$ that intersect the singularities for which $r>r_s$ (marked by ``I'' in figures 8a--8c). For domains that are `close' to the singularities in which $r\approx r_s$ we can evaluate the sign of $\hq$ by using the analytic solutions (\ref{cthyp})--(\ref{ctell}) and the scaling laws (\ref{LGdef}), (\ref{slaw_mk}) and (\ref{Gamma}) to obtain the following expansions for the integrands in the definition (\ref{avedef}):
\begin{equation} \fl \frac{\HH  a^3 \Gamma r^2}{\FF} \approx \frac{-4\pi\rho_{q0}r^3 \,\tbb'(r_s)}{3\FF},\qquad \frac{a^3 \Gamma r^2}{\FF} \approx \frac{4\pi\rho_{q0}r^3}{\FF}[\tbb'(r_s)]^2(r-r_s),\label{expbb2}\end{equation}
which are bounded as $r\to r_s$ (the minus sign is consistent with the regularity conditions (\ref{noshxGh})--(\ref{noshxGe})). Notice that for $r\approx r_s$ in the case $t\approx \tcoll$ we obtain the same expansions  (\ref{expbb2}), with $\tcoll'(r_s)$ instead of $\tbb'(r_s)$ and without the minus sign (consistent with the regularity conditions (\ref{noshxGe}) that require $\tcoll'>0$). Inserting the first order expansion $S\approx S(r_s)+S'(r_s)(r-r_s)$ for $S=m_{q0}r^3/\FF=M/\FF$ into the integrands (\ref{expbb2}) and evaluating the integrals in (\ref{avedef}) and (\ref{aveqdef}) the ratio $\HH_p/\HH_q$ becomes at first order in $r-r_s$
\begin{equation} \frac{\HH_p}{\HH_q}\approx  1+\left[\frac{\FF'}{6\FF}\right]_s(r-r_s),\label{expbb3}\end{equation}
where ${}_s$ denotes evaluation at $r=r_s$. Considering that $\hOm\approx 1+O(L)$ and expanding  $1+\Dm$ in (\ref{slaw_mk}) with $\Dim\approx [\Dim]_s$, together with (\ref{expbb3}), yields a deceleration factor $\hq$ in (\ref{newApos}) at first order in $r-r_s$:
\ba \fl \hq \approx \frac{1}{3}\left[-\frac{\FF'}{\FF}+\frac{3M'}{2M}\right]_s(r-r_s)=\frac{1}{3}\left[\frac{(\KK_{q0}r^2)'}{2[1-\KK_{q0}r^2]}+\frac{9(1+\Dim)}{2r}\right]_s(r-r_s),\label{hqrs}\nonumber\\\ea
where we used (\ref{Ridef}), (\ref{MF}) and (\ref{Dadef}). Since (\ref{noshxGh}) and (\ref{noshxGe}) imply $1+\Dim\geq 0$, then the necessary (not sufficient) condition for $\hq<0$ above is just the necessary condition (\ref{FFr}) evaluated at $r=r_s$. Considering the allowed form of $\FF$ and the results of lemmas 1 and 2 (see section 5 and \cite{sussBR}), we have the following possibilities:          
\begin{description}
\item[Domains in hyperbolic models.] For domains $r\approx r_s$ with very large $r_s$ in models converging to a vacuum LTB state, the asymptotic convergence forms (\ref{asconvm})--(\ref{asconvk}) (see table 2) transforms (\ref{hqrs}) into the asymptotic form:
\ba\fl \hq \approx \frac{1}{r_s}\left[-\frac{2-\beta}{6}|k_0|r_s^{2-\beta}+\frac{3[1-\alpha/3]}{2}\right](r-r_s),\qquad 0<\beta\leq 2,\quad 0<\alpha\leq 3,\nonumber\\ \label{hqrsh}\ea
Considering the values of the parameters $\alpha,\,\beta$ in table 2, the following cases arise for domains $r\approx r_s$ in the asymptotic radial range:
\begin{enumerate}
\item There is necessarily effective acceleration in all models in which $\FF\to\infty$ and $\FF'\to\infty$ hold, corresponding to $0<\beta<2$ (see table \ref{tabla2}). In particular, $\hq$ can be negative and large if $0<\beta\leq 1$.
\item There is necessarily effective deceleration in all models in which $\FF\to \Fty>1$ and $\FF'\to 0$ hold (corresponding to $\beta=2$), with the exception of VD models with $\alpha=3$, as in this case both terms inside brackets in (\ref{hqrsh}) vanish. 
\end{enumerate}
It is important to remark that the results above are fully consistent with the results of sections 9 and 10. However, a comparison with section 9 and table 2 reveals that an effective acceleration may exist in slices $t\approx \tbb(r_s)$ in some hyperbolic models in which this acceleration in latter times was either proven not to exist (MD models with $\FF\to\infty$) or be very constrained (G models with $\FF\to\infty$). While the asymptotic expansion (\ref{hqrsh}) is strictly valid only for large $r_s$, this expansion indicates that it is likely that a large effective acceleration may exist in domains with $r_s$ in the intermediate radial range, or even in complete slices $t>\tbb$ with $t\approx\tbb$.  As we show in the numeric example of figure 10a, this effective acceleration does occur (even in complete slices $t\approx \tbb$) in a mixed elliptic/hyperbolic model converging to a negatively curved FLRW state. 

\item[Domains in elliptic models.] In domains contiguous to the center in which (\ref{FFr}) does not hold (see lemma 2) and section 5), we have necessarily $\hq>0$. However, if there is a TV (turning value) of $\FF$ at some $r=\rtv$ (see lemma 2b), then $\hq<0$ may hold in more ``external'' domains with $\{r_s,r\}>\rtv$. However, since $\FF\leq 1$ holds in these domains, we have the same qualitative situation as in hyperbolic models with $\FF\to\Fty>1$. This can be appreciated by writing the asymptotic expression analogous to (\ref{hqrsh}):
\ba\fl \hq \approx \frac{1}{r_s}\left[\frac{2-\beta}{6} k_0 r_s^{2-\beta}+\frac{3[1-\alpha/3]}{2}\right](r-r_s),\qquad \beta\geq 2,\quad 0<\alpha\leq 3.\nonumber\\ \label{hqrse}\ea
Since we have $\beta>2$ when (\ref{FFr}) holds, the first term inside the brackets is negative but tends to zero as $r_s$ grows. Also, (\ref{tbb}) and (\ref{tcoll}) imply $\tcoll'>|\tbb'|$ for every $r$, hence the curve $\tcoll(r)$ will show a large deviation from the lines of constant $t$ (slice $t=t_2$ in figure 9a), and thus the condition $1+\Dm\approx 0$ (which implies $\Oml\approx 0$) is far less likely to hold in domains that approach or hit the collapse singularity than in domains that approach or hit the big bang (see the caption of figures 9a and 9b). Since $\hq>0$ holds for all models with $\alpha<3$ as $r_s$ becomes large, it is not likely that $\hq$ could be negative for smaller values of $r_s$. The situation may improve if $\alpha=3$ (models converging to Schwarzschild, case (D)(ii) of section 9), so that $\hq<0$ may hold in some mid--range domains (see figure 7a). Considering the arguments expressed above, the conditions for an effective acceleration near the expanding and collapsing singularities are (at best) similar to those of hyperbolic models with $\FF\to\Fty$, and far more restrictive than in the hyperbolic cases with $\FF\to\infty$. As shown in the numeric example of figure 10b, there is effective deceleration in domains hitting the collapse singularity.
\end{description}
\begin{figure}[htbp]
\begin{center}
\includegraphics[width=3.5in]{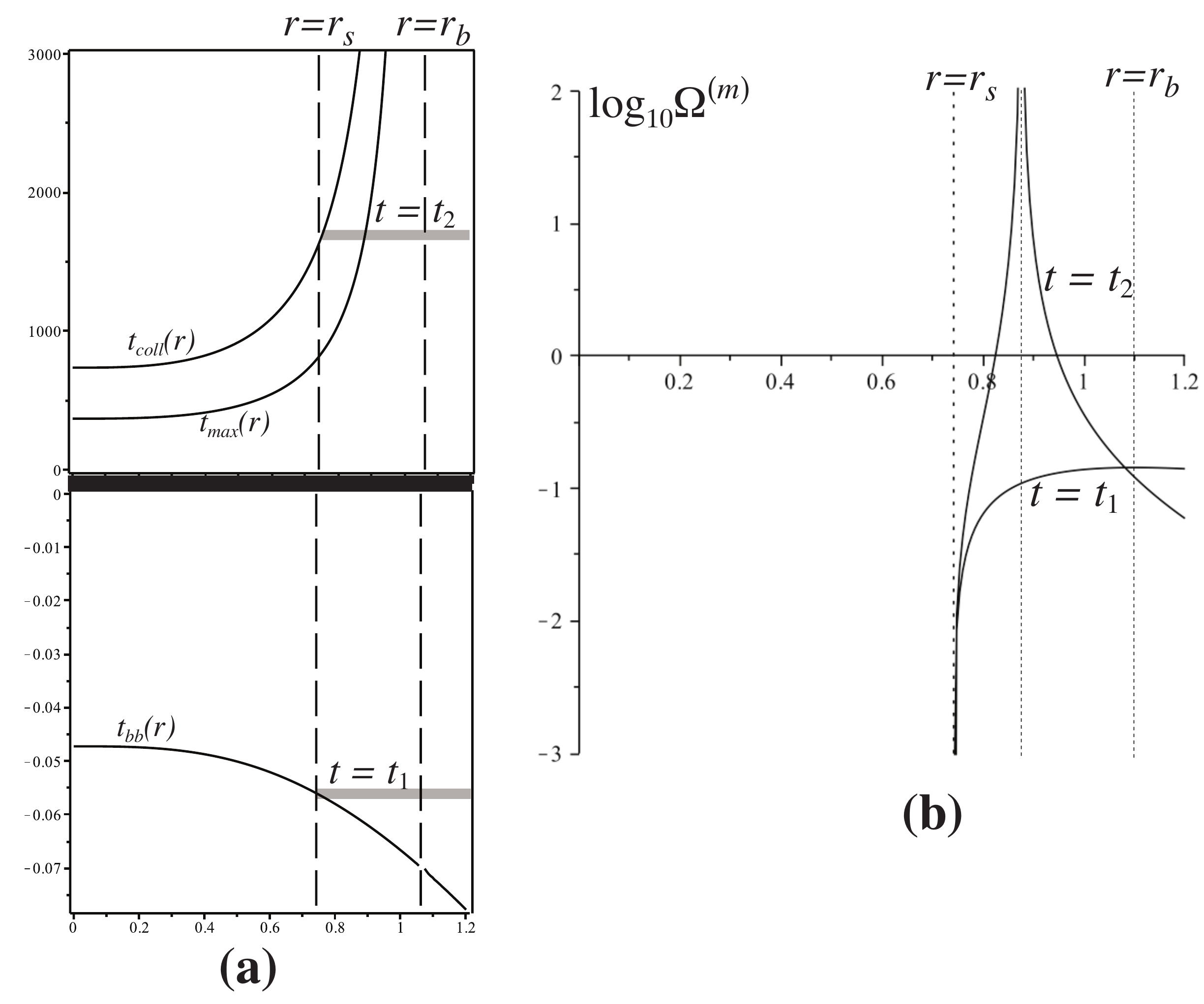}
\caption{{\bf  Local density near the singularities of the spherical collapse model.} Panel (b) depicts the radial profile of $\log_{10}\Oml$ for two incomplete slices marked by $t=-0.056$ (label ``$t=t_1$'') and $t=1629.01$ (label ``$t=t_2$'') that hit, respectively, the big bang $\tbb(r)$ and the big crunch $\tcoll(r)$ singularities in a mixed elliptic/hyperbolic model like that of figure 6a. The parameters of the model are the same $\KK_{q0}$ used in figure 6a, and $m_{q0}=(4\pi/3)\rho_{q0}=0.1+99.9/(1+r^3)$. Panel (a) shows each one of the slices in the $(t,r)$ plane. Notice that the vertical axes in the plot of panel (a) has very different scales for positive and negative times, so that $\tbb$ has a negligible time variation in comparison with $\tcoll$. While we have $\Oml\to 0$ as $r\to r_s$ for both slices, for the slice $t=t_2$ that hits $\tcoll(r)$ the function $\Oml$ diverges as the slice hits the bounce surface $\tmax(r)$, whereas for the other slice $\Oml\ll 1$ holds for all the radial range. Since $\Oml\ll 1$ facilitates $\hq<0$, then it is evident that conditions for an effective acceleration are much more likely to be met for slices $t=t_1$ near the big bang than $t=t_2$ near the collapsing singularity (a large central density $\Omega_{q0}(0)=\Oml(0)=100$ was selected to emphasize this feature). }
\label{f9}
\end{center}
\end{figure}
\begin{figure}[htbp]
\begin{center}
\includegraphics[width=3.5in]{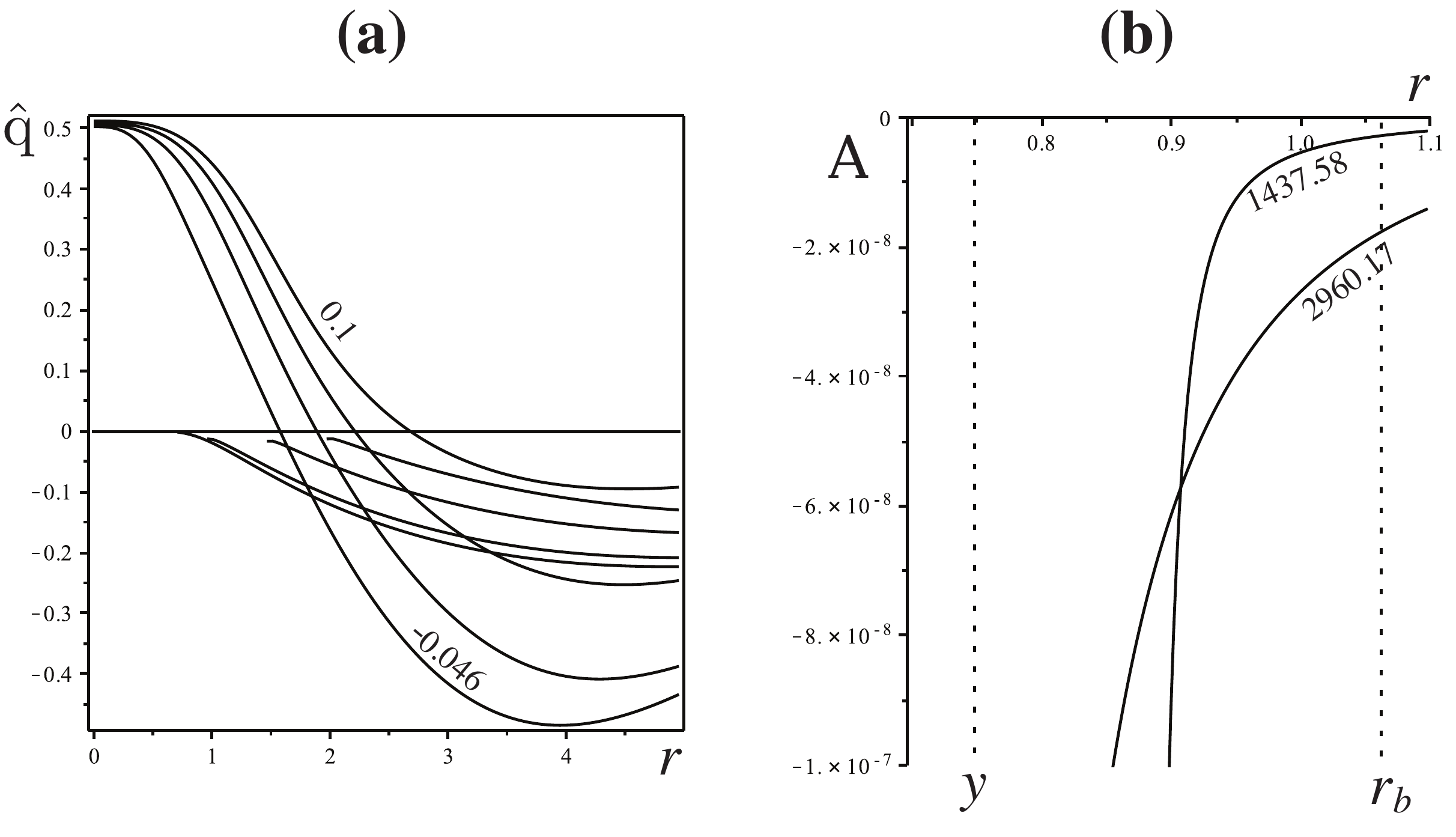}
\caption{{\bf  Effective acceleration near the singularities of the spherical collapse model.} Panel (a) depicts the deceleration parameter $\hq$ as a function of $r$ for slices near the big bang for the same model of figure 9. The complete slices are marked by $t= -0.046,-0.4,0,0.1$ (with $t_0=0$).  These curves extend all the way to $r=0$ and reach negative values only in the intermediate and asymptotic radial range well inside the hyperbolic exterior ($r_b=1.06$). The slices hitting the singularity are marked by $t=-0.055,-0.064,-0.096,-0.137$ and reach $\hq=0$ at various $r=r_s$. Panel (b) shows the acceleration function $\A(r)$ for the slices $t=1437.58,2960.17$ that hit the collapse at various $r=r_s$ (as $t=t_2$ in figure 9a). Since $\A(r)<0$, then $\hq>0$ holds for these slices and domains in them effectively decelerate. }
\label{f10}
\end{center}
\end{figure}
\section{Back--reaction and effective acceleration in structure formation scenarios.}

Useful toy models (known as ``spherical collapse models'') of cosmic structure in an expanding background follow from a mixed LTB configuration in which a collapsing elliptic core region ($0\leq r<r_b$) is smoothly matched ($\KK_q$ passes smoothly from positive to negative at $r=r_b$) to an expanding hyperbolic exterior ($r>r_b$). A configuration of this type is illustrated by figures 8c and 9a. Back--reaction have been studied for the spherical collapse model by R\"as\"anen \cite{ras2,ras3} (who did not assume a smooth transition) and Paranjape and Singh \cite{ParSin3}. Contrary to the claim expressed in \cite{ras2,ras3}, collapsing conditions do not favor an effective acceleration (this fact has been noted in the more recent references \cite{ras4,ParSin3}). However, an effective acceleration does arise in the intermediate radial range of the spherical collapse model in the example in figure (6a).   

The results of lemmas 1 and 2 apply for spherical collapse model configurations (see \cite{sussBR}). Since $\FF^2\leq 1$ must hold in the elliptic core and $\FF\geq 1$ holds in the  hyperbolic exterior, it is evident that the elliptic region must have a TV (turning value) of $\FF$ at some $\rr=\rtv<r_b$, so that $\FF$ must increase in the outer layers of the core (see figure 11a). The sign of $\QQ[r]$ depends on whether the domain $\vartheta[r]$ lies entirely within the core or reaches into the exterior region. In the latter case the sign of $\QQ[r]$ is the same as that of domains in hyperbolic models (lemma 1), while in the former case the sign is the same as in domains of elliptic models with a TV of $\FF$ (lemma 2(b)). Even if there are TV's of $\HH$, we have $\QQ[r]>0$ in all domains with sufficiently large $r\gg r_b$  extending in the hyperbolic region. The same remarks apply if we have a parabolic (instead of hyperbolic) exterior \cite{sussBR}, since in this case $\FF=1$ must hold for $r\geq r_b$, hence $\FF$ must have a TV and grow to reach $\FF=1$ inside the elliptic core, which means that $\QQ[r]>0$ holds already for domains reaching the external layers of the core.  

While the expansions (\ref{hqrsh}) and (\ref{hqrse}) are respectively valid in the elliptic core and the hyperbolic exterior, some important differences emerge in comparison with the pure hyperbolic and elliptic models: the big bang occurs in both regions, but the collapse singularity must occur only in the elliptic core. Hence, $\tcoll(r)$ only reaches up to $r=r_b$ and we have $\tcoll\to\infty$ as $r\to r_b$ (compare the curve $\tcoll$ in figures 8c and 9a with the same curve in figure 8b). As a consequence, if we consider domains $r\approx r_s$ with $r_s>\rtv$  near the collapse singularity, we have $\FF(r_s)\leq \FF(r_b)=1$ and $\FF'(r_s)\leq \FF'(r_b)$ for all $r_s$, and thus  the term $\FF'/\FF$ in (\ref{hqrse}) cannot be arbitrarily large. As a consequence, conditions for $\hq<0$ in the collapsing regime are (at best) similar to those of hyperbolic models with $\FF\to\Fty>1$, and thus are far more restrictive than near the big bang.  

Another nuance in domains near the collapsing singularity comes from the fact that every slice hitting $t=\tcoll$ will also hit the locus of the bouncing surface $\HH_q=0$: the curve $t=\tmax(r)$ in figures 8c and 9a, with $\tmax(r)\to\infty$ as $r\to r_b$. As shown in figure 9b, $\hq$ diverges at the value $r=\rmax$ such that $t=\tmax(\rmax)$ is reached along any slice $\T[t]$ with $t>\tmax(0)$, which includes the slices $t>\tcoll(0)$. However, this effect is simply an artifact of the definition of $\hq$ in (\ref{newApos}), as there is nothing peculiar in the back--reaction term in the numerator of (\ref{newApos}) as $\Q(r)\to \Q(\rmax)$. In principle, it could be possible (if $\rmax>\rtv$) that $\hq\to-\infty$ could occur as $r\to\rmax$ in these slices, though this possibility is quite contrived as we also have (by construction) $\hOm\to\infty$ as $r\to\rmax$. Since $\tcoll$ becomes very large (the curve $\tcoll(r)$ bends upwards and grows rapidly in figures 8c and 9a) in the outer parts of the elliptic core ($\rtv<r<r_b$), then in the slices that hit $\tcoll$ the domains $\vartheta_s[r]$ rapidly reach values of $r$ away from the curve $\tcoll(r)$ where the quantity $1+\Dm$ may significantly differ from zero (its strict limit (\ref{DmOmsing})), hence the term $\Oml$ in (\ref{Om}) that contains the positive contribution to $\hq$ in (\ref{newApos}) can easily become large enough to offset the negative contribution given by $\OmC$ in (\ref{OmC}). This possibility is depicted graphically in figure 9b. 

Since we can only prove analytically the value of $\hq$ in domains in which $r-r_s\approx 0$, we tested numerically the presence of an effective acceleration in domains in which  $r-r_s$ is not small in slices near the big bang and near the collapsing singularities (figures 10a and 10b), for the configuration depicted by figure 9b. Bearing in mind that $\hq$ diverges as $r\to\rmax$ in slices hitting the collapse singularity (see figure 9b), we computed for this case the scalar $\A$ in (\ref{Apos}) instead of $\hq$ [notice that $\A>0$ is equivalent to $\hq<0$]. As shown by figure 10a, slices near the big bang and intersecting it exhibit large effective acceleration (up to $\hq \approx -0.5$) in mid range domains extending into a hyperbolic exterior that is asymptotically FLRW with negative curvature ($\hOm = 0.3$). On the other hand, as shown in figure 10b, domains hitting the collapse singularity exhibit effective deceleration, just as domains in all complete slices (not displayed) near the collapsing singularity ($t\approx\tcoll(0)$ with $t<\tcoll(0)$). These numerical results are consistent with the analytic results and inferences obtained in this section, which clearly show that the conditions for $\hq<0$ are much more relaxed for $t\approx \tbb$ than for $t\approx \tcoll$. These results are also consistent with the numerical results in \cite{LTBave3}, as well as \cite{ras4,ParSin3}.          
       
\section{Effective acceleration in the time asymptotic range.}

Looking at the effective acceleration in the asymptotic limit $t\to\infty$ for finite comoving radii $r$ is only possible for hyperbolic models, as the evolution time range of dust layers in elliptic models is necessarily finite because they collapse in a ``big crunch'' singularity at finite $t$. The existence of an effective acceleration in this ``late time'' evolution was discussed in \cite{LTBave2}, but as we show in this section the results of this reference are mistaken.      

The asymptotic radial dependence of $\rho_q,\,\KK_q$  along the $\T[t]$ only depends on the initial conditions (the initial value functions $\rho_{q0},\,\KK_{q0}$, see \cite{RadAs}), which can be chosen freely (as long as they comply with regularity conditions). On the other hand, the asymptotic time dependence of these scalars along comoving worldlines is completely determined by the solutions (\ref{cthypv})--(\ref{cthyp}) of the Friedmann--like  equation (\ref{slaw_H}) (or (\ref{eqR2t}) in the standard variables). As a consequence, the conditions for the fulfillment of (\ref{newApos}) in the asymptotic ``late time'' regime are far more restrictive than those in the asymptotic radial regime.

Since we are interested in the limit $t\to\infty$ for comoving observers ($r$ finite) of hyperbolic models, we can consider the functions $x_i(r),\,y_i(r)$ in (\ref{cthyp}) as finite constants as we look at the late time regime from the asymptotic expansion of (\ref{cthyp}) on the variable $a$ (the scale factor defined in (\ref{LGdef})). Up to the leading term,  (\ref{cthyp}) yields the following late time ($t\gg t_0$) forms for the scale factors $a$ and $\Gamma$:
\ba a \approx a_{_\infty} = 1+|\KK_{q0}|^{1/2}\,(t-t_0), \label{L0}\\
 \Gamma \approx \Gamma_{_\infty} = 1 + \frac{3\,|\KK_{q0}|^{1/2}\,\Dik\,(t-t_0)}{2\,a_{_\infty}}\approx 1+\frac{3}{2}\Dik +O(t_0/t),\label{G0}\ea
The scale factors $a_{_\infty}$ and $\Gamma_{_\infty}$ approach the exact forms of $a$ and $\Gamma$ associated with the solution (\ref{cthypv}) for the vacuum LTB models. Therefore, for sufficiently large $t$ along comoving worldlines all non--vacuum hyperbolic models should converge to these models. As a consequence, applying (\ref{L0})--(\ref{G0}) to the scaling laws (\ref{slaw_Dh}), (\ref{Omdef}) and (\ref{slaw_mk}) yields the following asymptotic forms ($\tau\equiv (t-t_0)/t_0\gg 1$):
\bse\ba
\fl \rho =\rho_q(1+\Dm)\approx \frac{\rho_{q0}\,(1+\Dim)}{|\KK_{q0}|^{3/2}\,(1+\frac{3}{2}\Dik)\,\tau^3}+O(\tau^{-4}),\label{tasm}\\
\fl \tilde \HH_q \approx \frac{1}{\tau}+\frac{\frac{4\pi}{3}\rho_{q0}-2|\KK_{q0}|}{2|\KK_{q0}|^{3/2}\,\tau^2}+O(\tau^{-3}),\label{tasHq}\\
\fl \HH \approx \frac{1}{\tau}+\frac{\frac{4\pi}{3} \rho_{q0}(1+2\Dim-\frac{3}{2}\Dik)-2|\KK_{q0}|(1+\Dik)}{2|\KK_{q0}|^{3/2}\,(1+\frac{3}{2}\Dik)\,\tau^2}+O(\tau^{-3}),\label{tasHH}\ea\ese
It is straightforward to prove that $\A\to 0$ as $\tau\to\infty$. The limits $\HH_q\to 0$ and $\HH\to 0$ as $\tau\to\infty$ follow from (\ref{tasHq}) and (\ref{tasHH}),  and these limits imply $\HH_p\to 0$ (the proof is analogous to that of the limit as $r\to\infty$ in (\ref{A0}), replacing $r$ by $\tau$). Since $\HH,\,\HH_q,\,\HH_p$ tend to zero, we have $\Q\to 0$, and since $\rho\to 0$ follows from (\ref{tasm}), then $\A\to 0$ holds. However, it is not trivial to estimate if conditions exist (as claimed in \cite{LTBave2}) for $\A$ to be positive as it tends to zero in this limit. To address this issue we need to obtain the time asymptotic form of $\HH_p$, which  follows from integrating (over $r$) the time asymptotic  forms of the integrands of (\ref{avedef}) obtained from (\ref{L0}), (\ref{G0}) and (\ref{tasHH}), leading to: 
\begin{equation} \fl \HH_p \approx \frac{1}{\tau}+\frac{\int_0^r{[\frac{4\pi}{3}\rho_{q0}(1+2\Dim-\frac{3}{2}\Dik)-2|\KK_{q0}|(1+\Dik)]\FF^{-1}\rr^2\dd\rr}}{2\,\tau^2\,\int_0^r{|\KK_{q0}|^{3/2}\,(1+\frac{3}{2}\Dik)\FF^{-1} \rr^2\dd\rr}}+O(\tau^3),\label{tasHp}\end{equation}
which, by looking at the time asymptotic forms (\ref{tasHq})--(\ref{tasHH}), implies that the fluctuations $\HH-\HH_p$ and $\HH-\HH_q$ both decay (at least) as $\tau^{-2}$, so that $\Q(r)$ decays as $\tau^{-4}$. The sign of $\Q(r)$ in the late time regime will depend on the values of the coefficients of order $\tau^{-2}$ in the expansions (\ref{tasHq})--(\ref{tasHH}) and (\ref{tasHq}) (or on coefficients of higher order on $\tau^{-1}$ if the coefficients of order $\tau^{-2}$ vanish for a special choice of initial conditions). Therefore, for all initial conditions (compatible with regularity) and irrespective of the sign of $\Q(r)$ at late times, we have from (\ref{Cpos}) and (\ref{tasm}):
\begin{equation} \rho(r)\approx O(\tau^{-3})\gg \tilde\Q(r)\approx O(\tau^{-4}),\end{equation}
and thus, as long as $\rho>0$ holds for finite times (non--vacuum models), we have necessarily $\A<0$ (effective deceleration) as $\A\to 0$ (or, equivalently, $\hq>0$ as $\hq\to 0$) in the late time regime, thus violating conditions (\ref{effe_acc}), (\ref{Apos}) and (\ref{newApos}) in this regime. We remark that this result is valid even if density is very low (but nonzero), because as small as it may be for any finite time, it will still decay slower (as $\tau^{-3}$) than $\Q(r)$. As a consequence, it is evident that the results of reference \cite{LTBave2} are mistaken.  We may have $\A>0$ as $\A\to 0$ in the late time regime only for the vacuum LTB models (see figure 2)  or for vacuum regions (such as the vacuum central region and the Schwarzschild exterior of section 7.2).                
\begin{table}
\begin{center}
\begin{tabular} {|l|l|l|l|l|}
\hline
    Models     & limit of $\FF$   & type   & Parameters        & Effective \\
converging to  &  as $r\to\infty$ &        &  $\alpha,\,\beta$ & acceleration\\ 
\hline
    Vacuum LTB & $\to \Fty>1$ & MD & $\beta=2,\,0<\alpha<2$ & No\\
\cline{3-5}
               &              & G  & $\beta=\alpha =2$& No \\
\cline{3-5}
               &              & VD & $\beta=2,\,2<\alpha\leq 3$ & restricted \\
               &              &    &                            & ($\hq\to 0$) \\
\cline{2-5}
               & $\to\infty$  & MD & $0<\beta<2,\, \alpha<\beta<2$ & No \\
\cline{3-5}
               &              & G  & $0<\beta=\alpha<2$ & restricted \\
               &              &    &                    & ($\hOmty<0.065$) \\
\cline{3-5}
               &              & VD & $0<\beta<2,\,\beta<\alpha\leq 3$ & Yes: $\hq\to \hqty<0$\\ 
\hline
    Milne      & $\to\infty$  & VD & $\beta=0,\,0<\alpha\leq 3$ & restricted \\
               &              &    &                            & ($\hq\to 0$) \\
\hline
    open FLRW  & $\to\infty$  & G  & $\beta=\alpha=0$ & No\\ 
\hline
    Einstein  & $\to \Fty>1$ & MD & $\beta=2,\,\alpha=0$ & No\\
\cline{4-5}
 de Sitter    & $\to\infty$  &    & $0<\beta<2,\,\alpha=0$ &No\\ 
\hline
\end{tabular}
\caption{{\bf {Effective acceleration in the radial asymptotic range of hyperbolic models}}. The ``type'' classification ``MD'' (matter dominated),``G'' (generic) and ``VD'' (vacuum dominated) is given in equation (\ref{x}). The parameters $\alpha$ and $\beta$ in (\ref{asconvm})--(\ref{asconvk}) characterize the asymptotic convergence forms of $\rho_{q0}$ and $\KK_{q0}$, with the asymptotic forms of $M$ and $\FF$ given by (\ref{asMF}). All hyperbolic models comply with $\QQ[r]>0$ in their radial asymptotic range (see lemma 1).}
\end{center}
\label{tabla2}
\end{table}
\begin{table}
\begin{center}
\begin{tabular} {|l|l|l|l|l|}
\hline
    Models     & limit of $\FF$   & type   & Parameters        & Effective \\
converging to  &  as $r\to\infty$ &        &  $\alpha,\,\beta$ & acceleration\\ 
\hline
    Vacuum LTB & $\to \Fty<1$ & MD & $\beta=2,\,0<\alpha<2$ & No, ($\QQ[r]<0$)\\
\cline{4-5}
               &              & G  & $\beta=\alpha =2$& No, ($\QQ[r]<0$)\\
\cline{2-5}
               & $\to\Fty=1$  & MD & $\beta>2,\, \beta>\alpha$ & No for $\alpha<3$, \\
               &              &    & $0<\alpha\leq 3$          & restricted for  \\
\cline{3-4}
               &              & G  & $2<\beta=\alpha\leq 3$ &  $\alpha=3$ \\
               &              &    &                        &  ($\hq\to 0$)  \\ 
\hline
    Einstein  & $\to \Fty<1$ & MD & $\beta=2,\,\alpha=0$ & No, ($\QQ[r]<0$)\\
\cline{4-5}
 de Sitter    & $\to\Fty=1$  &    & $\beta>2,\,\alpha=0$ &No\\ 
\hline
\end{tabular}
\caption{{\bf {Effective acceleration in the radial asymptotic range of elliptic models}}. The table follows the same definitions and conventions of Table 1. All elliptic models listed in the table in which ``$\QQ[r]<0$'' is not stated comply with $\QQ[r]>0$ in their radial asymptotic range (see lemma 2).}
\end{center}
\label{tabla3}
\end{table}
\begin{figure}[htbp]
\begin{center}
\includegraphics[width=3.5in]{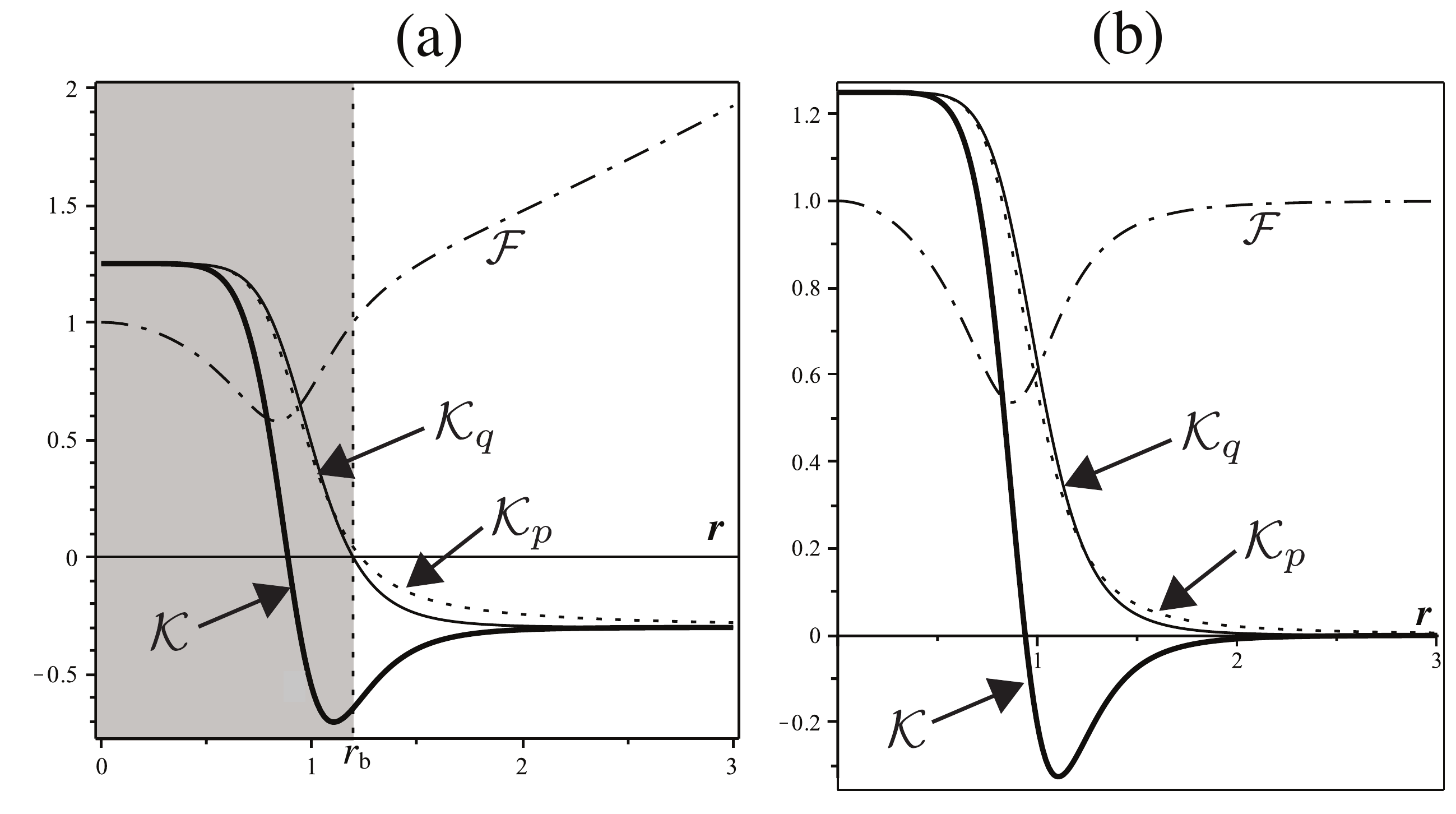}
\caption{{\bf  Spatial curvature in elliptic models and regions.} The figure displays the radial profile (slice $\tau=0$) of the local spatial curvature $\KK=\RR/6$ (thick solid curve), together with the quasi-local and average spatial curvature $\KK_q,\,\KK_p$ and the function $\FF$, for a mixed elliptic/hyperbolic model like that of figure 6a (panel (a)) and a pure elliptic model converging to Schwarzschild like that of figure 7a (panel (b)). For the model of panel (a) we used the same $(4\pi/3)\rho_{q0}$ of figure 6a with $\KK_{q0}=-0.3+1.55/(1+r^8)$, for panel (b) we used the same parameters of figure 7a. Notice that we have in both cases $\KK<0$ in regions where $\hq<0$ holds, even if both $\KK_q$ and $\KK_p$ are positive. The scalars $\KK,\,\KK_q,\,\KK_p$ exhibit the same qualitative behavior in all complete slices in these models. }
\label{f11}
\end{center}
\end{figure}
\section{Back--reaction, effective acceleration and spatial curvature.}

It is evident from table 1 that the conditions for $\QQ[r]>0$ are far less restrictive if $\KK_q<0$ (hyperbolic models), though a positive back--reaction is also possible if $\KK_q>0$ in domains of elliptic models in which $\FF'>0$, so that and $\KK_q$ is decreasing rapidly along $r$ (see section 5). While the compatibility between $\QQ[r]>0$ and a positive spatial curvature that is decreasing has been highlighted previously \cite{ras3}, it is important to discuss the following caveats.

The ``quasi--local'' spatial curvature, $\KK_q=\RR_q/6$ in (\ref{qvars}), which distinguishes elliptic ($\KK_q\geq 0$) and hyperbolic ($\KK_q\leq 0$) models or regions, is a different function from the averaged ($\kav=\KK_p=\RR_p/6$) and local  ($\KK=\RR/6$) spatial curvatures, the latter given by (\ref{ThetaRR}). Hence the signs of the three curvatures may not always coincide in a given domain. The averaged spatial curvature is always positive/negative in domains of elliptic/hyperbolic models. This follows in elliptic models from the fact that regularity conditions require $\KK'_q\leq 0$ to hold in these models \cite{RadProfs}, so that (from (\ref{Apr}) and (\ref{Phi}) applied to $\KK$, see also figure 1) we have $\KK_p\geq \KK_q\geq 0$, whereas for hyperbolic models or regions $\KK_p \leq \KK_q\leq 0$ holds when $\KK_q$ tends to zero (the only situation in which a change of sign of $\KK_p$ could occur). As a consequence, the sign of $\KK_q$ determines the sign of $\KK_p$, but not (necessarily) the sign of the local curvature $\KK$. This follows from the relation: 
\begin{equation}\KK=\KK_q \left[1+\Dk\right]= \KK_q \left[\frac{1}{3}+\frac{2/3+\Dik}{\Gamma}\right],\label{kvskq}\end{equation}
where we used (\ref{Dadef}) and (\ref{slaw_mk}), and $\Gamma$ is the scale factor defined in (\ref{LGdef}). Considering the regularity conditions (\ref{noshxGh}) and (\ref{kvskq}), we have $\KK_q<0\,\,\Rightarrow\,\, \KK<0$, so that the local spatial curvature is also negative everywhere in hyperbolic models, but the converse of this implication is not true: $\KK<0$ may hold in elliptic models or regions in which $\KK_q>0$ and $\KK_p>0$ hold everywhere.   

The implication $\KK_q>0\,\,\Rightarrow\,\, \KK>0$ is only valid for elliptic models in which  $\FF$ is monotonously decreasing (from (\ref{FFr})), and it necessarily leads to $\Dik>-2/3$ holding everywhere (it follows from (\ref{kvskq})). Since in this case $\QQ[r]<0$ holds (case (a) of lemma 2), then positive local spatial curvature implies effective deceleration. However, $\KK_q>0$ (and thus $\KK_p>0$) do not (necessarily) imply $\KK>0$ for elliptic models in which $\FF$ has a TV at $r=\rtv$ (case (b) of lemma 2), since (\ref{FFr}) implies $\Dik(\rtv)= -2/3$ and $\Dik<-2/3$ must hold for $r>\rtv$ (the regularity conditions (\ref{noshxGe}) do not place lower bounds on $\Dik$), hence (\ref{FFr}) and (\ref{kvskq}) imply that $\KK<\KK_q/3$ holds for $\Dik<-2/3$, and thus the local curvature $\KK$ becomes in these domains increasingly small and may become negative as $\KK_q>0$ decreases. Considering the asymptotic form (\ref{asconvk}) for elliptic models, (\ref{kvskq}) becomes asymptotically
\begin{equation}\KK \sim k_0\left(1-\frac{\beta}{3}\right)\,r^{-\beta},\qquad \beta\geq 2,\end{equation}
so that $\KK<0$ holds asymptotically if $\beta>3$, which is consistent with $\Dk<-1$ holding for $r>\rtv$, from (\ref{Dadef}), (\ref{FFr}) and (\ref{kvskq}).  In particular, it is possible to prove that $\KK<0$ necessarily holds in regions of an elliptic core matched to a hyperbolic exterior at some $r=r_b$. Since we have $\KK_q(r_b)=0$ with $\KK_q<0$ for $r>r_b$, and $\KK'_q(r_b)<0$, then (\ref{Apr}), (\ref{VVr}), (\ref{Ridef}) and (\ref{LGdef}) yield:
\begin{equation}\KK(r_b) = \frac{r_b}{3}\frac{\KK'_q(r_b)}{\Gamma(r_b)}<0,\end{equation}
so that local spatial curvature must be already negative in the outer regions of the elliptic core in which $\KK_q>0$ holds. The fact that spatial curvature can be negative in elliptic models is depicted graphically by figures 11a and 11b. 

While $\KK> 0$ may still hold in domains of elliptic regions or models compatible with $\QQ[r]>0$ (case (b) of lemma 2) if $\KK_q$ decays sufficiently slowly (as in (\ref{asconvk}) with $2< \beta<3$), we must necessarily have $\KK<0$ for $\hq<0$ to occur in elliptic models.  This follows from the parameters of the elliptic models converging to Schwarzschild examined in sections 9.1 and 10 (case (D)(ii)), which are the only elliptic models compatible with an effective acceleration (in domains in their intermediate and asymptotic radial range, see figure 7a and table 3). The conditions for $\hq<0$ in these models are precisely the same conditions in which $\KK<0$ holds, namely: a decay of $\KK_q$ with $\beta>3$. This is depicted graphically by comparing figures 7a and 11a--11b. As a consequence, $\hq<0$ only occurs in domains (hyperbolic or elliptic) if $\KK<0$ holds, even in elliptic models in which the averaged and quasi--local spatial curvature is positive.

As mentioned in various references \cite{buchrev,buchcar,buchert,buchlet}, the coupling between back--reaction and spatial curvature implies that the scaling law for $\kav[r]=\KK_p(r)$ in terms of the average scale factor $a_\DD^3=\Vp$ deviates from the FLRW scaling law $\KK\propto a^{-2}$, where $a$ is the FLRW scale factor. It is interesting to verify the dependence of $\KK_p$ on $a_\DD$ for LTB models that comply with $\hq<0$. Considering two classes of models admitting accelerating domains in the radial asymptotic range (section 9), we have:
\begin{itemize}
\item Elliptic models converging to Schwarzschild (case (D)(ii)): Since $\FF\to 1$, then $\Vp\propto r^3$, so that $a_\DD\propto r$. Hence, we have $\KK_p\propto a_\DD^{-\beta}$ with $\beta\geq 2$.  
\item Hyperbolic VD models with $\FF\to\infty$ (case (A)). Since $\FF\propto r^{1-\beta/2}$, with $0<\beta<2$, then $\Vp\propto r^{2+\beta/2}$ and $a_\DD\propto r^{(4+\beta)/6}$. Hence, we have $|\KK_p|\propto r^{-6\beta/(4+\beta)}$, with $0<6\beta/(4+\beta)<2$. 
\end{itemize}
Evidently, the average spatial curvature does deviate from the FLRW scaling law. In the elliptic case $\KK_p$ decays as a power law but faster than $a_\DD^{-2}$ (consistent with $\kav$ rapidly tending to zero), while in the hyperbolic case $|\KK_p|$ decays slower than $a_\DD^{-2}$. In both cases above (hyperbolic and elliptic) we obtain the same FLRW scaling law only in the case $\beta=2$, which corresponds to the asymptotic limit $\FF\to \Fty>1$ (limit (\ref{limHpHqF0})) for which $\hq>0$ holds (cases (B) in section 9, see tables 2 and 3), though back--reaction is positive (at least in the radial asymptotic range) for the hyperbolic models (see tables 1 and 2).  

\section{Final discussion.}

We have examined in a comprehensive manner the sufficient conditions for the existence of a positive back--reaction term (condition (\ref{Cpos})) and an ``effective'' acceleration (conditions (\ref{Apos}) and (\ref{newApos})) in the context of Buchert's averaging formalism for generic open LTB models admitting one symmetry center (see \cite{sussBR} for a discussion on ``closed'' models with spherical topology). We have looked at these conditions by evaluating the involved scalars at the boundary of arbitrary simply connected spherical comoving domains $\vartheta[r]$. It is important to remark that we have only examined sufficient conditions for the existence of an effective acceleration, hence we cannot rule out that such an acceleration may arise in cases not considered here.

We reviewed the following results proven rigorously in \cite{sussBR} (see lemmas 1 and 2 in section 5 and table 1): back--reaction is non--negative ($\QQ[r]\geq 0$) in the radial asymptotic range in all open models, except elliptic models in which $\FF$ is monotonous, so that spatial curvature decays asymptotically as $R^{-2}$. Following qualitative guidelines (section 6) describing specific conditions favoring the existence of an effective acceleration ($\hq<0$ or $\A>0$), we proved the existence of accelerating domains in various scenarios in sections 7--13. These scenarios are summarized in table 4 and the details identifying the corresponding models and regions are fully stated and justified in these sections.      
\begin{table}
\begin{center}
\begin{tabular} {|l|l|l|l|}
\hline
Scenario     & Type of models   & Numerical    & Reference  \\
             &            & estimate     &   \\ 
\hline
Central vacuum & Vacuum LTB region matched to&  $\hq\approx -0.03$     &  section 7.2 \\
               & a hyperbolic model with &                  &  figure 2 \\
               & a simultaneous bang and &                &   \\
               & converging to open FLRW &                &    \\
\hline
Schwarzschild  & Schwarzschild exterior of a dust&      not       & section 7.2\\
vacuum         & ball. Hyperbolic and elliptic& estimated   & figure 3     \\
               & models compatible with $\Q>0$&                &              \\
\hline
Central void   & Under dense non--vacuum central & $\hq \approx -0.003$ & section 8 \\
               & region of a hyperbolic model &                      &  figure 4  \\
\hline
Asymptotic  & Hyp VD and G  converging to& $\hq\approx -0.035$ & section 9 \\
radial range   &  vacuum LTB  with $\FF\to\infty$ &              & figure 5  \\
 \cline{2-4}
          & Hyp VD converging to   & $\hq\to 0$          &  section 9\\
          & vacuum LTB with $\FF\to\Fty>1$ &             &     \\
\cline{2-4}
          & Hyp converging to Milne  & $\hq\to 0$       &  section 9\\
\cline{2-4}
          & Ell converging to Schwarzschild  & $\hq\to 0$  &  section 9\\
\hline
Intermediate  & Hyp converging to open FLRW & $\hq \approx -0.1$ & section 10 \\
radial range   &                             &                    & figure 6a \\
\cline{2-4}
              & Hyp and Ell converging to EdS       & restrictive  & section 9  \\
              &                             &  & figure 6b \\
\cline{2-4}     
          & Hyp converging to vacuum LTB  & $\hq \approx -0.1$ & section 10\\
          & with $\FF\to\infty$           &                    &   figure 7a \\
\cline{2-4}
          & Hyp converging to vacuum Milne & not     & section 10 \\
          &                                & estimated & \\
\cline{2-4}
          & Ell converging to Schwarzschild & $\hq \approx -0.1$ & section 10  \\
          &                                 &                    &  figure 7b  \\
\hline
Near                 & Hyp converging to vacuum LTB  & $\hq \approx -0.5$ & sections 11,12 \\
non                  & with $\FF\to\infty$ and open FLRW  &               &  figures 9,10 \\
\cline{2-4}
-simultaneous        & Ell converging to Schwarzschild   &  not       & section 11\\
big bang             &                                   &  estimated &          \\ 
\hline
\end{tabular}
\caption{{\bf {Scenarios with effective acceleration}}. Further detail and information on the characteristics of these scenarios are found in the appropriate sections and figures. ``Hyp'' and ``Ell'' stand for hyperbolic and elliptic, ``EdS'' is short for Einstein de Sitter, ``open'' FLRW refers to negatively curved FLRW, \, ``VD'' and ``G'' correspond to the classification given by (\ref{x}) and ``TV'' stands for a turning value (see section 5). The numeric estimates are the maximal values found in the examples.}
\end{center}
\label{tabla4}
\end{table}
\begin{figure}[htbp]
\begin{center}
\includegraphics[width=3.5in]{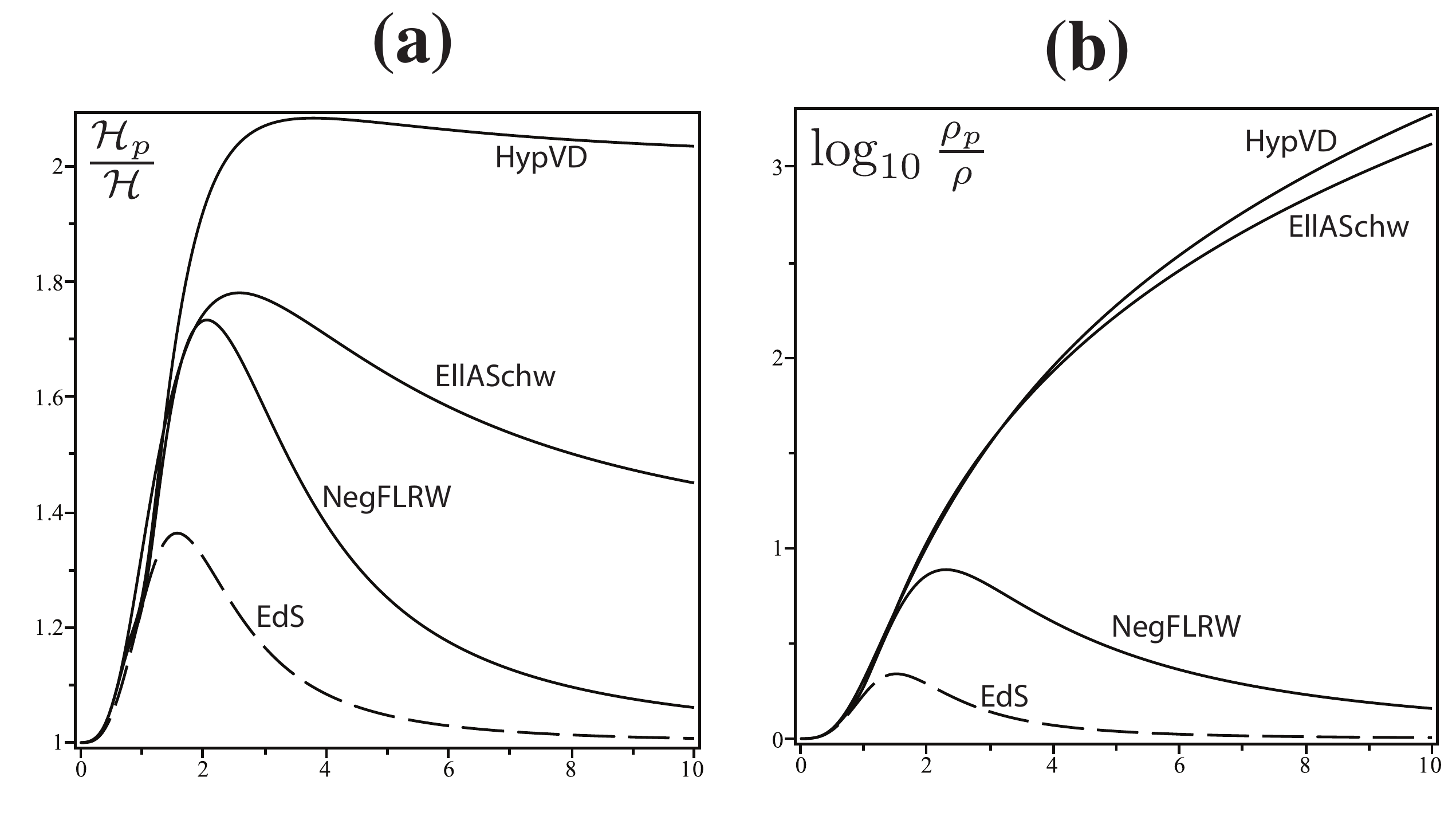}
\caption{{\bf  Averaged vs local Hubble rate and density.} Panel (a) depicts the ratio $\HH_p/\HH$ as a function of $r$ along the slice $t=0$ for the hyperbolic models asymptotic to negatively curved FLRW of figure 6a (NegFLRW), asymptotic to Einstein de Sitter of figure 6b (EdS), VD asymptotic to vacuum LTB of figure 7a (HypVD) and the elliptic model asymptotic to Schwarzschild  of figure 7b (EllASchw). Panel (b) depicts the ratio $\log_{10} \rho_p/\rho$ for the same models. Notice that the models HypVD and EllASchw exhibit the largest values in both ratios, denoting the largest deviation from FLRW conditions. On the other hand, the least deviation occurs for the EdS model for which $\HH_p/\HH\approx 1$ and $\rho_p/\rho\approx 1$ holds for all $r$, while the NegFLRW model deviates from FLRW conditions only in the intermediate radial range. }
\label{f12}
\end{center}
\end{figure}

It is evident, from looking at sections 7--13 and tables 1--4, that the conditions for an effective acceleration are far more restrictive than the conditions for a positive back--reaction (section 5).  While condition (\ref{FFr}) is necessary for both $\QQ[r]\geq 0$ and $\hq\leq 0$, the sufficient conditions for a positive back--reaction are far more relaxed than the sufficient conditions for an effective acceleration. Practically all domains in the asymptotic radial range of hyperbolic models and of elliptic models in which (\ref{FFr}) holds comply with $\QQ[r]\geq 0$, but, as shown in the various scenarios examined in sections 7--12 (compare table 1 with tables 2--4), an effective acceleration exists only in a restricted subset of these models. In general, the conditions that (on top of $\QQ[r]\geq 0$) are most favorable for an effective acceleration are a combination of negative  spatial curvature (hyperbolic models), regions with low density (or vacuum) and large gradients of density and spatial curvature, with $\FF\to\infty$ being a key factor in the radial asymptotic range. 

As argued in section 14, even if effective acceleration may also arise in domains where the averaged spatial curvature is positive (the intermediate and asymptotic radial range of elliptic models converging to Schwarzschild, case (D)(ii) in section 9), the local spatial curvature in these domains is necessarily negative (see figures 7a and 11a--b). 

While accelerating domains exist in vacuum and non--vacuum central voids (sections 7.2 and 8, figures 2 and 4), a density void profile ($\rho'_q\geq 0,\,\Dm\geq 0$), as used in most applications of LTB models to fitting observations,  is not necessary at all, as accelerating domains also exist when density has a clump profile ($\rho'_q\leq 0,\,\Dm\leq 0$), provided a low density region  occurs from a sufficiently fast decay of the over--density in the mid--range and asymptotic range, with the growth of back--reaction mostly driven by fluctuations (gradients) of spatial curvature (see sections 6, 7, 9 and 10). In fact, the estimated numerical values of $\hq$ are more negative in the intermediate and asymptotic range associated with these clump profiles (figures 5--7) than in the ``inner'' void profiles of central vacuum or low density regions (figures 2 and 4).

While all models radially converging to FLRW decelerate in their radial asymptotic range (section 7.1), the hyperbolic models and hyperbolic exteriors with this asymptotic convergence are compatible with scenarios in which there is effective acceleration, either in a central void (section 8, figure 4) or in the intermediate radial range (section 10, figure 6). The conditions for this are more favorable if the asymptotic FLRW state has negative curvature, and this fact is consistent with the existence of accelerating domains in the models examined numerically by Chuang {\it {et al}} in \cite{LTBave3} that have this radially asymptotic convergence.             

The restrictive conditions for accelerating domains in models asymptotically convergent to Einstein de Sitter (consistent with the results of \cite{ras4,ParSin4,ParSin3}) have been interpreted as a signal that back--reaction effects should be negligible when ``realistic'' models are considered \cite{ParSin4,Wald,zibin}. However, the contention that only an Einstein de Sitter background is ``realistic'' has been criticized on the grounds that it imposes the assumptions of the standard treatment of cosmic structure of the concordance model (quasi--Newtonian perturbative approach and asymptotic spatial flatness \cite{ParSin4,Wald}) on configurations in which such assumptions may not be applicable (see \cite{buchrev,buchcar,ellisR2,antiWald,ras5}). In particular, from an observational point of view, asymptotic spatial flatness is not (necessarily) required to fit supernovae and CMB constraints under a non--perturbative approach to inhomogeneity (see \cite{BW09,ClRe2010}). While the relation between averaging inhomogeneities and the fitting of observations is not well understood and requires further examination (see \cite{wiltshireF,ellisR1,ellisR2,kolbetal,buchfocus,BRobs}), the results of a perturbative approach (as considered in \cite{ParSin4,Wald}) seem to be insufficient to justify a forceful conclusion that non--linear effects associated with back--reaction should be negligible.

It is still an open question whether models fitting cosmic observations are also compatible with accelerating domains (in the context of the present work). Bolejko and Andersson \cite{BolAnd} have looked at this issue by means of specific examples: two of their models admit accelerating domains (models 5b and 8) but fail to satisfy age constraints (these two models have shell crossings and are not asymptotic to a negatively curved FLRW state). However, the results of these authors are based on a reduced sample of models and do not rule out the possibility that fitting observations can be compatible with models exhibiting mid--range accelerating domains and converging to a negatively curved FLRW state (as described in section 10). 

Some of the results proved here directly contradict previously published results. In particular, our proof (section 13) that all domains in the asymptotic time range of non--vacuum hyperbolic models necessarily decelerate proves that the condition for the existence of ``late times'' accelerating domains (found by Paranjape and Singh \cite{LTBave2}) is mistaken. We have also shown (sections 11 and 12) that the existence of effective acceleration is very restrictive for domains in collapsing configurations, which  disproves an earlier claim by R\"as\"anen in \cite{ras2,ras3} that gravitational collapse favors this acceleration (this was also proven to be mistaken in  \cite{ParSin3,ras4,mattsson}).          

We have obtained numerical estimates of the value of the deceleration parameter $\hq$ defined in (\ref{newApos}) (see table 4 and figures 2, 4, 5, 6, 7 and 10).  It is important to clarify that the main aim of this article is to study the conditions for the existence of accelerating domains, not to obtain ``realistic'' estimates for this acceleration. Therefore, we emphasize that the parameters used in these numerical examples (described in the figure captions) were selected in order to convey a rough comparative inference of the order of magnitude of typical values of the effective acceleration in the various scenarios that were examined.  Since the value of the effective acceleration is given by $\qef$ defined by (\ref{qef}), not by $\hq$ in (\ref{newApos}), the obtained values of $\hq$ should be considered as an approximated estimation of the effective acceleration. Considering domains in the radial asymptotic range of the VD hyperbolic models of case (A) in section 9, the definitions of $\qef$ and $\hq$ in (\ref{qef}) and (\ref{newApos}) yield the ratio
\begin{equation}\frac{\hq}{\qef}=\frac{\A}{\ACal}\frac{\HH_p^2}{\HH_q^2}\approx \frac{\A}{\A_p}\left(1+\xi\right)^2,\end{equation}
where $\A$ is defined in (\ref{Apos}) and $\xi$ is the constant appearing in (\ref{xias}) with $n=\beta/2$ and $0<\beta<2$, so that $\HH_p^2/\HH_q^2 \approx 1+ \hbox{O}(10^{-2})$. However, $\A_p>\A$ (considering from (\ref{Phi}) that $\A'<0$ and $\A'\to 0$ must hold as $r\to\infty$) and $\A_p$ and $\A$ must have the same asymptotic limit (from (\ref{A0})). Hence, the numeric values of $\qef$ and $\hq$ in the asymptotic radial range are for these VD models of the same order of magnitude.

Excluding domains near a non--simultaneous big bang, where values up to $\hq =-0.5$ were found (figure 10a), the numerical values of $\hq$ in the other scenarios range between $-0.003$ and $-0.1$, and thus are roughly consistent with estimates in a perturbative context (see \cite{buchrev} and references quoted therein). While these values are small, they imply a significant difference between local and averaged scalars, which is an immediate consequence of inhomogeneity. As shown by the curves plotted in figures 12a--b, the existence of an effective acceleration is a good indicator of how much (and in which radial range) the averaged and local basic scalars of a given LTB model deviate from each other, and thus deviate from FLRW conditions. This can be appreciated by comparing  the averaged and the local densities and Hubble rates (the latter approximately given by the averaged expansion scalar $\HHav[r]=\HH_p(r)$ and its local counterpart $\HH$). As shown in figures 12a and 12b displaying the ratios $\HH_p/\HH$ and $\log_{10} \rho_p/\rho$ as functions of $r$, the largest deviation of $\HH_p$ and $\rho_p$ with respect to $\HH$ and $\rho$ occurs precisely in the scenarios favoring $\hq<0$. In  the hyperbolic VD model case (A) and the elliptic model asymptotic to Schwarzschild in section 9 the averaged Hubble rate can be up to 75\% (elliptic model) and 100 \% (hyperbolic model) larger than the local rate, while for both models the average density decays at a much slower rate than local density (both decay to zero). For the hyperbolic models converging to a negatively curved FLRW and Einstein de Sitter backgrounds the averaged Hubble rate can be up to 70\% and 40\% larger than the local rate and the average density can be up to 10 and 3 times larger than the local density, with these values taking place precisely in the intermediate radial range where $\hq<0$ occurs (compare the curves for these models in figures 12a--b with the range where $\hq<0$ holds in figure 6a). Evidently, a comparison of all configurations implies that an asymptotic radial convergence to an Einstein de Sitter background yields a lesser deviation from FLRW conditions in all domains.           

The possibility that the back--reaction contribution $\OmC$ in the equation for the deceleration parameter (\ref{newApos}) could explain dark energy by ``competing'' with the value $\Omega^\Lambda\approx 0.7$ of the cosmological constant in the concordance model was suggested by Kolb {\it et al} \cite{kolbetal}. However, this proposal has been criticized in \cite{buchrev,buchlet} by arguing that the condition $\OmC\approx  \Omega^\Lambda$ cannot be reconciled with standard structure formation scenarios based on plausible initial conditions for an inhomogeneous cold dark matter source in a FLRW background (for further discussion on this issue, see \cite{wiegand}). This criticism is consistent with the fact that the only scenario in which we found a sufficiently large effective acceleration ($\hq \approx -0.5$) is in domains near the big bang (figure 10a), but such domains would be associated with early universe conditions that cannot be accommodated into structure formation scenarios in such a cosmological context (they imply an excessively large decaying mode that cannot be reconciled with the accepted early universe theories and phenomenology). Other scenarios compatible with $\hq<0$ (models of section 8 and figure 4 or those of section 10 and figure 6) are not (at least in principle) incompatible with standard structure formation scenarios in a FLRW background, but the maximal values $\hq\approx -0.1$ are too small to explain dark energy. In fact, in all models converging to a FLRW state we have $\hq\to\hqty>0$ in the asymptotic radial range, and thus $\hqty$ can be associated with the positive deceleration factor $q_0$ of the FLRW dust background. 

On the other hand, the models converging to a vacuum LTB state (see sections 9 and 10) do not converge asymptotically to any FLRW background, and thus $\qef$ and $\hq$ cannot converge to the observable $q_0$ of some FLRW model. The same remark applies to the relation between the Hubble rates $\HH$ and $\HH_p$ and the FLRW Hubble factor, $H_0$, though  averages of observational parameters should follow from an averaging procedure in the the light cone \cite{voids3}, not along the rest frames (the slices $\T[t]$).   

In fact, models converging to a vacuum LTB state for which we found $\hqty\approx -0.035$ or $\hqty= 0$ negatively (cases (A) and (D) of section 9) can be understood in terms of the notion of ``globally stationary'' spacetimes intruduced by Buchert in \cite{buchrev,buchlet,buchlet2}, and defined by the ``global stationarity conditions'' given by $\qef =0$ applied to domains encompassing a whole slice $\T[t]$ (assumed compact). Since $\hq=0\;\;\Rightarrow\;\; \qef=0$, then the models of case (D) in section 9 can be regarded as exact spherically symmetric (but non--compact) realizations of such class of spacetimes, with models of case (A) being approximate realizations (since $0.035\ll 1$). As shown in section 14, the scaling law of average spatial curvature in terms of the average scale factor $a_\DD$ significantly deviates from the FLRW scaling law $a_\DD^{-2}$ (it is a power law $a_\DD^{-n}$ with $0<n<2$), though it does not coincide with the scaling reported in \cite{buchlet} (its equation 15), which likely occurs because the slices $\T[t]$ in these LTB models are not compact.  

Another interpretation for models asymptotic to Minkowski (either asymptotic to Milne or to vacuum LTB or elliptic models asymptotic to Schwarzschild) follows from the notion of ``finite infinity'' (``fi''), introduced by Ellis \cite{fi1} and considered more recently by Wiltshire \cite{wiltshire1,wiltshireF}, which defines a scale in which cosmological structures approximate bound systems that are asymptotically flat. Under this interpretation, the Minkowskian asymptotic vacuum field of these models can be considered as an approximation to a large cosmic void surrounding a spherically symmetric structure, and approximately describing the conditions in an intermediate scale between the galactic cluster scale and a cosmological FLRW far field that emerges statistically by considering ensembles of similar bound structures and voids \cite{wiltshire1} (see also \cite{ellisR1,ellisR2,clifton}). The existence of an effective acceleration associated with an effective deceleration parameter $\qef \approx 0$ in the intermediate ``fi'' scale could provide important information on the dynamics of larger scales encompassing ensembles of bound structures. 

While the results obtained in this article are restricted by the spherical symmetry of the LTB models, it is plausible that they may apply, or at least provide important theoretical clues, in more realistic models that depart from this symmetry.  Whether considering possible extensions of the present article to non--spherical models (for example the quasi--spherical Szekeres solutions \cite{szeknew}), or if we remain working within the framework of LTB models, it is necessary to obtain numerically the precise values of an effective acceleration in domains of models that best fit inhomogeneities of astrophysical interest, as well as exploring the connection between this acceleration and the fitting of observational data. Evidently, the results of the present article provide an effective theoretical guideline for further work along these lines. This work is currently under elaboration, and will be submitted in the near future.

\begin{appendix}

\section{}

\subsection{An initial value formulation for LTB models.}

Considering the radial coordinate gauge
\footnote{This gauge freedom to set the radial coordinate follows from the fact that the metric (\ref{ltb}) is invariant under an arbitrary rescaling $r=r(\bar r)$.  This specific coordinate choice is adequate for open models (hyperbolic and elliptic) for which $R_i$ can be any monotonous function such that $R_i(0)=0$.}  
\begin{equation} R_0 = R(t_0,r)=r,\label{Ridef}\end{equation}
where the subindex ${}_0$ denotes evaluation at an arbitrary fiducial slice $t=t_0$. The conventional LTB metric (\ref{ltb}) can be parametrized as the following FLRW--like form:
\begin{equation} \fl\dd s^2=-\dd t^2+a^2\left[\frac{\Gamma^2\,\dd r^2}{1-\KK_{q0}\,r^2}+r^2\left(\dd\theta^2+\sin^2\theta\dd \phi^2\right)\right].\label{ltb2}\end{equation}
where we have rephrased the metric functions $R$ and $R'$ in (\ref{ltb}) as dimensionless scale factors  
\begin{equation} a \equiv \frac{R}{R_0}=\frac{R}{r},\qquad
 \Gamma = \frac{R'/R}{R'_0/R_0}=1+\frac{r\,a'}{a},\label{LGdef}\end{equation}
so that $a(t,r)=0$ and $\Gamma(t,r)=0$ respectively mark the central and shell crossing  singularities. The conventional free parameters $M$ and $\FF$ in (\ref{qvars}) can be expressed in terms of initial value functions as
\begin{equation} M=\frac{4\pi}{3}\rho_{q0}\,r^3,\qquad  \FF=[1-\KK_{q0}\,r^2]^{1/2}\label{MF},\end{equation}
while the ``bang time'' is given in terms of $\rho_{q0},\,\KK_{q0}$ (see (\ref{tbb})). Since $M=M(r)$ and $\FF=\FF(r)$, then (\ref{qvars}), (\ref{Dadef}) and (\ref{Hqsq}) lead to the scaling laws 
\ba \fl \rho_q=\frac{\rho_{q0}}{a^3},\quad \KK_q=\frac{\KK_{q0}}{a^2},\quad 1+\Dm =\frac{1+\Dim}{\Gamma},\quad \frac{2}{3}+\Dk = \frac{2/3+\Dik}{\Gamma},\label{slaw_mk}\\
\fl \HH_q^2=\frac{\dot a^2}{a^2}=\frac{8\pi\rho_{q0}}{3\,a^3}-\frac{\KK_{q0}}{a^2},\quad\Rightarrow\quad \dot a^2 = \frac{8\pi\rho_{q0}}{3\,a}-\KK_{q0},\label{slaw_H}
\ea
with similar scaling laws for $\hOm,\,\Dh$ following from (\ref{slaw_Dh}) and (\ref{Omdef}), and the local scalars $S=\{\rho,\,\KK,\,\HH\}$ given by $S=S_q(1+\Da)$. 

\subsection{Analytic solutions.}

For hyperbolic ($\KK_{q0}<0$) and elliptic ($\KK_{q0}>0$) models the scale factor $a$ can be obtained by the implicit solutions of the Friedman equation (\ref{slaw_H}) (equivalent to (\ref{eqR2t})):
\ba \fl \hbox{vacuum LTB:}\qquad\qquad\qquad a = 1+|\KK_{q0}|^{1/2}(t-t_0), \label{cthypv}\\
\fl \hbox{hyperbolic (non--vacuum):}\qquad y_0\, (t-t_0) = Z_h(x_0\,a) - Z_h(x_0),\label{cthyp}\\
 \fl\hbox{elliptic:}\qquad y_0\, (t-t_0)+Z_e(x_0)  = \left\{ \begin{array}{l}
 Z_e(x_0\,a) \qquad\qquad \hbox{expanding phase}\\ 
  \\ 
 2\pi-Z_e(x_0\,a) \qquad \hbox{collapsing phase}\\ 
 \end{array} \right.\label{ctell}\ea
where $x_0=|\KK_{q0}|/m_{q0},\,y_0=|\KK_{q0}|^{3/2}/m_{q0}$ with $m_{q0}\equiv (4\pi/3)\rho_{q0}$, and $Z_h$ and $Z_e$ are
\ba
  u  \mapsto Z_h(u)=u ^{1/2} \left( {2 + u } \right)^{1/2}  - \hbox{arccosh}(1 + u ),\label{hypZ1a}\\
 u  \mapsto  Z_e(u)= \arccos(1 - u )-u ^{1/2} \left( {2 - u } \right)^{1/2}.\label{ellZ1a} 
\ea
\subsection{Regularity conditions in terms of initial value functions}

The condition to avoid shell crossing singularities (which must hold for all $(t,r)$ for which $a>0$) is
\begin{equation}\fl \Gamma = 1+3(\Dim-\Dik)\left(1-\frac{\HH_q}{\HH_{q0}}\right)
-3\HH_q\,(t-t_0)\,\left(\Dim-\frac{3}{2}\Dik\right) >0,\label{Gamma}\end{equation}
which follows from the implicit derivation of (\ref{cthyp}) with respect to $r$ and using (\ref{ctell}) and (\ref{Dadef}) to obtain $\Gamma$ in terms of $a'$. The condition (\ref{Gamma}) leads to the Hellaby--Lake conditions  \cite{ltbstuff,suss10a,HLconds} that can be given analytically in the initial value formulation as restrictions on the initial value functions and their fluctuations \cite{RadAs,RadProfs,suss10a}:  
\ba\fl  \tbb'\leq 0,\quad \Dik+\frac{2}{3}\geq 0,\quad \Dim+1\geq 0,\qquad \hbox{Hyperbolic models or regions},\label{noshxGh}\\
\fl  \frac{\tbb'}{3R'_0/R_0}\leq 0,\quad  \frac{\tcoll'}{3R'_0/R_0}\geq 0,\quad \Dim+1\geq 0,\qquad \hbox{Elliptic models or regions},\label{noshxGe}\ea
where $\tbb'$ and  $\tcoll'$ follow by differentiating the bang time and collapse time functions
\ba \tbb = t_0-\frac{Z_h(x_0)}{y_0},\qquad \tbb = t_0-\frac{Z_e(x_0)}{y_0},\label{tbb} \\
\tcoll = \tbb+\frac{2\pi}{y_0}=t_0+\frac{2\pi-Z_e(x_0)}{y_0}.\label{tcoll}\ea
The coordinate surface $\Gamma=0$ marks a shell crossing singularity only in non--vacuum models. In the Schwarzschild--Kruskal and vacuum LTB models, it is a coordinate singularity made by caustics of worldlines of test observers (radial geodesics in the Schwarzschild--Kruskal models). The surface $a=0$ does mark a curvature singularity in the Schwarzschild--Kruskal models, but not in vacuum LTB models (since $\rho=0$). In terms of initial value functions, the regularity conditions for the Schwarzschild--Kruskal and vacuum LTB models are just (\ref{noshxGh}) and (\ref{noshxGe}) with $\Dim=-1$ and $\Dim=0$. As shown in \cite{RadAs}, the coordinate gauge (\ref{Ridef})  together with absence of shell crossings (\ref{Gamma}) is sufficient to guarantee that 
\begin{equation} R'>0\quad \hbox{and}\quad \FF>0, \label{RrFpos} \end{equation}
hold everywhere, hence a radial asymptotic range is well defined, as the radial coordinate is qualitatively analogous to the proper radial length along radial rays.

\section{Proof of equation (\ref{HHavp}).} 

We need to prove that $\langle\W\rangle[r]=\langle\W(\rr,r)\rangle=0$ for 
\begin{equation} \W(\rr,r) \equiv \left(\HH(\rr)-\HHav[r]\right)^2-\left(\HH(\rr)-\HH_p(\rr)\right)^2,\label{newC}\end{equation}
where $\HH_p(\rr)$ is the p--function associated with the proper volume average $\HHav[r]$. Expanding (\ref{newC}) and applying (\ref{avedef}) and (\ref{pqfun}) we obtain
\begin{equation} \langle\W\rangle[r]=-\HHav[r]^2+\frac{1}{\VV_p(r)}\int_0^r{[2\HH\HH_p-\HH_p^2]\,\VV_p'\,\dd\rr},\end{equation}
where we used the variance relation $\langle (\HH(\rr)-\HHav[r])^2\rangle[r]=\langle\HH^2\rangle[r]-\HHav[r]^2$.  
Inserting $\Theta=3\HH=\dot\VV_p'/\VV_p'$ and $\Theta_p=3\HH_p=\dot\VV_p/\VV_p$ in the integrand above, and bearing in mind that $\HHav$ and $\HH_p$ coincide at the domain boundary $x=r$, leads to the desired result:
\begin{equation} \fl \langle\W\rangle[r]=-\HHav^2[r]+\frac{1}{\VV_p(r)}\,\int_0^r{[\dot\VV_p^2/\VV_p]'\,\dd\rr}=-\HHav^2[r]+\HH_p^2(r)=0.\end{equation}
An analogous result follows for the quasi--local average acting on a scalar like $\W$ with $\langle\hskip 0.1 cm\rangle_q$ and $\HH_q$ instead of $\langle\hskip 0.1 cm\rangle$ and $\HH_p$.


\section{Proof of the limits (\ref{limHpHqF0}) and (\ref{limHpHq}).}

We prove these limits in full generality without making assumptions on the radial asymptotic convergence of the scalars in (\ref{newApos}) (other than $\{\FF,\,R\}\to\infty$ and $\{\HH_p,\,\HH_q\}\to 0$). 

\begin{description} 
\item[The limit (\ref{limHpHqF0}).] From (\ref{A0}) (see also \cite{RadAs}) we have $\HH_p\to 0$ and $\HH_q\to 0$ as $r\to\infty$ for models radially asymptotic to vacuum LTB. The definitions of $\HH_p$ and $\HH_q$ in (\ref{avedef}) and (\ref{aveqdef}) can be rewritten as
\begin{equation}  \fl \HH_p(r)=\frac{\int_0^r{\HH\,\FF^{-1}\Vq'\,\dd\rr}}{\int_0^r{\FF^{-1}\Vq'\,\dd\rr}},\qquad \HH_q(r)=\frac{\int_0^r{\HH\,\Vq'\,\dd\rr}}{\Vq(r)},\label{pq_map}\end{equation}
Since $\FF\to\FF_0$ as $r\to\infty$, then for every $\epsilon>0$ there is a value $\rr=y(\epsilon)$ in the integration range of (\ref{pq_map}) so that $[\FF_0+\epsilon]^{-1} <\FF^{-1}<[\FF_0-\epsilon]^{-1}$ holds for all $\rr>y$. Using (\ref{VVr}), and applying this constraint to the definitions (\ref{pq_map}), we obtain the inequality 
\ba\fl \frac{\FF_p(r)}{\FF_0+\epsilon}\left[\HH_q(r)-\HH_q(y)\frac{\Vq(y)}{\Vq(r)}\right]<\HH_p(r)-\HH_p(y)\frac{\Vp(y)}{\Vp(r)}\nonumber\\ 
\fl <  \frac{\FF_p(r)}{\FF_0-\epsilon}\left[\HH_q(r)-\HH_q(y)\frac{\Vq(y)}{\Vq(r)}\right].\label{FF01} \ea
As $r\to\infty$ we have $\Vp(r)\to\infty,\,\Vq(r)$ and $\FF_p(r)\to\FF_0$ (by Lemma 4), therefore, if we keep $y$ fixed and let $r\to\infty$ we get $\Vp(y)/\Vp(r)\to 0$ and $\Vq(y)/\Vq(r)\to 0$ in this limit, so that (\ref{FF01}) becomes
\begin{equation} \frac{\FF_0}{\FF_0+\epsilon}<\frac{\HH_p(r)}{\HH_q(r)}<\frac{\FF_0}{\FF_0-\epsilon}, \label{FF02}\end{equation}
Since $\epsilon$ can be arbitrarily small as $r$ grows, we obtain the result (\ref{limHpHqF0}) when $\FF$ tends asymptotically to a constant $\FF_0$. Notice that the converse of (\ref{limHpHqF0}) is not true, as we always obtain $\HH_p/\HH_q\to 1$ if $\HH\to \HH_0\ne 0$ without $\FF$ tending asymptotically to a constant. 
\item[The limit (\ref{limHpHq}).] Considering (\ref{VVr}), we have
\begin{equation}\frac{\HH_p(r)}{\HH_q(r)} = \FF_p(r)\,J(r),\qquad J(r) = \frac{\int_0^r{\HH\FF^{-1}\Vq' \dd\rr}}{\int_0^r{\HH\Vq' \dd\rr}}.\end{equation}
If $\FF\to\infty$, then it is straightforward to prove that $\FF_p\to\infty$, while the limit $\FF^{-1}\to 0$ implies that for every $\epsilon>0$ there is a value $x=y(\epsilon)$ such that $\FF^{-1}< \epsilon$. Hence,  $0<J(r)< \epsilon$ and (as long as $\FF\to\infty$) we have 
\begin{equation} J(r)< 1\qquad\hbox{and}\qquad \mathop {\lim }\limits_{r \to \infty } J(r) = 0.\label{limitJ}\end{equation}
Since $\FF_p> 1$ for all $r>0$ and it diverges as $r\to\infty$, then (\ref{limitJ}) implies that $\HH_p(r)/\HH_q(r)\to 1$ can only happen if $\FF_p^{-1}$ and $J(r)$ have the same rate of convergence to zero, which is not true in general (unless $\FF\to\Fty$ finite). Hence 
\begin{equation} \mathop {\lim }\limits_{r \to \infty } \FF_p(r)J(r) = \chi \ne 1,\end{equation}
but if we assume expanding models then $\HH\to 0$ as $r\to\infty$ implies $\HH'< 0$ in this limit, and thus we get from (\ref{Phi}) that $\HH_p(r)>\HH_q(r)$. Therefore, $\chi>1$ holds, leading to:
\begin{equation} \mathop {\lim }\limits_{r \to \infty } 1-\frac{\HH_p(r)}{\HH_q(r)} = 1-\chi =-\xi<0\qquad \xi>0,\label{D7}\end{equation}
which is the result given in (\ref{limHpHq}).
\end{description} 
The asymptotic form of the constant $\xi$ in the limit (\ref{limHpHq}) and the asymptotic value of $\hq$ can be estimated under the assumptions on asymptotic convergence given by the power law forms (\ref{asconvm})--(\ref{asconvk}). Since $\HH=\HH_q(1+\Dh)$, we assume the generic convergence law $\HH\sim \HH_0\,r^{-n}$, where the values of $n$ and $\HH_0$ depend on the asymptotic form of $\HH_q$ and $\Dh$ that follow from (\ref{asconvm})--(\ref{asconvk}). Evaluating the asymptotic forms of the integrals in the numerators of (\ref{avedef}) and (\ref{aveqdef}), we obtain after some tedious algebra:
\begin{equation} 1-\frac{\HH_p(r)}{\HH_q(r)} \to -\frac{n(2-\beta)}{3(4+\beta-2n)}<0,\label{HpHq2}\end{equation}
which coincides with (\ref{limHpHq}) and (\ref{D7}), since $\FF\to\infty$ in (\ref{asMF}) with $k_0<0$ requires $0<\beta<2$. Notice that the limit $1-\HH_p/\HH_q\to 0$ in the cases $\FF\to\Fty$ in (\ref{limHpHqF0}) follows by taking $\beta=2$ (see (\ref{asMF})), while the limit $1-\HH_p/\HH_q\to 0$ of models asymptotically FLRW and Milne is recovered if $n=0$. The constant $\xi$ in the limit (\ref{limHpHq}) takes the form
\begin{equation}  \xi =\frac{n(2-\beta)}{3(\beta-2n+4)},\quad n = \left\{ \begin{array}{l}
 \alpha/2,\quad \hbox{(MD models)} \\  
 \gamma/2,\quad \hbox{(G models)}  \\
 \beta/2,\quad \hbox{(VD models)}\\ 
 \end{array} \right.,\label{xias}\end{equation} 
where we used the asymptotic forms (\ref{asconvm})--(\ref{asconvk}).             
 
\end{appendix}

\section*{Acknowledgments}
  I acknowledge financial support form grant PAPIIT--DGAPA IN-119309. I am thankful to Thomas Buchert for useful suggestions and encouragement. 
  
\section*{References}

\end{document}